\documentclass[
aps,nofootinbib,
showpacs,showkeys,preprint
tightenlines,preprintnumbers,] {revtex4}

\usepackage{epsf,epsfig,subfigure,
graphicx,amsmath,amssymb}
\usepackage{color}
 
\newcommand{\dis}[1]{\begin{equation}\begin{split}#1\end{split}\end{equation}}

\newcommand{\gev}{\,\textrm{GeV}}

\newcommand{\eV}{\,\mathrm{eV}}
\newcommand{\fde}{f_{\rm DE}}
\newcommand{\Mp}{M_{\rm P}}
\newcommand{\ndw}{N_{\rm DW}}
\newcommand{\Mpt}{$M_{\rm P}$}
\newcommand{\Mgt}{$M_{\rm GUT}$}
\newcommand{\Mg}{M_{\rm GUT}}
\newcommand{\vew}{v_{\rm ew}}
\newcommand{\Ude}{U(1)$_{\rm de}$}

\newcommand{\UPQ}{U(1)$_{\rm PQ}$}
\newcommand{\ULA}{U(1)$_{\rm ULA}$}
\newcommand{\UG}{{U(1)$_{\Gamma}$}}

\newcommand{\ie}{{\it i.e.~}}
\newcommand{\etal}{{\it et al.}}

\newcommand{\Uanom}{U(1)$_{\rm anom}$}

\newcommand{\Qem}{Q_{\rm em}}
\newcommand{\Qanom}{Q_{\rm anom}}
 \newcommand{\cagg}{c_{a\gamma\gamma}}

\newcommand{\thb}{\overline{\theta}}
 \newcommand{\delq}{\delta_{\rm CKM}}
 \newcommand{\dell}{\delta_{\rm PMNS}}
 \newcommand{\LQCD}{\Lambda_{\rm QCD}}
    \newcommand{\NDW}{N_{\rm DW}}

\def\EE8{{\rm E_8\times E_8'}}

\def\Z{{\bf Z}}

\begin{document}

\title{Fate of global symmetries in the Universe: QCD axion, quintessential axion and trans-Planckian inflaton 
decay-constant 
}

\author{Jihn E. Kim$^{(1,2,3)}$, Soonkeon Nam$^{(3)}$, Yannis K. Semetzidis$^{(3)}$}
 
\address{$^{(1)}$Department of Physics and Astronomy, Seoul National University,
 Seoul 08826,  Republic of Korea,\\
 $^{(2)}$Center for Axion and Precision Physics Research (IBS),  
KAIST Munji Campus,
 Munjiro 193, Yuseong-Gu, Daejeon 34051, Republic of Korea,\\
 $^{(3)}$Department of Physics, Kyung Hee University, 26 Gyungheedaero, Dongdaemun-Gu, Seoul 02447,  Republic of Korea.
}
 
\begin{abstract}
Pseudoscalars appearing in particle physics are reviewd systematically. From the fundamental point of view at an ultra-violat completed theory, they can be light if they are realized as pseudo-Goldstone bosons of some spontaneously broken global symmetries. The spontaneous breaking scale is parametrized by the decay constant $f$.  The global symmetry is defined by the lowest order terms allowed in the effective theory consistent with the gauge symmetry in question. Since any global symmetry is known to be broken at least by quantum gravitational effects, all pseudoscalars should be massive. The mass scale is determined by $f$ and the explicit breaking terms $\Delta V$ in the effective potential and also anomaly terms  $\Delta\Lambda^4_G$ for some non-Abelian gauge groups $G$. The well-known example by non-Abelian gauge group   breaking  is the potential for the ``invisible'' QCD axion, via the Peccei-Quinn symmetry, which constitutes a major part of this review. Even if there is no breaking terms from gauge anomalies, there can be explicit breaking terms $\Delta V$ in the potential in which case  the leading term suppressed by $f$ determines the pseudoscalar mass scale. If the breaking term is extremely small and the decay constant is trans-Planckian, the corresponding pseudoscalar can be a candidate for a `quintessential axion'. In general, $(\Delta V)^{1/4}$ is considered to be  smaller than $f$, and hence the pseudo-Goldstone boson mass scales are considered to be smaller than the decay constants.  In such a case,   the potential of the pseudo-Goldstone boson at the grand unification scale is sufficiently flat near the top of the potential that it can be a good candidate for an inflationary model, which is known as  `natural inflation'. We review all these ideas in the bosonic collective motion framework. 

\keywords{Bosonic collective motion, Discrete symmetries,  ``Invisible'' axion, Domain walls,  String axion,  Quintessential axion, Natural inflation, Trans-Planckian decay constant.}
\end{abstract}

\pacs{14.80.Va, 11.15.Ex,  11.25.Wx, 11.30.Ly.}

\maketitle
  
\section{Introduction}

Spontaneously broken global symmetries have been proved extremely successful in understanding particle phenomena at the low energy  regime. The vacuum expectation value (VEV), $f$, of an effective scalar field $\Phi$ signals such a breaking.  Below $f$  the symmetry is realized in the Nambu--Goldstone manner \cite{NambuJona61, Gold61,Nambu60}. The breaking effect is very significant if the canonical dimension of the field $\Phi$ is 1, \ie for a fundamental spin 0 boson. So it is remarkable that the so-called Higgs boson has been discovered for the first time as a spin 0 fundamental scalar  \cite{HiggsDisc12,HiggsDiscCMS}. Therefore, it is very probable that a fundamental pseudoscalar boson can be present also in the Universe with its prominent cosmological role in the Universe.

 The astrophysical evidence in favor of the existence of DM has grown over the years \cite{PopoloA}.  If cold dark matter (CDM) has provided the dominant mass at the time when CDM fluctuation enters into the horizon  after inflation, the fluctuation scale given by that horizon scale is typically the scale of galaxies.   The time scale corresponding to this is about $3\times 10^{-5}$ times the time of $z\simeq10$, \ie at $z\simeq 3000$ which is close to the time when ``matter~=~radiation''.  Before the time of ``matter~=~radiation'' the density perturbation has grown only logarithmically.  After the time of ``matter~=~radiation'' the desity perturbation has grown linearly, which became nonlinear around $z=10$.  Since the time of $z=10$ on, `galaxy formation' has started. Thus, in the study of the Universe evolution, in particular in the N-body simulations, CDM attracted a great deal of attention since 1984 \cite{Blum84}.

The Planck data confirms the domination of the Universe energy by the invisibles:  dark energy (DE) 68\,\% and  dark matter (DM)  27\,\% \cite{Planck15}. It has been known that bosonic collective motions (BCMs) can describe both DE and  DM  \cite{KimYannShinji}, which are the main emphases of this review. Suggestion of DM first given by Zwicky in 1933 was from the observation of velocity dispersion of the galaxies in the Coma Cluster \cite{ZwickyF33,ZwickyF2}. The so-called rotation curves in the halo \cite{Rubin70} and the image of the collision of Bullet Cluster \cite{Bradley08} form the most convincing arguments for the existence of DM.   

In particle physics, the first example for CDM was introduced in an effort to constrain the mass scale of another (heavy) lepton doublet which interacts weakly \cite{HLCDM,LeeWein77}. This showed that the multi-GeV weakly interacting particles could have closed the Universe, and opened the idea of weakly interacting massive particles (WIMPs). The symmetry for most WIMPs is ``parity'' or ${\bf Z}_2$ symmetry in which the SM particles carry parity or ${\bf Z}_2$ even particles and the CDM candidate particle is the lightest one among  the parity or ${\bf Z}_2$ odd particles. This got tremendous interest in supersymmetric (SUSY) models where the R parity works for the needed symmetry \cite{Goldberg83}. The second  example for CDM was found \cite{Preskill83,AbbottSik83,DineFish83} for the ``invisible'' axion \cite{KSVZ1,KSVZ2,DFSZ,DFSZ2}, which belongs to the BCM scenario \cite{KimYannShinji}.  For a decade, the CDM dominated Universe seemed not contradicting the observation. However, the Type 1A supernova candle discovered in 1998 \cite{RiessA98,Perlmutter98} changed the CDM scenario completely such that the dark energy (DE) is dominant, composing 68\,\% of the energy of the Universe, and the CDM fraction is moved down to a mere 27\,\%.

A simple  interpretation of DE can be related to the cosmological constant, originally introduced by Einstein \cite{Einstein17}. But there has been a theoretical prejudice that the cosmological constant (CC) at the true vacuum must be zero, which is considered to be one of the most important problems in theoretical physics \cite{WeinbergRMP}. After the introduction of fundamental scalars in particle physics, the vacuum energy of scalar fields has been appreciated to contribute to the cosmological constant \cite{Veltman75}. Therefore, the problems related to the CC must be considered together with the vacuum energy of scalar fields.

A CC dominated universe is called the LeMa\^itre universe \cite{LeMaitre31}, where the scale factor expands exponentially. This idea of exponential expansion was used in early 1980's to solve the horizon and flatness problems and to dilute  monopoles (arising in grand unification (GUT) models \cite{PreskillMono}), which is now called  ``inflation'' \cite{Guth81}.
Inflation predicts that the total mass-energy density of the Universe is very close to the critical closure density.  The  Planck data \cite{Planck15} confirm
that the energy density of the Universe is nearly $\rho_c$ (spatially
flat) and that the present DE is about 68\% of the critical energy density of the Universe.
 
Under this circumstance, I present the BCM idea which can be applicable to all the problems mentioned above, CDM by the ``invisible'' QCD axion \cite{KSVZ1,KSVZ2,DFSZ}, DE by a quintessential axion \cite{KimNilles14,KimDE14},  the cusp/core problem by ultra light axions (ULAs) \cite{Hu00,MarshSilk14}, and inflation by ``natural'' inflation \cite{Freese90,KNP05}.

BCM is described by a scalar field.
The Brout--Englert--Higgs--Guralnik--Hagen--Kibble (BEHGHK) boson, simply called the Higgs boson, seems to be a fundamental scalar field. So, we can imagine that the QCD axion and an inflaton may be  fundamental fields also.  Compared to  spin-$\frac12$ fermions of the canonical dimension $\frac32$, these bosons with canonical dimension 1  can affect more importantly to low energy physics.
This has led to the Higgs boson acting as a portal to the high energy scale \cite{HiggsPortal,Patt06,Wells05}, to the axion scale or even to some standard model (SM) singlets in the   GUT  scale. Can these singlets explain both DE, CDM, and even inflation in the evolving Universe? In this short review of global symmetries, we attempt to answer  all these questions in the affirmative direction. 

In Sect. \ref{sec:Global}, the importance of spontaneously broken global symmetries and the resulting pseudo-Goldstone bosons are discussed. In particular, the so-called 't Hooft mechanism is discussed in Subsect. \ref{subsec:Hooft}. In Sect. \ref{sec:CP}, the CP symmetry and its violation in weak interaction is discussed. In Sect.\ref{sec:QCDaxions},  the CP violation in QCD  and its ``invisible'' axion solution is discussed.  Especally, the importance of the axion-photon-photon coupling toward discovery in the cavity type experiments is emphasized in  \ref{subsubsec:cagg}. In Sect.\ref{sec:CosAxion}, effects of  ``invisible'' axions in cosmology are presented. In Sect. \ref{sec:GlAndNonA}, possible non-Abelian groups related to axions are discussed.  
Cousins of  ``invisible'' axions are discussed in Sect. \ref{sec:axionde} for dark energy, in Sect.  \ref{sec:ULA} for the ULA, and in Sect. \ref{sec:Inflation} for gravity waves from natural inflation.
Section \ref{sec:DisConl} is a discussion on the content of this review and a conclusion.

\section{Global symmetries}\label{sec:Global}

For a BCM, we introduce a corresponding global symmetry. The global symmetry is that present in the Lagrangian.  However,
 global symmetries are known to be broken in general by the quantum gravity effects, especially via the Planck scale  wormholes \cite{QuantumGr,Gilbert89}.   To resolve this dilemma, we can think of two possibilities of discrete symmetries below \Mpt\, 
  in the top-down approach: (i) The discrete symmetry arises as  a part  of  a gauge symmetry \cite{KraussWil88,Ibanez92,Banks92,Wise91}, 
and (ii) The string selection rules directly give the discrete symmetry \cite{Kim13worm}.  As far as discrete symmetries are concerned, a bottom-up approach can be useful also \cite{KimPLB17}. The interesting cases are
the discrete gauge symmetries allowed in string compactification. Even though the Goldstone boson directions, \ie the longitudinal directions, of spontaneously broken {\it GAUGE} symmetries are flat, the Goldstone boson directions of spontaneously broken {\em GLOBAL} symmetries are not flat. Namely, global symmetries are always approximate. The question is how approximate it is. 
 
\begin{figure}[t]
\centerline{\includegraphics[width=6cm]{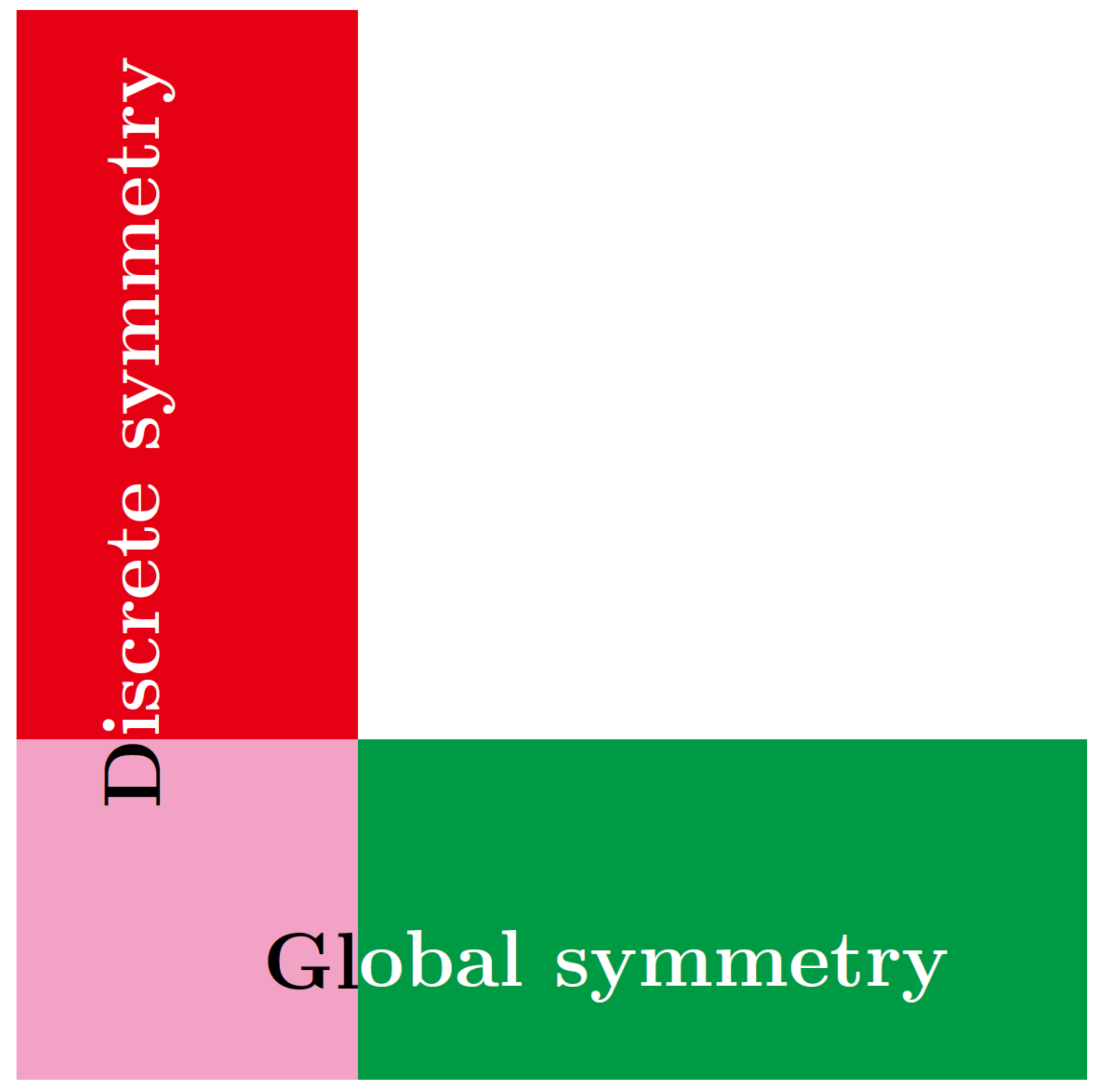}
} 
\caption{Terms respecting and breaking discrete and global symmetries. Terms in the red color break the global symmetry and terms in the green color break the discrete symmetry. }\label{Fig:discrete}
\end{figure}

In Fig. \ref{Fig:discrete}, we present a cartoon separating effective terms according to (string-allowed for example) discrete symmetries. The terms in the  vertical column represent exact symmetries such as gauge symmetries and string allowed discrete symmetries. If we consider a few terms in the lavender part, we can consider a {\em global symmetry}. With the global symmetry, we can consider the global symmetric terms which are in the lavender and green parts of Fig. \ref{Fig:discrete}. However, the global symmetry is broken by the terms in the red part in Fig. \ref{Fig:discrete}.

\subsection{The Nambu--Goldstone boson and the Brout--Englert-Higgs--Guralnik--Hagen--Kibble mechanism}\label{subsec:Goldstone}

\begin{figure}[t]
\centerline{\includegraphics[width=10cm]{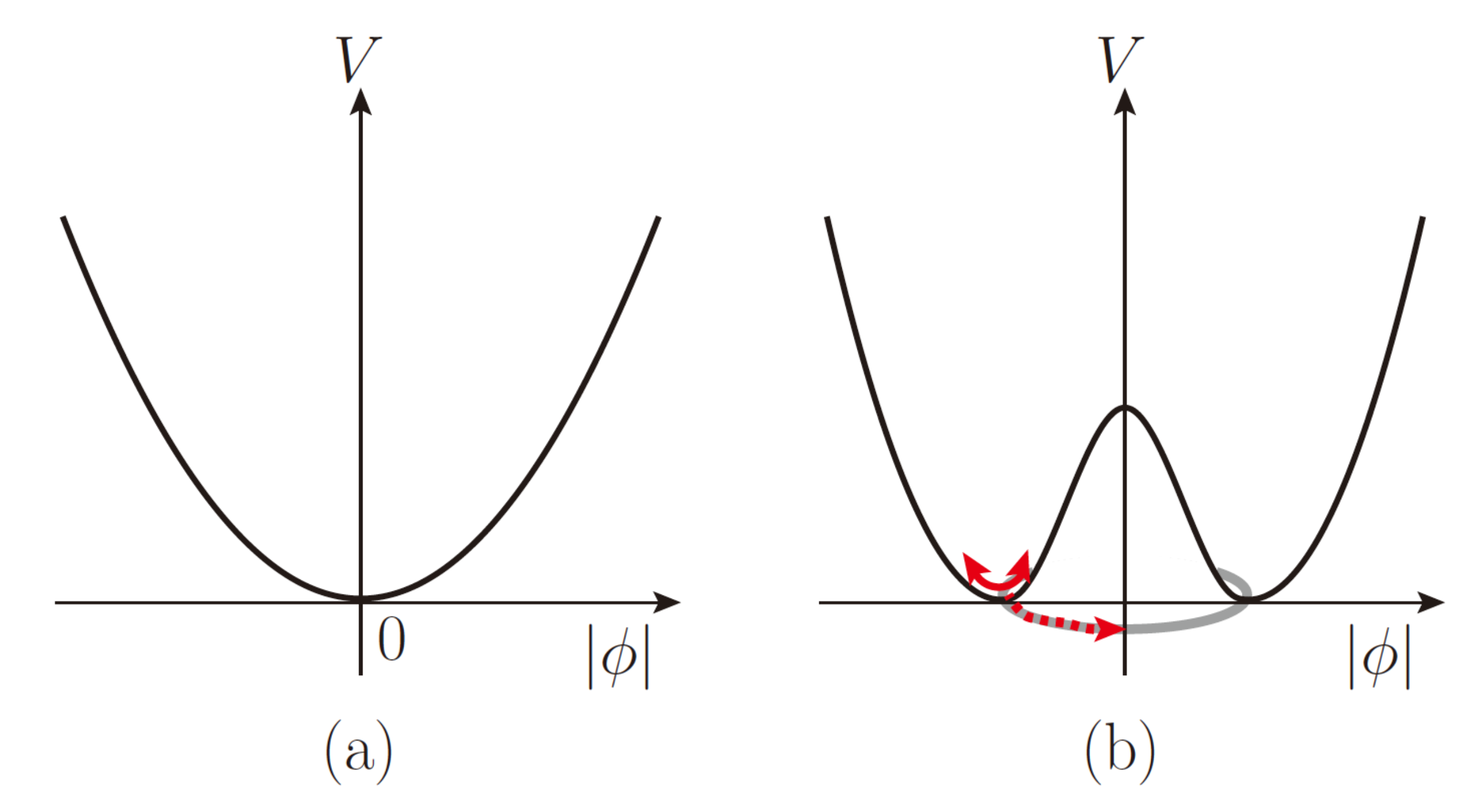}
 }
\caption{Two cases of $V(\phi^*\phi)$: (a) $\mu^2>0$, and (b) $\mu^2<0$.}\label{Fig:Goldstone}
\end{figure}

 Consider a U(1) global transformation on a complex field $\phi$,
\dis{
\phi\to e^{i\alpha Q}\phi \label{eq:U1Trans}
}
where $Q$ is the charge generator of U(1) symmetry. Spontaneous symmetry breaking of the global symmetry is concisely studied by the Goldstone Lagrangian,
\dis{
{\cal L}=(\partial_\mu\phi^*)(\partial^\mu\phi) -V(\phi^*\phi) \label{eq:LagGlobal}
}
where the vacua of
\dis{
V(\phi^*\phi)= \lambda  (\phi^*\phi)^2+ \mu^2 \phi^*\phi,
}
was studied by Goldstone \cite{Gold61} for a global U(1) and used by Higgs \cite{Higgs64} for a gauged U(1) simply by changing $\partial_\mu$ of (\ref{eq:LagGlobal}) to $\partial_\mu\to D_\mu=\partial_\mu+({\rm gauge~boson~term})$. In both cases of Goldstone \cite{Gold61} and Higgs \cite{Higgs64}, the potential is the same as shown in Fig. \ref{Fig:Goldstone}, (a) for $\mu^2>0$ and (b) for $\mu^2<0$. For $\mu^2>0$, the minimum of the potential is unique and the vacuum is a singlet.  For $\mu^2<0$, the minima of the potential are degenerate as shown in Fig.  \ref{Fig:Goldstone}\,(b) for the value 
\dis{\langle \phi\rangle=\frac{v}{\sqrt2}e^{i\alpha}, ~~{\rm with}~~v=\sqrt{\frac{-\mu^2}{\lambda}}
} 
where  $\alpha$ is a continuous parameter denoting the degeneracy. The oscillating directions of two real fields of $\phi$ are shown in  Fig.  \ref{Fig:Goldstone}\,(b) as the radial oscillation and the valley shown as gray.  The oscillation along the dashed arrow is flat and there is no potential barrier along this direction. This mode represented by $\alpha$ is massless. This massless mode is the Goldstone boson in case of spontaneously broken {\it global} U(1) \cite{Gold61}, and  in case of spontaneously broken {\it gauge} U(1) it provides the longitudinal mode of the gauge boson \cite{Higgs64}. In both cases, (a) and  (b), the numbers of continuous degrees are the same.

\subsection{The 't Hooft mechanism}\label{subsec:Hooft}

Let us now consider two phase directions. The case in this subsection is different from  that of Subsec. \ref{subsec:Goldstone} in that both of two degrees are phases here while only one is the phase in  Subsec. \ref{subsec:Goldstone}. Two phases accompany two generators $Q_1$ and $Q_2$ for which we can consider three cases,
\begin{itemize} 
\item[(i)] both  $Q_1$ and $Q_2$ are generators of two global symmetries,
\item[(ii)]   both  $Q_1$ and $Q_2$ are generators of two gauge symmetries,
\item[(iii)]  one is a gauge generator and the other is a global generator.
 \end{itemize}
 
In the case (i), if both global symmetries are spontaneously broken, there result  two Goldstone bosons. If one global symmetry is spontaneously broken and the other unbroken, there results only one Goldstone boson. 
 
In the case (ii), if both gauge symmetries are spontaneously broken,  two gauge bosons obtain masses. If one gauge symmetry is spontaneously broken and the other unbroken, only one gauge boson obtains mass and the other remains massless, which is pointed out by Kibble \cite{Kibble67} that has been used in the SM with $Z$ and photon. 

For the case (iii), there arises the 't Hooft mechanism \cite{Hooft71} which  is discussed below. It is a very simple and elementary concept, but it seems that it is not known widely in the community. For the case (iii), if one Higgs vacuum expectation value (VEV)  breaks two continuous symmetries then there results the  't Hooft mechanism. It is obvious that the gauge symmetry is broken because the corresponding gauge boson obtains mass.  Namely, only one phase or pseudoscalar is absorbed to the gauge boson, and there remains one continuous direction. To see this clearly, let us introduce a field $\phi$ on which charges $Q_{\rm gauge}$ and  $Q_{\rm global}$ act. The gauge transformation parameter is a local $\alpha(x)$ and the global transformation parameter is a constant $\beta$. Transformations are
\begin{eqnarray}
 \phi\to e^{i\alpha(x) Q_{\rm gauge}} e^{i\beta Q_{\rm global}}\phi,
\end{eqnarray}
which can be rewritten as
\begin{eqnarray}
 \phi\to e^{i(\alpha(x)+\beta) Q_{\rm gauge}} e^{i\beta (Q_{\rm global}-Q_{\rm gauge} )}\phi.
\end{eqnarray}
Redefining the local direction as $\alpha'(x)=\alpha(x)+\beta$, we obtain the transformation
\begin{eqnarray}
 \phi\to e^{i \alpha'(x) Q_{\rm gauge}} e^{i\beta (Q_{\rm global}-Q_{\rm gauge} )}\phi.
\end{eqnarray}
So, the $\alpha'(x)$ direction becomes the longitudinal mode of heavy gauge boson. 
Now, the charge $Q_{\rm global}-Q_{\rm gauge}$ is reinterpreted as the new global charge and is not broken by the VEV, $\langle\phi\rangle$, because out of two continuous directions one should remain unbroken. Basically the direction $\beta$ remains as the unbroken continuous direction.
 This  is  the essence of the 't Hooft mechanism: ``If both a gauge symmetry and a global symmetry are broken by one scalar VEV, the gauge symmetry is broken and a global symmetry survives''.  The resulting global charge is a linear combination of the original gauge and global charges as shown above. 
This theorem has a profound effect in obtaining the intermediate scale of the ``invisible'' axion from string compactification \cite{Kim88}.

\subsection{Breaking scales}
 
The red part potential in  Fig. \ref{Fig:discrete} breaks the global symmetry shown as the horizonal bar. In some cases, the global symmetry is broken by the anomalies of non-Abelian gauge groups. The well-known example is the Peccei-Quinn (PQ) global symmetry, broken by the quantum chromodynamics (QCD) anomaly, the {\UPQ}--SU(3)$_c$--SU(3)$_c$ anomaly.  The PQ proposal assumes more in that any term is not present in the red part of Fig.  \ref{Fig:discrete}. Suppose a global symmetry \UG. If \UG~is spontaneously broken by a VEV, \ie by a SM singlet VEV $\langle \sigma\rangle\equiv f/\sqrt2$, the pseudo-Goldstone boson mass corresponding to the spontaneously broken  \UG~is
\begin{eqnarray}
m_{\rm pseudo}^2=\frac{(\rm symmetry~breaking~
scale)^4}{f^2}.
\end{eqnarray}

If the non-Abelian anomalies are the sole contribution in  breaking the  global symmetry, the $m_{\rm pseudo}\simeq \Lambda^2/f$ where $\Lambda$ is the scale of the non-Abelian gauge group. Depending on the scale of $\Lambda$, the pseudo-Goldstone is called
\begin{eqnarray}
\begin{array}{lll} 
 \textrm{Name} \qquad & \qquad  \Lambda & m_{\rm pseudo} \\[0.6em]
\textrm{QCD axion}\qquad  &\rm   \sim 380\, MeV\qquad &  10^{-4}\textrm{ eV} \\[0.4em]
\textrm{KNP inflaton}\qquad &\rm 10^{16}-10^{17}\, GeV\qquad &  10^{14-15}\textrm{ GeV} \\[0.4em]
\textrm{ULA}  \qquad &\rm  3\times10^{-7}\,  GeV\qquad &  10^{-22}\textrm{ eV} 
\end{array}\label{eq:LambdaPseudo}
\end{eqnarray}
For the case of ULA in Eq. (\ref{eq:LambdaPseudo}), the SU(2)$_W$ can work for the non-Abelian gauge group if one accepts a few orders of magnitude discrepancy.
 
Even if there is no non-Abelian anomalies,  the global symmetry \UG~is broken  by the terms in the red part of  Fig. \ref{Fig:discrete}. Then, the $m_{\rm pseudo}\simeq \varepsilon^2/f$ where $\varepsilon^4$ in $V$ is the leading term in the red part. Assuming that there is no non-Abelian anomalies,  the pseudo-Goldstone boson mass is estimated as
\begin{eqnarray}
\begin{array}{lll} 
 \textrm{Name} &\qquad \varepsilon \qquad & m_{\rm pseudo} \\[0.6em]
\textrm{ALPs}\qquad  &\rm   Any ~value\qquad &  \textrm{any value} \\[0.4em]
\textrm{N-flaton}\qquad &\rm 10^{16} \, GeV\qquad &  10^{13}\textrm{ GeV} \\[0.4em]
\textrm{ULA}  \qquad &\rm  3\times10^{-7}\,  GeV\qquad &  10^{-22}\textrm{ eV} 
\\[0.4em]
\textrm{Quintessential axion}  \qquad & \rm  1.7\times10^{-12}\,  GeV\qquad & 10^{-33}\textrm{ eV} 
\end{array}\label{eq:VPseudo}
\end{eqnarray}

The most interesting BCM is the QCD axion of Eq. (\ref{eq:LambdaPseudo}) which will be discussed in detail in the following section as a solution of the strong CP problem and as a candidate for CDM in the Universe. A BCM for inflation can arise with non-Abelian anomalies of Eq. (\ref{eq:LambdaPseudo}) \cite{KNP05}
or without non-Abelian anomalies corresponding to Eq. (\ref{eq:VPseudo}) \cite{Nflation06}. A quintessential axion for the DE scale should not have the QCD anomaly and the height of the potential must be very small. Also, the ultra-light axion (ULA) should have a very small height of the potential and most probably it is in the form of Eq. (\ref{eq:VPseudo}).
But, the ULA is also included in Eq. (\ref{eq:LambdaPseudo}) because SU(2)$_W$ anomaly can provide such a small height within a few orders. Axion-like particles (ALPs) behave like axions in the detection experiments, \ie having the pseudoscalar--photon--photon couplings but their masses and decay constants are not related as in the QCD axion. Most probably, they arise in the form of  Eq. (\ref{eq:VPseudo}).
 
\begin{figure}[t!]
\centerline{\includegraphics[width=11cm]{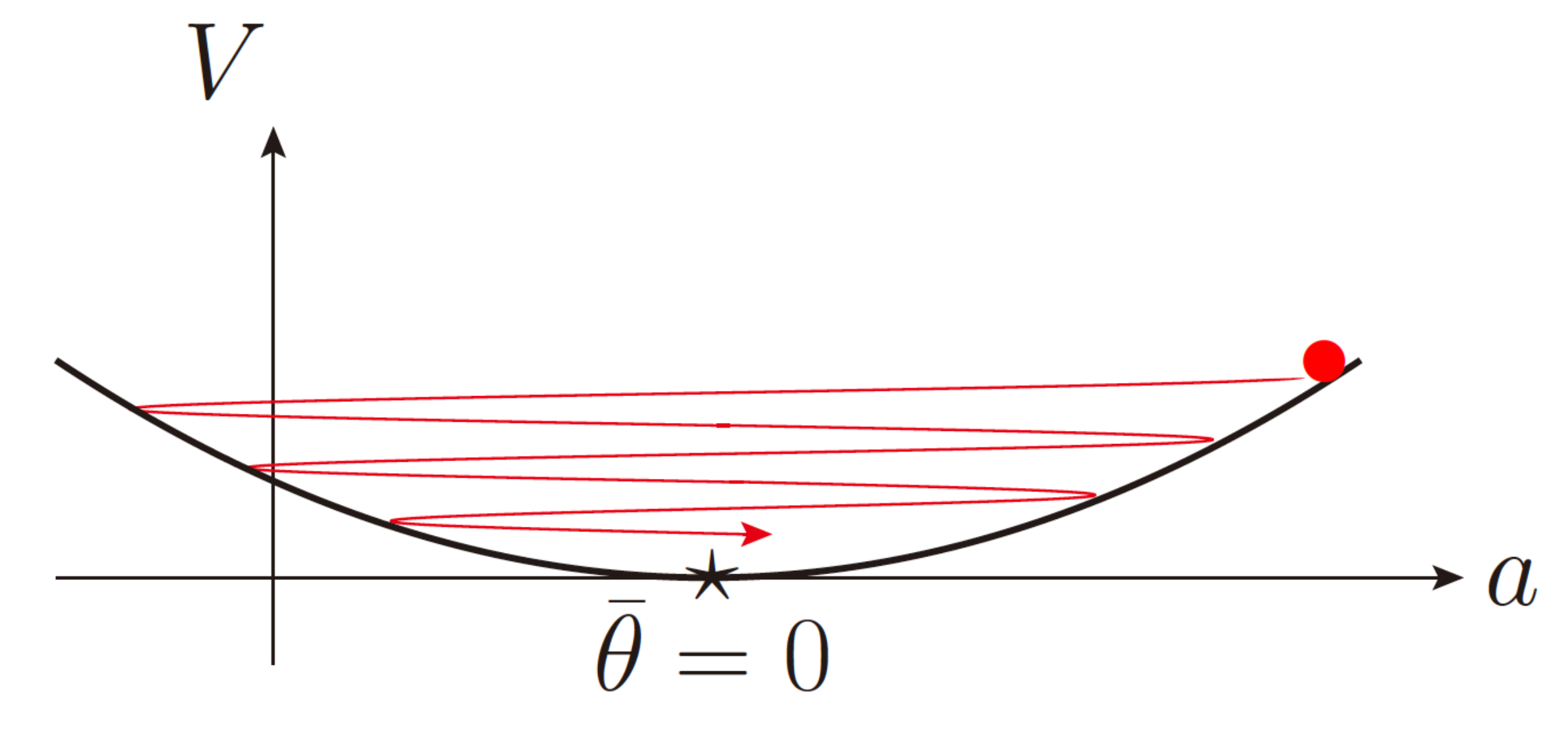}
 }
\caption{Rolling of $\thb$ on the almost flat axion potential.}\label{Fig:PotAxion}
\end{figure}

All pseudoscalar bosons listed in Eqs.  (\ref{eq:LambdaPseudo})  and (\ref{eq:VPseudo}) use the idea of collective/coherent motion of the classical part of the pseudoscalar field. The classical potential can be visualized as a very shallow potential $V$ as shown in  Fig. \ref{Fig:PotAxion}. The classical vacuum shown as the red bullet starts to rolls down the hill ``late'' in the history of the Universe. The amplitude decreases (as illustrated with the red curve) due to the Hubble expansion.  At present, the amplitude is non-zero, which is the basis for attempting to detect QCD axions \cite{KimYannShinji}. The names listed  in Eqs.  (\ref{eq:LambdaPseudo})  and (\ref{eq:VPseudo})  are made to represent the origin or usefulness of the corresponding pseudoscalar boson.

\subsection{Global symmetries inspired by string-compactification}\label{subsec:GlFromSt}
It is generally known that 10 dimensional (10D) string theories do not allow global symmetries. Gauge fields fulfilling the local transformations are 10D graviton $g_{MN}$, anti-symmetric tensor $B_{MN}$ and gauge potential $A_M$. Upon compactification of 10D down to 4D, with the compactified six internal spaces, some of these fields can exhibit global symmetries. In the compactification of the heterotic string based on $\EE8$ and SO(32), the model-independent (MI) axion becomes the global shift direction \cite{WittenMI},
\dis{
a_{\rm MI}\to a_{\rm MI}+ {\rm constant}.
}
In the compactification of type I and type II string models, even 3 global symmetries are obtained \cite{Ibanez99}.  Model-dependent (MD) axions can be used for some 4D global symmetry directions \cite{WittenMD}, which however is known to be approximate \cite{WenWitten86}. 

\section{CP violations}\label{sec:CP}

The CP invariance is guaranteed if the Lagrangian remains the same even after the operation of CP,
\dis{
CP:~~{\rm (CP)\,}{\cal L}\,{\rm (CP)}^{-1}={\cal L}.\label{eq:defCP}
}
In this definition, the transformation of the field operators under CP are included. So, if we find any one set on the definition of CP on the field operators that satisfies Eq. (\ref{eq:defCP}), the CP is the invariance of the Lagrangian. Here, the phase redefinitions of the complex field operators are of course included. General CP transformations on the fields has been presented recently based on the   classification of {\it discrete} groups \cite{Ratz14}: 
(i) (discrete) group-theoretic origin of the complexity of some Clebsch--Gordan coefficients, (ii) groups that allow all the real Clebsch--Gordan coefficients, and (iii) groups that do not admit real Clebsch--Gordan coefficients but possess a class-inverting automorphism that can be used to define a proper (generalized) CP transformation. The most frequently used case is (ii) on which we focus here. In this case, the CP transformation is   defined as 
\dis{
\phi_i(x)\stackrel{CP}\longrightarrow  X_{ij}\phi_j^*({\cal P}x),~~{\rm with}~{\cal P}(t, {\bf x})= (t, -{\bf x}),
}
where $X$ is a unitary matrix.

\subsection{Models for the weak CP violation}\label{subsec:WeakCP}

With this understanding, let us consider the weak CP violations of Kobayashi and Maskawa \cite{KM73} and   Weinberg  \cite{Weinberg76}. We can list  the following possibilities for the weak CP violation,
\begin{itemize}
\item[(i)]  by light  colored scalar,
\item[(ii)]  by right-handed current(s),
\item[(iii)]  by three left-handed families,
\item[(iv)]  by propagators of light color-singlet scalars,
\item[(v)]  by an extra U(1) gauge interaction.
 \end{itemize}
The first three are those of Kobayashi and Maskawa(KM) \cite{KM73}, among which the second was presented earlier by Mohapatra also \cite{Mohapatra72}.  But this second model has the flavor-changing neutral current problem. The third one is the so-called KM model.
 
In 1976, the third quark family was not known in which case Weinberg introduced the weak CP  violation in the Higgs potential with multi Higgs doublets \cite{Weinberg76},
\begin{eqnarray}
&V_{\rm W} =  -\frac12\sum_{I} m^2_{I}\phi^\dagger_I\phi_I+\frac14\sum_{IJ}\big[a_{IJ} \phi^\dagger_I\phi_I
\phi^\dagger_J\phi_J +b_{IJ}  \phi^\dagger_I \phi_J
\phi^\dagger_J  \phi_I &\nonumber \\[0.5em]
 & +c_{IJ} \phi^\dagger_I\phi_J
\phi^\dagger_I\phi_J\big]+{\rm H.c.},\quad\quad   &
\label{eq:WeinV}
\end{eqnarray}
where the reflection symmetry $\phi_I\to-\phi_I$ is imposed. Equation (\ref{eq:WeinV}) is a good example for illustrating Fig. \ref{Fig:discrete}. The discrete symmetry implied by the lavender part of Fig.  \ref{Fig:discrete} is this reflection symmetry for which fermions possess the appropriate transformation properties which we do not list here. 
The mass parameters $m^2_{I}$ are at the electroweak scale such that the electroweak symmetry is broken at the electroweak scale. With the potential (\ref{eq:WeinV}), three Higgs doublets are needed to introduce CP violation \cite{Weinberg76}. Not to introduce flavor changing neutral currents, Weinberg \cite{Weinberg76} required that only one Higgs doublet, $\phi_1$, couples to $\Qem=-\frac13$ quarks, and another Higgs doublet, $\phi_2$, couples to $\Qem=+\frac23$ quarks \cite{GW77}. The above reflection symmetry achieves  this goal. For all scalars and pseudoscalars of $\phi_I$ to obtain mass, all parameters in (\ref{eq:WeinV}) are required to be nonzero. Right after this potential was known, Peccei and Quinn discovered a global symmetry if the terms $c_{IJ}$ are excluded \cite{PQ77}. It corresponds to the case that the vertical reds of Fig. \ref{Fig:discrete} are excluded.

\subsection{CP violation and global symmetries}\label{subsec:CPaGlobal}

The relation between discrete and global symmetries are already pointed out in Fig. \ref{Fig:discrete}. Considering the gauge and Yukawa interactions, the global symmetry of the SM in the limit of vanishing neutrino masses is U(1)$_B\times$U(1)$_e\times$U(1)$_
\mu\times$U(1)$_\tau$ where $B$ is the baryon number,  and $L$ are the lepton numbers, $L=e,\mu,\tau$. If we include the Weinberg operator \cite{Weinberg79}
\dis{
{\cal L}_{\nu~\rm mass}\propto~ \ell^T{C}^{-1}\ell\, H_u^2\label{eq:WeinNu}
}
where $\ell$ are lepton doublets and $H_u$ is a Higgs doublet with $Y=+\frac12$, the the global symmetry is reduced to U(1)$_B$. If we allow all terms in the  vertical-column of Fig. \ref{Fig:discrete}, $B$ is also broken. Suppose, we  keep only a few terms symbolized by the lavender color. Then, there can be a global symmetry shown as the horizontal-column in Fig. \ref{Fig:discrete}. The PQ {\it global} symmetry commented at the end of Subsec. \ref{subsec:WeakCP} belongs here. Even if the CP invariance is not imposed in weak interactions (allowing the KM CP violation), the quark Yukawa couplings of the Glashow--Weinberg type and the Higgs potential assumed at the end of Subsec. \ref{subsec:WeakCP} show a global symmetry at the tree level. Being a global symmetry, it is destined to be broken at some level, in this case by the QCD anomaly. Another example is Eq. (\ref{eq:WeinNu}) which breaks the lepton numbers $L_{e,\mu,\tau}$ and renders the SM neutrinos mass. The discrete symmetry we implied in Fig.  \ref{Fig:discrete} can be considered as the one allowed in models from string compactification.

Suppose, we impose a CP invariance in the theory instead of the general discrete symmetry allowed in string compactifications. In this case also, only a few terms symbolized by the lavender color may exhibit a global symmetry. The calculable models, in the solutions of the strong CP problem starting with the CP invariance of the Lagrangian, may create such a global symmetry as the example shown by Georgi, Nelson and Shin \cite{Georgi85}.

If a CP phase is related to a global symmetry, the phase can be dynamically determined in theories where the phase is represented by a pseudoscalar field. The PQ  global symmetry was introduced in relation to the strong CP problem and the phase $\thb$ is determined at 0 \cite{VW84}, which is the case where a non-Abelian anomaly breaks the global symmetry.  If the potential $\Delta V$ in the red part of  Fig.  \ref{Fig:discrete} breaks the global symmetry, the determined phase depends on which discrete symmetry is used in the vertical column.

In the deterministic principle, the weak CP violation is introduced by the spontaneous breaking mechanism \cite{LeeTD73} such that   $\delta_{\rm CKM}$ is dynamically determined as the phase of a VEV-generating scalar field.\footnote{In the example of  Georgi, Nelson and Shin, the CKM phase was determined at $\frac{\pi}{2}$ \cite{Georgi85}, but an ansatz was used there.} The CP phases  introduced in the Type-I \cite{FY86} and Type-II \cite{CoviII16} leptogenses can be also determined.  In contrast to the one parameter  $\thb$ in QCD, weak interactions introduce many parameters, which makes it difficult to obtain  $\delta_{\rm CKM}$ in simple forms.  

\subsection{Magnitude of the weak CP violation}\label{subsec:JarlsJ}

\begin{figure}[t!]
\centerline{\includegraphics[width=6cm]{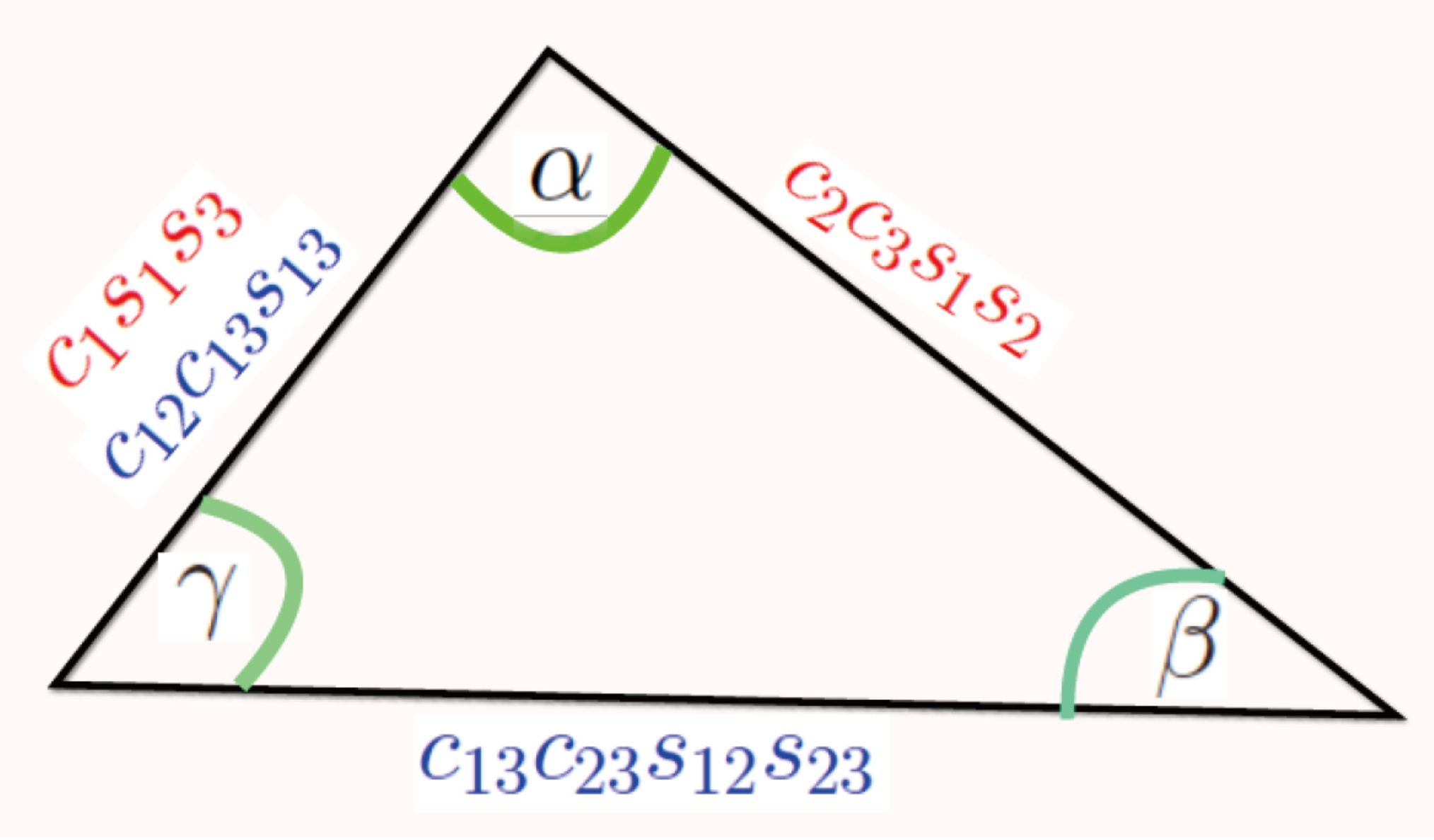}
 }
\caption{The unitarity triangle for a $B$ meson decay to a strange meson plus pions, with $i=b$ and $j=d$ in Eq. (\ref{eq:Unitarity}).}\label{Fig:Jarlskog}
\end{figure}

As shown above, even if the CPT symmetry is preserved in QFT, in the SM any of its subgroups is not guaranteed to be preserved. Some continuous symmetry can guarantee/break the invariance of some discrete subgroup of CPT. The parity invariance, for example, is broken by the chiral nature of the fermion representations in the SM. For CP, phase transformations of independent complex fields are the continuous symmetries we imply. We note again that CP violation is an interference phenomenon. In the SM with three families, there appears an unremovable CP phase $\delq$. So, if we consider only two families, there is no observable effects on the weak CP violation. Also, if two quark masses are degenerate, there is no observable effects. So, the observable CP violation effects are products of quark mass differences: $ (m_u-m_c)(m_c-m_t)(m_t-m_u)  (m_d-m_s)(m_s-m_b)(m_b-m_d)$. The phase symmetry is a continuous symmetry we consider. For this symmetry, the preserved discrete symmetry is the CP symmetrry. The invariant magnitude for the weak CP violation is denoted by the product of quark mass differences and the Jarlskog determinant $J$ \cite{JarlsPRL85}, which is defined from the unitarity of $V$,
\begin{eqnarray}
(V^\dagger V)_{ik}=\sum_j V^*_{ji}V_{jk}=\delta_{ik}.\label{eq:Unitarity}
\end{eqnarray}
A typical Jarlskog triangle for a $B$ meson decay is shown in Fig. \ref{Fig:Jarlskog}, and $J$ is two times the area of the triangle.

For examples to be presented concretely here, let us choose the quark mixing matrix such that the first row is real \cite{KimBrazil},
\begin{eqnarray}
&&V_{\rm KS}= \left(\begin{array}{ccc} c_1,&s_1c_3,&s_1s_3 \\ [0.2em]
 -c_2s_1,&e^{-i\delq}s_2s_3 +c_1c_2c_3,&-e^{-i\delq} s_2c_3+c_1c_2s_3\\[0.2em]
-e^{i\delq} s_1s_2,&-c_2s_3 +c_1s_2c_3 e^{i\delq},& c_2c_3 +c_1s_2s_3 e^{i\delq}
\end{array}\right),\label{eq:KSexact}\\[0.4em]
&&{\rm  with~Det\,}V_{\rm KS}=1, 
\nonumber
\end{eqnarray}
where $c_i=\cos\theta_i$ and $s_i=\sin\theta_i$ for $i=1,2,3$. In this form, the invariant quantity for the CP violation, the Jarlskog determinant is directly seen from $V_{\rm KS}$ itself \cite{KimBrazil},\footnote{For the PMNS matrix, we use different parameters by replacing $\theta_i\to\Theta_i$ and $\delq\to \dell$.}
\begin{eqnarray}
J=|{\rm Im}\,V_{13} V_{22} V_{31} |\simeq O(\lambda^6),\label{eq:JfromV}
\end{eqnarray}
where  $\lambda\simeq 0.22$ is the Cabibbo parameter $\lambda=\sin\theta_C$ \cite{Cabibbo63}, which has been used as a small expansion parameter \cite{Wolfenstein83}. Note that all three families participate in the evaluartion of $J$, fulfilling the claim that CP violation is an ``interference phenomenon''.  
   
There are many different parametrization schemes.  Different parametrizations give different CP phases $\delq$ \cite{KimBrazil}. This can be understood by the Jarlskog triangle for B-meson decay, shown in Fig. \ref{Fig:Jarlskog}.
For the parametrization of Eq. (\ref{eq:KSexact}), $J$ with the red parameters for the sides is given by
\begin{equation}
J=c_1 c_2c_3 s_1^2 s_2 s_3 \sin\alpha,
\end{equation}
while the PDG parametrization \cite{PDG17} with the blue parameters for the sides gives $J=c_{12}c_{31}^2c_{23}s_{12} s_{23}s_{13}\sin\gamma$. 

\begin{figure}[!t]
\centerline{\includegraphics[width=6cm]{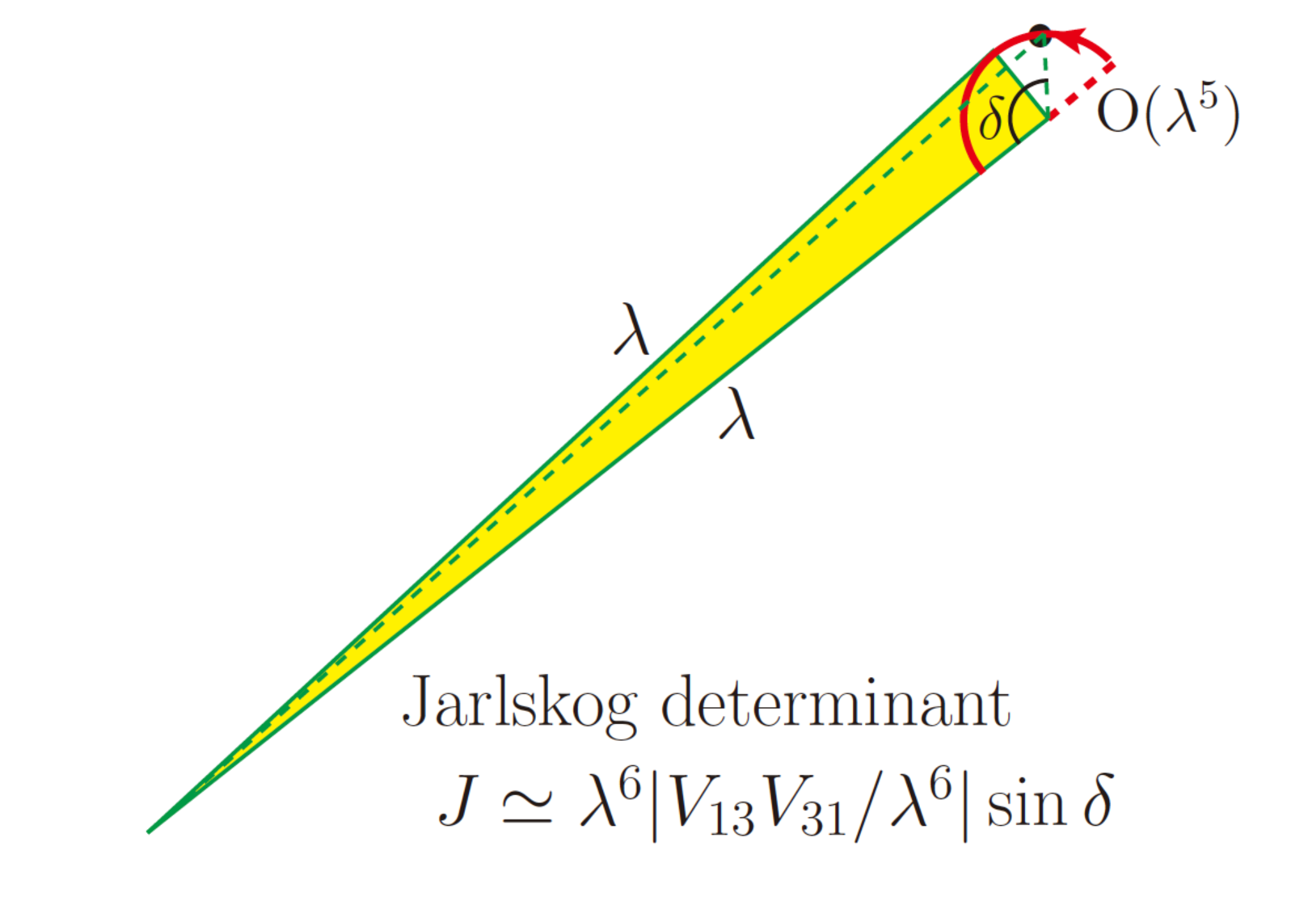}
 }
\caption{Same as Fig. \ref{Fig:Jarlskog} but without a strange meson in the final state.}\label{Fig:Long}
\end{figure}
 
The invariant nature of the weak CP violation is that the areas of any triangle are the same, \ie the areas of Figs. \ref{Fig:Jarlskog} and \ref{Fig:Long} are the same.  So, consider another Jarlskog triangle shown in Fig. \ref{Fig:Long}. Each length of the sides of the triangle is a product from two elements of the CKM matrix $V: V_{ji}^*V_{jk} $($j$ not summed). Note that there is one common angle in all these triangles, \ie $\alpha,\beta,$ or $\gamma$ of Fig. \ref{Fig:Jarlskog} and $\delta$ (any other angle) of Fig. \ref{Fig:Long}. The essence of weak CP violation in the quark sector is given by this area.   Since $\lambda$ is known to be small,  by rotating the side of O$(\lambda^5)$ in Fig. \ref{Fig:Long}, one notes that the maximal area occurs when $\delta\simeq \frac{\pi}{2}$. Going back to Fig. \ref{Fig:Jarlskog}, the knowledge on $\alpha,\beta,$ and $\gamma$ can pin down which parameter is more useful to use. With the Chau--Keung--Maiani (CKM) parametrization \cite{CKMpara84,Maiani76,Maiani77}, colored blue in Fig. \ref{Fig:Jarlskog}, $\gamma$ is the invariant angle $\delq$.  With the Kim--Seo (KS) parametrization, colored red in Fig. \ref{Fig:Jarlskog}, $\alpha$ is the invariant angle $\delq$. In fact, one can use any parametrization and can calculate $\alpha,\beta,$ and $\gamma$ by
\begin{eqnarray}
 &\alpha={\rm Arg.}\left(-\frac{V_{td}V_{tb}^*}{V_{ud}V_{ub^*}}
 \right),\nonumber\\[0.5em]
&\beta={\rm Arg.}\left(-\frac{V_{cd}V_{cb}^*}{V_{td}V_{tb^*}}
\right),\label{eq:betagamma}
\\[0.5em]
&\gamma={\rm Arg.}\left(-\frac{V_{ud}V_{ub}^*}{V_{cd}V_{cb^*}} \right).\nonumber
\end{eqnarray}
In general, the phase appearing in the quark mixing matrix is not $\alpha,\beta,$ or $\gamma$. But, the phase in the CKM matrix can be chosen as one of  $\alpha,\beta,$ or $\gamma$ if Det.$V_{\rm CKM}$ is made real, as presented by Kim and Seo \cite{KimBrazil}. With the phase $\delta'$ in the KM parametrization \cite{KM73}, we calculate $\alpha$ from Eq. (\ref{eq:betagamma}), 
\begin{eqnarray}
\alpha\simeq\pi-\delta'.
 \end{eqnarray}
Thus, the phase $\delta'$ appearing in the matrix of the original KM paper can be chosen as the invariant phase $\delq$ which can be $\alpha$ up to $\pi$.  In the quark sector alone without any assumptions, any parametrization is allowed.

\subsection{Determination of CP phases}

The CP phases in the weak interactions of quarks and leptons are derived from the Yukawa couplings. In Subsec. \ref{subsec:Inv}, it will be shown that the strong CP phase $\thb$ is determined as 0. This will be shown possible because $\thb$ will be promoted to a dynamical field and the point $\thb=0$  is chosen as the minimum of the  vacuum energy of this one dynamical field. This leads to a question, ``Can one determine the weak CP phases $\delq$ and $\dell$?'' If we attempt to determine them dynamically, similarly they must be made dynamical fields participating in weak interactions.   To determine $\delq$ and  $\dell$, these phases are promoted to dynamical pseudoscalar fields,
\begin{eqnarray}
a_q=f_{\rm quark} \delq,~~{\rm and/or}~~a_l=f_{\rm lepton} \dell.
\end{eqnarray}
Note that if the CP phase is determined to be 0 or $\pi$, there is no observable CP violation. Therefore, the exact discrete symmetry in the lavender part of Fig. \ref{Fig:discrete} cannot be $\Z_2$ which would determine the phase as 0 or $\pi$.  The exact discrete symmetry in the lavender part of Fig. \ref{Fig:discrete} implies, from the hermiticity of the potential with a $\Z_N$ symmetry,
\begin{eqnarray}
V= V_0 +\Delta V=\frac12(A+A^*)\propto\pm  \cos\frac{\delta}{N},\label{eq:formV}
\end{eqnarray}
where $\Delta V$ is the contribution from the red part of  Fig. \ref{Fig:discrete}. If + sign is realized, $\delta=0$ is a maximum of the potential. But if --  sign is realized, $\delta=0$ is a minimum of the potential. A non-zero $\delta$ for the weak CP violation is a cosmological choice.  Some examples can be found in References \cite{Georgi85,
KimPLB11,KimNamEP15,CoviII16}.

Determining  $\delta_{\rm CKM}$ belongs to the problem on the texture of quark mass matrix. Similarly, determining  $\dell$ belongs to the problem on texture of lepton mass matrices, including the heavy leptons. But determinination of $\delq$ and $\dell$ is much more involved than the determination of $\thb$.   Determination of $\thb$ is  simple because there is only one class of parameters among the anomalous terms \cite{Bardeen69}. 
But, three quark families, for example, introduce  many parameters, and determination of $\delta_{\rm CKM}$ by minimizing an effective potential must consider the ``texture'' of quark mass matrix.  A hope toward this direction is to parametrize the CKM matrix in the form of Ref. \cite{KimSeo11} such that  $\delta_{\rm CKM}$ is the phase in the CKM matrix itself, which may help finding out the needed texture. Note that Eq. (\ref{eq:formV}) involves an unknown phase $\delta$ which can be present if one allows complex Yukawa couplings. So, to determine $\delq$, one should consider only real Yukawa couplings, \ie the Lagrangian possesses the CP invariance. Then, the needed CP phase $\delq$ is determined by the spontaneous breaking mechanism of the weak CP invariance \cite{LeeTD73}. 

\section{QCD axions}\label{sec:QCDaxions}
 
 The modern theory of strong interactions is the SU(3)$_c$ gauge theory QCD. Gluons $G^a_\mu\,(a=1,2,\cdots,8)$ interact as
 \begin{eqnarray}
 {\cal L}= \frac{-1}{2g_3^2}\,{\rm Tr}\,G_{\mu\nu} G^{\mu\nu}+\sum_{i=\rm quarks}\,\bar{q}^{\,i}  \gamma^\mu(\partial_\mu+i G^a_\mu T^a)q^i ,\label{eq:QCDglu}
 \end{eqnarray}
where $T^a$ is the QCD generator on the quark fields, and the matrix field strength is   \begin{eqnarray}
G_{\mu\nu}=\partial_\mu G_\nu-\partial_\nu G_\mu-i[G_\mu,G_\nu],\label{eq:Gmunu}
 \end{eqnarray}
 with $G_\mu=(\lambda^a/2)G^{a}_{\mu}$.
  The strong interaction coupling is $g_3$. It was a profound observation that the above gluon interactions, due to the nonlinear term in Eq. (\ref{eq:Gmunu}),  admit instanton solutions  \cite{Belavin75}. These gluon vacua  allow degenerate (discrete) vacua which can be labelled by the so-called instanton number or the Pontryagin index, calculated in the Euclidian space as,
\begin{eqnarray}
n= \frac{1}{32\pi^2\,g_3^2}\int d^4 x_E\,  {G}^{a}_{\mu\nu} \tilde{G}^{a\,\mu\nu},\label{eq:Pontyagin}
 \end{eqnarray}
where $\tilde{G}^{a}_{\mu\nu}\equiv \frac12  \epsilon_{\mu\nu\rho\sigma}{G}^{a\,\rho\sigma}  $ is the dual of $ {G}^{a}_{\mu\nu} $.

The gauge invariant combination of $n$ different vacua is parametrized by  a continuous  parameter $\theta_{\rm QCD}$ \cite{CDG76,Jackiw76} which is called the QCD vacuum angle, defined in the Euclidian space,
\begin{eqnarray}
|\theta \rangle=\sum_{n} e^{in\theta}|n\rangle,\label{eq:thetaV}
 \end{eqnarray}
 where  $|\theta\rangle=|\theta_{\rm QCD}\rangle$. $\theta_{\rm QCD}$ is a P and T violating parameter since its origin is $G^{a}_{\mu\nu} \tilde{G}^{a\,\mu\nu}$ in Eq. (\ref{eq:Pontyagin}). A gauge transformation $G_m$ with a gauge field having the instanton number $m$ shifts the $|n\rangle$ vacuum to the $|n+m\rangle$ vacuum. So, $\theta_{\rm QCD}$ is a good gauge invariant parameter, and the strongly interacting gluons have another strong interaction coupling $g_3^2\theta_{\rm QCD}$. If $\theta_{\rm QCD}$ is not small, the effects of the vacuum angle is strong and we introduced another strong interaction parameter. Calculating the transition amplitude between two $\theta_{\rm QCD}$ vacua, one obtains an effective interaction in the $\theta_{\rm QCD}$ vacuum,
\begin{eqnarray}
{\cal L}_{\theta_{\rm QCD}}
 = \frac{\theta_{\rm QCD}}{32\pi^2 }  \, G^{a}_{\mu\nu} \tilde{G}^{a\,\mu\nu}.\label{eq:Ltheta}
 \end{eqnarray}
Absorbing $g_3^2$ in the denominator of (\ref{eq:Pontyagin}) into the gluon fields, we note that the nonlinear term is proportional to $g_3^2$. So, it arises from QCD interactions at order $g_3^2$, which is in fact  the one loop anomaly term \cite{OneLoopAnom,BellJackw}. Since there is no quark field in ${\cal L}_{\theta_{\rm QCD}}$, the term is a flavor singlet if one mimicks the effect  by the quark fields.

\subsection{CP violation in strong interactions and the problem}\label{subsec:strongCP}

Chiral roation of quark fields introduces an anomaly term,
\begin{eqnarray}
q_i\to e^{i\gamma_5\alpha_i/2 }q_i:~\frac{\alpha_i}{32\pi^2}\,G^{a}_{\mu\nu} \tilde{G}^{a\,\mu\nu} .\label{eq:chiralTr}
\end{eqnarray}
The quark mass term is not invariant under the transformation (\ref{eq:chiralTr}). If any quark mass is zero, say for $q_0:~q_0\to e^{i\gamma_5\alpha_0 }q_0$, then the chiral transformation  is a good symmetry. Therefore, even if $\theta_{\rm QCD}$ is nonzero in the beginning, we can shift it to zero by choosing $\alpha_0=-\theta_{\rm QCD}$. 

\begin{figure}[t!]
\centerline{\includegraphics[width=6cm]{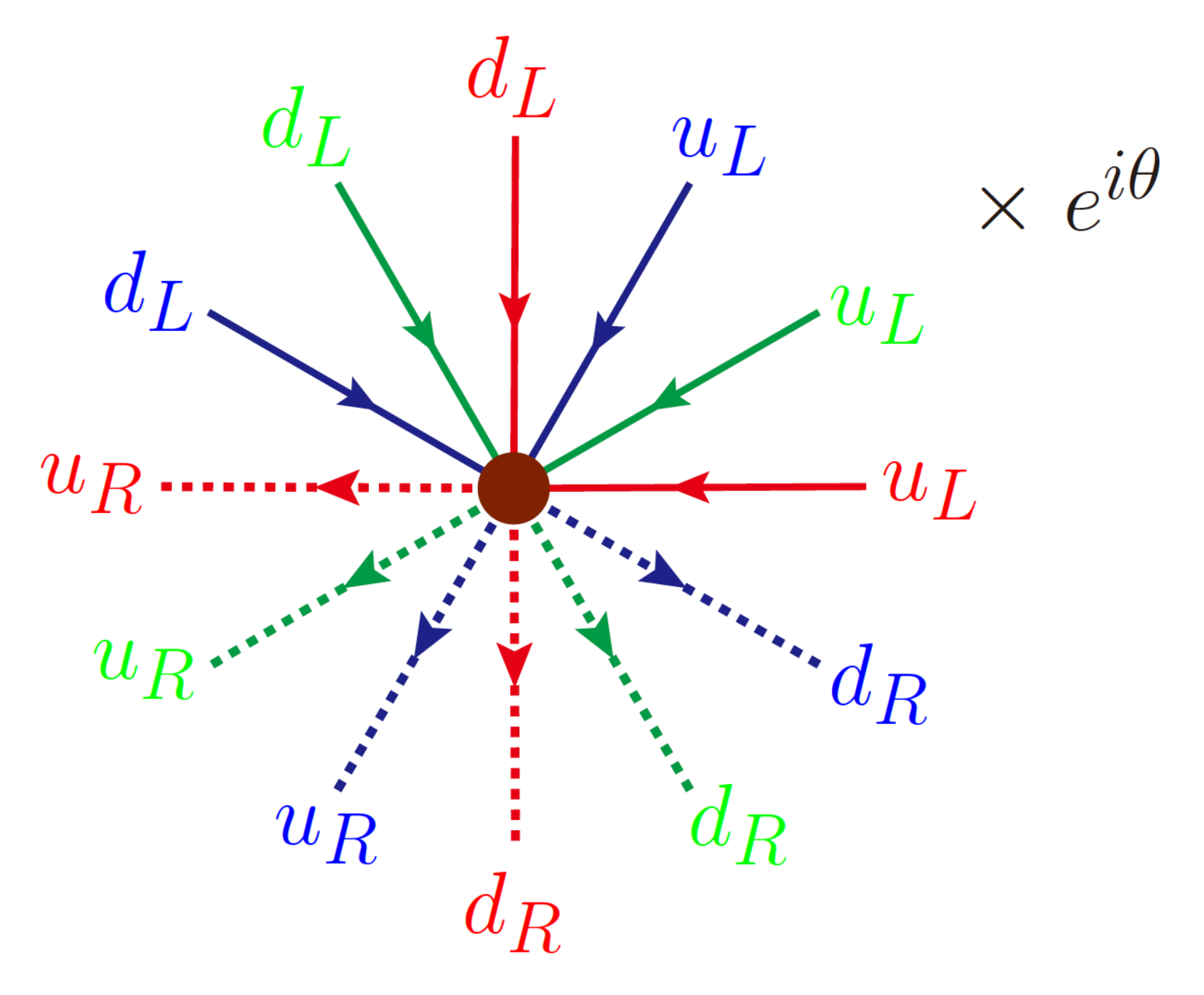}
 }
\caption{The determinental interaction of two light quarks. The determinental interaction changes the chiralities of quarks, \ie violating the chiral symmetry.}\label{Fig:DetInt}
\end{figure}

With the chiral transformations, the induced interaction (\ref{eq:chiralTr}) is a flavor singlet. But, the resulting quark mass matrix is transformed to $e^{i \alpha_i} M_q$. If the phase of the determinant of  $M_q$ is $e^{i\theta}$, we can remove the phase by the chiral rotation of quark fields such that $\sum_i\alpha_i=-\theta$, making Det.$M_q=$\,real. This makes the quark mass matrix free of the anomaly term. Thus, the interaction induced by instantons is a determinental interaction of quark fields, as sketched in Fig. \ref{Fig:DetInt}.
 Even if we start with real Yukawa couplings, the weak CP violation must be introduced by complex VEVs of scalar fields, which allows a complex determinant for $M_q$. In the KM model for the weak CP violation, the Yukawa couplings times VEV are complex, allowing a complex determinant for $M_q$. Thus,  making a real determinant for $M_q$ after one introduces the weak CP violation, one must shift the vacuum angle $\theta_{\rm QCD}$ to   $\theta_{\rm QCD}+\theta$. Namely,
the interaction (\ref{eq:Ltheta})  is necessarily intertwined with the observed weak CP violation, and a phenomenologically meaningful angle is  the $\overline{\theta} $ parameter,
\begin{eqnarray}
\overline{\theta}\equiv\theta_{\rm QCD}+\theta_{\rm weak}
\label{eq:defthetabar}
\end{eqnarray}
where  $\theta_{\rm weak}$ is the contribution generated when one introduces the weak CP violation at the electroweak scale, and given by Arg.Det.$M_q$. Since this $\overline{\theta}$ gives a neutron EDM (nEDM) of order $10^{-16}\overline{\theta} \,e$cm, the experimental upper bound on nEDM \cite{NEDMexp},
\begin{eqnarray}
|d_n| \le 2.8\times 10^{-26\,}e\,{\rm cm}~(\rm 90\% ~C.L.),
\end{eqnarray}
restricts $|\overline{\theta}|\lesssim 10^{-10}$. ``Why is $\overline{\theta}$ so small?" is the strong CP problem. 

When we consider the $\theta$ term with the determinental interaction of Fig. \ref{Fig:DetInt} together, we can transform one to another for the purpose of discussion. To find the minimum of the ``invisible''  axion potential, it is better to make $\thb=0$. To find out the mixing effects of the ``invisible'' axion with $\pi^0$ and $\eta'$ mesons, it is better to make $\thb=0$.
 
\subsection{The Peccei--Quinn symmetry}\label{subsec:PQ}

Peccei and Quinn  observed that if all $c_{IJ}$ in Eq. (\ref{eq:WeinV}) are zero, which is equivalent to considering only the terms in the lavender part in Fig.  \ref{Fig:discrete}, then the discrete symmetry is promoted to a global symmetry, according to our general scheme, which we now call the PQ symmetry \cite{PQ77},
\begin{eqnarray}
q_L&\to &   q_L,\label{eq:PQql}\\[0.2em]
u_R&\to & e^{-i\alpha} u_R, \label{eq:PQu}\\[0.2em]
d_R&\to & e^{-i\beta} d_R, \label{eq:PQd}\\[0.2em]
\phi_1&\to & e^{i\alpha} \phi_1, \label{eq:PQHu}\\[0.2em]
\phi_2&\to & e^{i\beta} \phi_2, \label{eq:PQHd}
\end{eqnarray}
where $q_L$'s are the left-handed quark doublets, $u_R$'s are the right-handed up-type quark singlets, and  $d_R$'s are the right-handed down-type quark singlets. 
Quarks obtain masses by
\begin{eqnarray}
 -  \bar{q}_Lu_R\phi_1-\bar{q}_Ld_R\phi_2 +{\rm H.c.}
\label{eq:qPQ}
\end{eqnarray}
which also respects the PQ symmetry.
This PQ transformation is a chiral transformation, Eqs. (\ref{eq:PQql}), (\ref{eq:PQu}) and (\ref{eq:PQd}),  creating  the QCD anomaly coefficient. Therefore, this simple phase transformation is thought to be equivalent to the gluon anomaly and in any physical processes, therefore, there will not appear the phase and there is no strong CP problem, which was the argument used in Ref. \cite{PQ77}.  
 
The above PQ symmetry (if exact) must lead to an exactly massless Goldstone boson because quarks must obtain masses by the VEVs of $\phi_1$ and $\phi_2$, \ie the PQ symmetry must be spontaneously broken. Note that the term (\ref{eq:qPQ}) is a part in defining the PQ symmetry. However, the PQ symmetry is explicitly broken at quantum level because the QCD anomaly is present. Thus, the Goldstone boson becomes a pseudo-Goldstone boson, obtaining mass of order $\approx\Lambda_{\rm QCD}^2/\vew$ by our general argument of Fig. \ref{Fig:discrete} \cite{Wein78,Wilcz78}. This pseudo-Goldstone boson was named {\it axion}. 
 From $V_{\rm W} $ of Eq. (\ref{eq:WeinV}), we can separate out this axion direction. If $c_{12}$ were present, which signifies the breaking of the PQ symmetry, there will appear the phase $e^{-i(a_1/v_1-a_2/v_2)}$. Thus, the axion direction is
 \begin{eqnarray}
a\propto \frac{a_u}{v_u}-\frac{a_d}{v_d}
 \end{eqnarray}
where $\langle\phi_1\rangle=(v_u+\rho_1)e^{i a_u/v_u}/\sqrt2$ and $\langle\phi_2\rangle=(v_d+\rho_2)e^{i a_d/v_d}/\sqrt2$. This electroweak scale axion is called the Peccei--Quinn--Weinberg--Wilczek (PQWW) axion, having mass $\sim 100\, $keV and lifetime $\sim  $\,second order \cite{Kimprp87} and cannot be the one considered in the cavity experiments for discovery. 

\subsection{The ``invisible'' axions}\label{subsec:Inv}

In 1978 this PQWW axion was known not to exist \cite{PecceiTokyo}, which had led to ideas on models without the PQWW axion in 1978 \cite{Cal78,Moha78,Segre79,Barr79,Georgi78}. Nowadays, these are used in the Nelson--Barr type models \cite{Nelson,NelsonBarr}, which are called ``calculable models'' \cite{Kimprp87}.
 The calculable models start with the CP symmetry in the Lagrangian such that $\theta_0=0$ in Eq. (\ref{eq:defthetabar}). But, one has to introduce the weak CP violation, \ie the CP violation is through the spontaneous mechanism \cite{LeeTD73}.
These calculable models seem not working in two aspects, firstly it is very difficult to build models in which the loop corrections allow  $\theta_{\rm weak}$ below O($10^{-10}$) and second the observed weak CP violation is of the Kobayashi--Maskawa form \cite{PDG16}.
 
 But the PQ-type solution of the strong CP problem was so attractive that the ``invisible'' axion with lifetime greater than the age of the Universe was reinvented with an SU(2)$_W\times$U(1)$_Y$ singlet scalar field \cite{KSVZ1}. Currently, this is the most studied global symmetry. For this \UPQ, the explicit breaking term is the QCD anomaly term
 \begin{eqnarray}  \frac{\overline{\theta}}{32\pi^2}G^a_{\mu\nu}
 \tilde{G}^{a\,\mu\nu}\label{eq:QCDax}
 \end{eqnarray}
where $G_{\mu\nu}^a\,(\tilde{G}^{a\,\mu\nu})$  is the gluon field strength (its dual), and $\overline{\theta}=a/f_a$. The symmetry in
(\ref{eq:PQu}), for example, changes $\overline{\theta}\to \overline{\theta}+\alpha$ and the anomaly term has the shift symmetry $a\to a+f_a\alpha$ if the anomaly term does not contribute to the effective potential. But, the anomaly contributes to the effective potential and the question is, ``What is the value $\overline{\theta}$ at the minimum of the potential\,?'' If the potential $V$ does not break the CP symmetry, it is easy to show that the potential generated by the anomaly term chooses  $\overline{\theta}=0$ as the minimum of the potential \cite{VW84}, which is a good solution of the strong CP problem. Even if the weak CP violation is introduced, the shift of the minimum is very tiny
\cite{GeoRand86}, $\Delta\overline{\theta}\simeq 10^{-16}$, which does not spoil the nature of the strong CP solution along this line.
 
In this solution, there is only one parameter, the value of the ``invisible'' axion field or $\thb$. The reason is that the form of the anomaly is completely specified at one loop and no more \cite{Bardeen69}. One parameter $\thb$ to this interaction is one coupling $G_F$ to the weak interaction. In both cases, details are more involved however, in the former case the detailed ``invisible'' axion models and in the latter case the CKM and PMNS weak gauge boson couplings to three family members. Because of this one coupling nature, one could have easily proved that $\thb=0$ is at the minimum of free energy \cite{VW84}. We have shown that  the phase 0 can be at the minimum of the potential if the sign in  Eq. (\ref{eq:formV}) is negative. So, what Vafa and Witten have shown \cite{VW84} is that the sign is -- if the CP violation in QCD occurs purely from the $\thb$ term without the $\Delta V$ contribution of Eq. (\ref{eq:formV}). 

\begin{figure}[!t]
\centerline{\includegraphics[width=10cm]{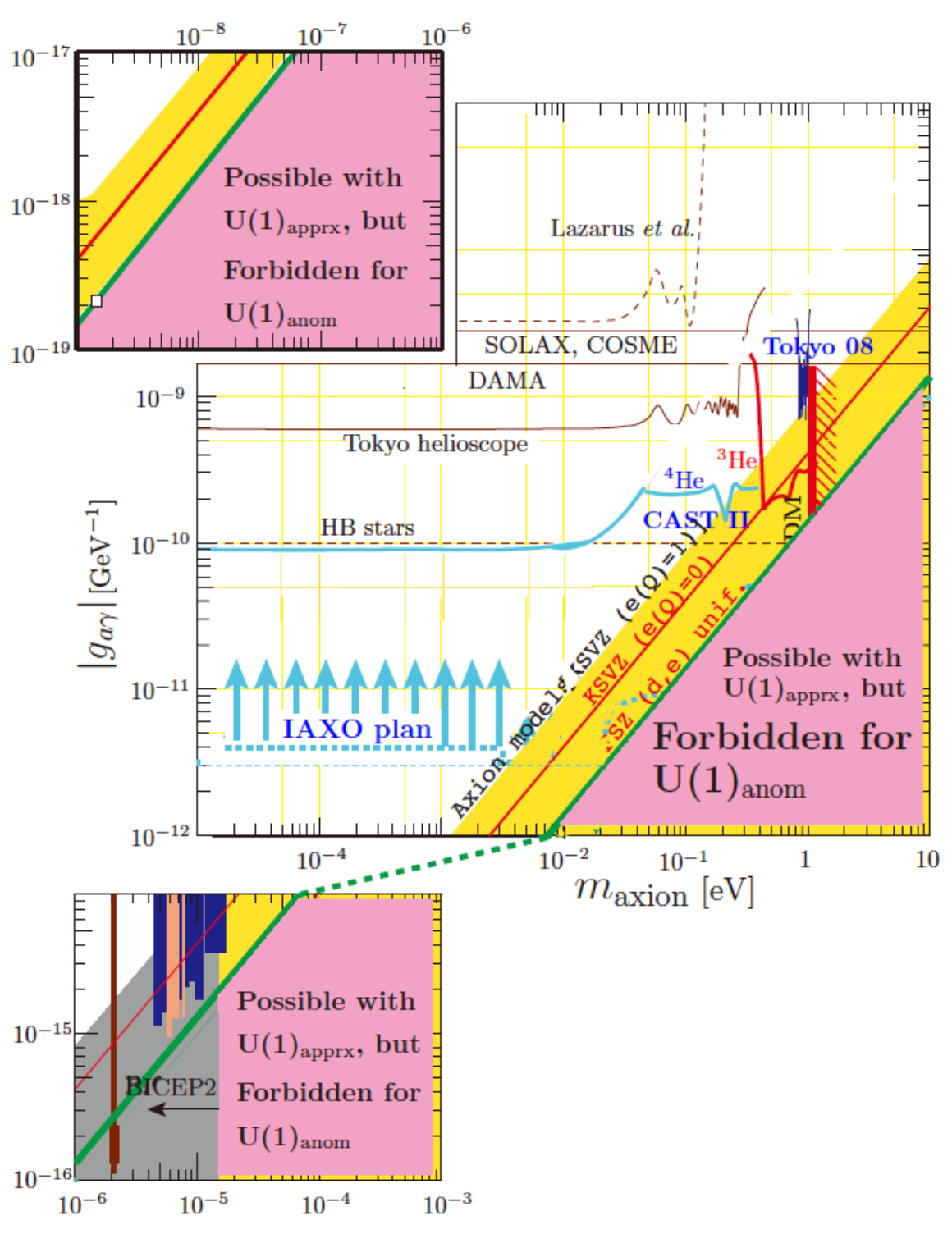} }
\caption{The $g_{a\gamma}(=1.57\times 10^{-10}c_{a\gamma\gamma})$ vs. $m_a$ plot \cite{KimPLB14,KimKyaeNam17}.}
\label{Fig:expbound}
\end{figure}

The axion detection  uses the following coupling  \cite{Primakoff51,Sikivie83},
  \begin{eqnarray}  
\frac{c_{a\gamma\gamma}}{16\pi^2}\,a\,F^{\rm em}_{\mu\nu}
 \tilde{F}^{\rm em\,\mu\nu}\label{eq:EMax}
 \end{eqnarray}
 where $F_{\mu\nu}^{\rm em}\,(\tilde{F}^{\rm em\,\mu\nu})$ is the electromagnetic field strength (its dual).  In the search of the QCD ``invisible'' axion, the axion--photon--photon coupling  $c_{a\gamma\gamma}$ is the key parameter. In a sense, the customary numbers presented in most talks are ad hoc, which can be glimpsed from those exclusion plots where the KSVZ lines are shown only for $\Qem=0$  \cite{Kim98}. My exclusion plot is presented in Fig. \ref{Fig:expbound}.

\subsubsection{Calculation of $\cagg$}
\label{subsubsec:cagg}

For a calculation of $c_{a\gamma\gamma}$ in any given model, one should pay attention  that the model must lead to an acceptable SM phenomenology with the electromagnetic charges defined in that model. In these calculations, the most important step is the definition of the PQ symmetry by the lowest order nontrivial interaction, \ie the term(s) in the lavender part of Fig. \ref{Fig:discrete}, connecting the weak interaction singlet $\sigma$ to a colored fermion(s): 
\begin{itemize}
\item{KSVZ models:}  The definition for the PQ symmetry is
\dis{
{\cal L}=-f_\sigma\,\bar{Q}Q\,\sigma+{\rm h.c.}\label{eq:KSVZ}
}
which is a renomalizable interaction. Different KSVZ models are just the differences of electromagnetic charges of the heavy quark $Q$.
\item{DFSZ models:}   The definition for the PQ symmetry in terms of renormalizable interactions needs both\footnote{Family indices for the SU(2) doublet $q_L$ and right-handed singlets $u_R$ and and $d_R$ are suppressed. Here, $\tilde{H}_{u,d}=i\sigma_2 H_{u,d}^*$.}
\dis{
{\cal L}_{\rm H_u~unif} &=-f_e \bar{\ell}_Le_R \tilde{H}_u-f_u \bar{q}_Lu_R \tilde{H}_d-f_d \bar{q}_Ld_R \tilde{H}_u 
+{\rm h.c.}\\
{\cal L}_{\rm H_d~unif} &=-f_e \bar{\ell}_Le_RH_d -f_u\bar{q}_Lu_RH_u-f_d\bar{q}_Ld_RH_d 
+{\rm h.c.} \label{eq:DFSZY}
}
and
\dis{
V =\frac{\lambda}{4}|\sigma^*\sigma|^2-\frac{\mu^2}{2}\sigma^*\sigma +\lambda_1\sigma^2H_uH_d+\cdots.
 \label{eq:DFSZV}
}
where $Y(H_u)=\frac12$ and $Y(H_d)=-\frac12$, and $\cdots$ represent the remaining terms consistent with the definition of the PQ symmetry by $\sigma^2H_uH_d$.
\end{itemize}
The breaking terms of the PQ symmetry is the red part of  Fig. \ref{Fig:discrete}, but only by the QCD anomaly term without any term from the potential $\Delta V$.   In the DFSZ model without SUSY, there must be a fine-tuning between $\lambda$ and $\lambda_1$ such that (electroweak scale)$/f_a$ is of order $10^{-8}$, and hence the fine-tuning on the squared mass parameters in $V$ of O($10^{-16}$) is required. But with SUSY, the definition of the PQ symmetry is by the dimension-5 $\mu$ term \cite{KimNillesPLB84} and ${\cal L}_{\rm H_d~unif}$,
\dis{
W_\mu\propto \frac{1}{M} H_uH_d\sigma^2,
}
and there is no fine-tuning problem with the SUSY extension.
In Table \ref{tab:KSVZDFSZ}, we present $c_{a\gamma\gamma}$ in several KSVZ and DFSZ models \cite{KimRMP10}.

\begin{table}[!t]
\caption{The coupling $c_{a\gamma\gamma}$ in the KSVZ and DFSZ models. For the $u$ and $d$ quark masses, $m_u=0.5 m_d$  is assumed for simplicity. $(m,m)$ in the last row the KSVZ means $m$ quarks of $Q_{\rm em}=\frac23 e$ and  $m$ quarks of $Q_{\rm em}=-\frac13 e$.  SUSY in the DFSZ includes contributions of color partners of Higgsinos. If we do not include the color partners, \ie in the MSSM without heavy colored particles, $c_{a\gamma\gamma}\simeq 0$.}
{\begin{tabular}{@{}cc|ccc@{}} \toprule
KSVZ:  $Q_{\rm em}$ & $c_{a\gamma\gamma}$ & DFSZ:  $(q^c$-$e_L)$ pair & Higgs &$c_{a\gamma\gamma}$ \\ \colrule
$0$ & $-2$ &  non-SUSY $(d^c,e)$ & $H_d$&$+\frac23$ \\[0.4em]
$\pm\frac13 $ &  $-\frac43$&  non-SUSY $(u^c,e)$ & $H_u^*$&$-\frac43$ \\[0.4em]
$\pm\frac23$ & $+\frac23$& GUTs & &$+\frac23$ \\[0.4em]
$\pm1$   & $+4$& SUSY & &$+\frac23$ \\[0.4em]
$(m,m)$ & $-\frac13$ & &  & \\ \botrule
\end{tabular} \label{tab:KSVZDFSZ}}
\end{table}

\begin{figure}[!t]
\centerline{\includegraphics[width=10cm]{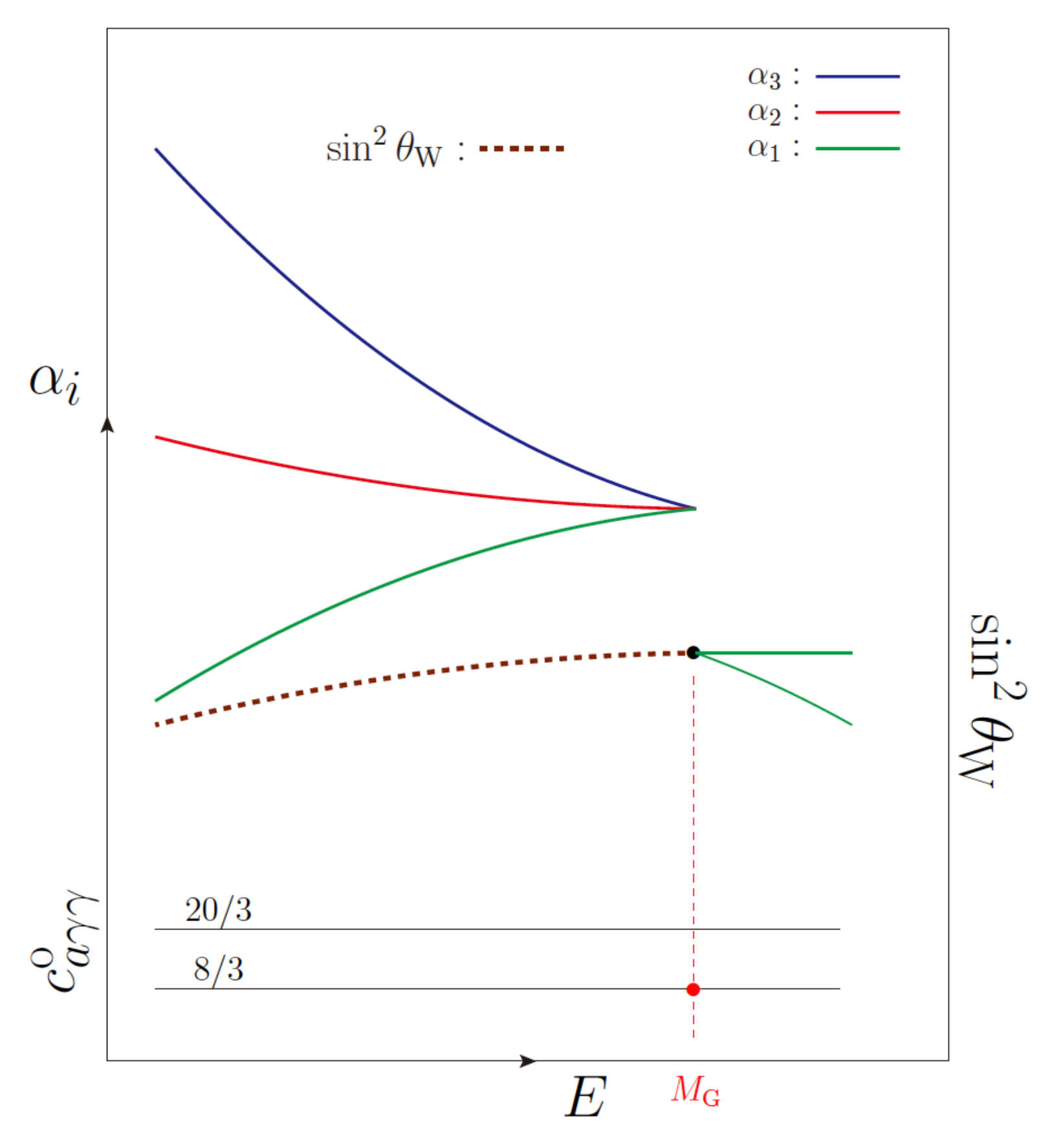}
 }
\caption{A schematic view on the gauge couplings, $\sin^2\theta_{\rm W}$, and  $\cagg^0$.  }\label{fig:caggGUT}
\end{figure}

In some GUT models, $\cagg^0$ is related to the weak mixing angle   $\sin^2\theta_{\rm W}$. A schematic view on the gauge couplings, $\sin^2\theta_{\rm W}$, and  $\cagg^0$ is presented in Fig. \ref{fig:caggGUT}. The evolution of $\sin^2\theta_{\rm W}$  are shown  in the middle part.   $\cagg^0$ is determined by quantum numbers and does not evolve above   as shown in the lower part of the figure. Axions from GUTs usually give $c_{a\gamma\gamma}^0= 
  \frac83$, which is not necessarily satisfied with unknown $\Qem$ charges above the GUT scale, and there exist models with $\cagg^0>\frac{8}{3}$, for example   $\frac{20}{3}$ in \cite{KimPRL80}.

Calculations of $\cagg$ from ultra-violet completed models are welcome. So far there exists only one calculation on $\cagg$ from string compactification \cite{KimPLB14,KimKyaeNam17}. In Table \ref{tab:StAxion}, we list possible ranges of  $\cagg^0$.  In a ${\bf Z}_{12-I}$ orbifold compactification present in Ref. \cite{HuhKK09}, the axion-photon-photon coupling has been calculated \cite{KimKyaeNam17},
\begin{equation} 
c_{a\gamma\gamma}= \cagg^0-c_{\rm chiral} \simeq \frac{-5406-1162-1960-784}{-3492} -2= +\frac23,\label{eq:TrQanom}
\end{equation}  
which is shown as the green line in the axion coupling vs. axion mass plot of Fig. \ref{Fig:expbound}. 

\begin{table}[!t]
\caption{String model prediction of $c_{a\gamma\gamma}$.  In the last line, $c_{a\gamma\gamma}= (1-2\sin^2\theta_W)/\sin^2\theta_W$ with $m_u=0.5 m_d$.}
{\begin{tabular}{@{}ccc@{}} \toprule
String:  & $c_{a\gamma\gamma}$  & Comments  \\ \colrule
Choi, Kim, and Kim\cite{IWKim07}\hphantom{00} & \hphantom{0}$-\frac13$& \hphantom{0}Approximate  U(1) global symmetry \\[0.5em]
 Kim\cite{Kim88},~Kim,Kyae, and Nam\cite{KimKyaeNam17},~Kim and Nam\cite{KimNamPLB16}\hphantom{00} & \hphantom{0}$+\frac23$ & \hphantom{0}Anomalous U(1) from string \\[0.1em] \botrule
\end{tabular} \label{tab:StAxion}}
\end{table}
 
 \subsection{QCD phase transition in the RD Universe and axion mass}\label{subsec:QCDphase}
 
From the early days on, breaking of the chiral symmetry has been the basis for estimating the axion mass \cite{BardeenTye78,Baluni79}. There are two kinds of phase transitions at the   QCD confinement scale: one is the transition from the quark--gluon phase (simply it will be called quark phase (q-phase)), to the hadronic phase (h-phase) and the other is the chiral symmetry breaking that quark condensates develop. We assume that these occur at the same scale, for simplicity.
 
 Let us consider $u$ and $d$  quarks for the study of QCD phase transition.\footnote{Addition of  $s$ would change parameters only at a 5 \%  level, viz. $m_d/m_s\simeq 1/20$.} 
The chiral symmetry breaking is proportional to light quark masses $m_um_d/(m_u+m_d)$  since it should vanish if any one quark is massless. Even if the QCD scale $\LQCD$ is a few hundred MeV, the axion mass should take into account the chiral symmetry breaking in terms of the current quark masses $m_u$ and $m_d$. At zero temperature, however, the parameters are those in the  h-phase and we do not use the constituent quark masses, but the axion mixing with $\pi^0$ is used to encode the chiral invariance in   the limit of  current quark mass $m_u=0$ or $m_d=0$. The result is a zero temperature value in the h-phase, even though the current quark masses are used. 

Recently, topological susceptibility has been presented  in the lattice calculation \cite{Lattice16,ICTP16}. On the other hand,  at high temperature around and above 1~GeV in the q-phase,  the QCD parameter $\LQCD$ is used for the non-Abelian gauge coupling via the dimensional tansmutation \cite{Pisarski81}. Again the chiral symmetry property should be valid, \ie  it vanishes in the limit of $m_u=0$ or  $m_d=0$.  In terms of the parameters in each case, we summarize them as 
\dis{
&\textrm{Quark and gluon phase~}\LQCD:~ \frac{m_u^2}{(1+Z)^2}\LQCD^2\left(\frac12 \thb^2\right) [107],\\[0.5em]
&\textrm{Hadronic phase~}f_\pi^2 m_\pi^2:~ \frac{Z}{(1+Z)^2}f_\pi^2 m_\pi^2\left(\frac12 \thb^2\right) [112,113],\\[0.5em]
&\textrm{Lattice calculation~}\chi_4:~\chi_4\left(\frac12 \thb^2\right) [114,115],\label{eq:AxPhmasses}
}
where $Z=m_u/m_d$ and $\chi_4\simeq (76~ \rm MeV)^4$ and $\left(\frac12 \thb^2\right)$ is simply denoting  the mass operator divided by $f_a^2$. If we use $m_u\simeq 2.5\rm ~MeV$, $Z\simeq \frac{1}{2}$~ \cite{Manohar16}  and $\LQCD\approx \LQCD^{(3)}\simeq 332~\rm MeV$ \cite{QCD16}, the values in the q-phase and in the  h-phase are  $(88~\rm MeV)^4$ and  $(77~\rm MeV)^4$, respectively. Note that the values calculated in the h-phase, the second and third lines of Eq. (\ref{eq:AxPhmasses}), are almost identical. We can consider this fact as a confirmation on the validity of the lattice calculation in the hadronic phase \cite{Lattice16,ICTP16,Klaer17}. 

The transitions from the q-phase to the h-phase and chiral symmetry breaking occur at the same time. These transitions are the first order phase transitions, starting at cosmic time $t_c$ of Fig. \ref{fig:rhoa}. This transition has been discussed extensively in the bag model \cite{DeGrand84}. It occurs somewhat below the critical temperature  $T_c$.  For one families, quarks before and pions after, the light degrees (considering light leptons and photon also) are 51.25 and 17.25, respectively.  Since  the lattice calculation is confirmed as shown in Eq. (\ref{eq:AxPhmasses}), the critical temperature $T_c$ from the lattice calculation \cite{ICTP16}  is used below
\begin{eqnarray}
T_c=165\pm 5\,\rm MeV.\label{eq:TcLatt}
\end{eqnarray}
In the q-phase at temperature above the $\rho$ meson mass scale, we will use the temperature dependence, $\propto T^{-8.16}$, as used in Ref.  \cite{Pisarski81}. Between $T_c$ and the $\rho $ mass scale, we will use  the temperature dependence, $\propto T^{-4.21}$,  to smoothly connect to $\chi$ at $T_c$.

\begin{figure}[!t]
\includegraphics[width=0.5\textwidth]{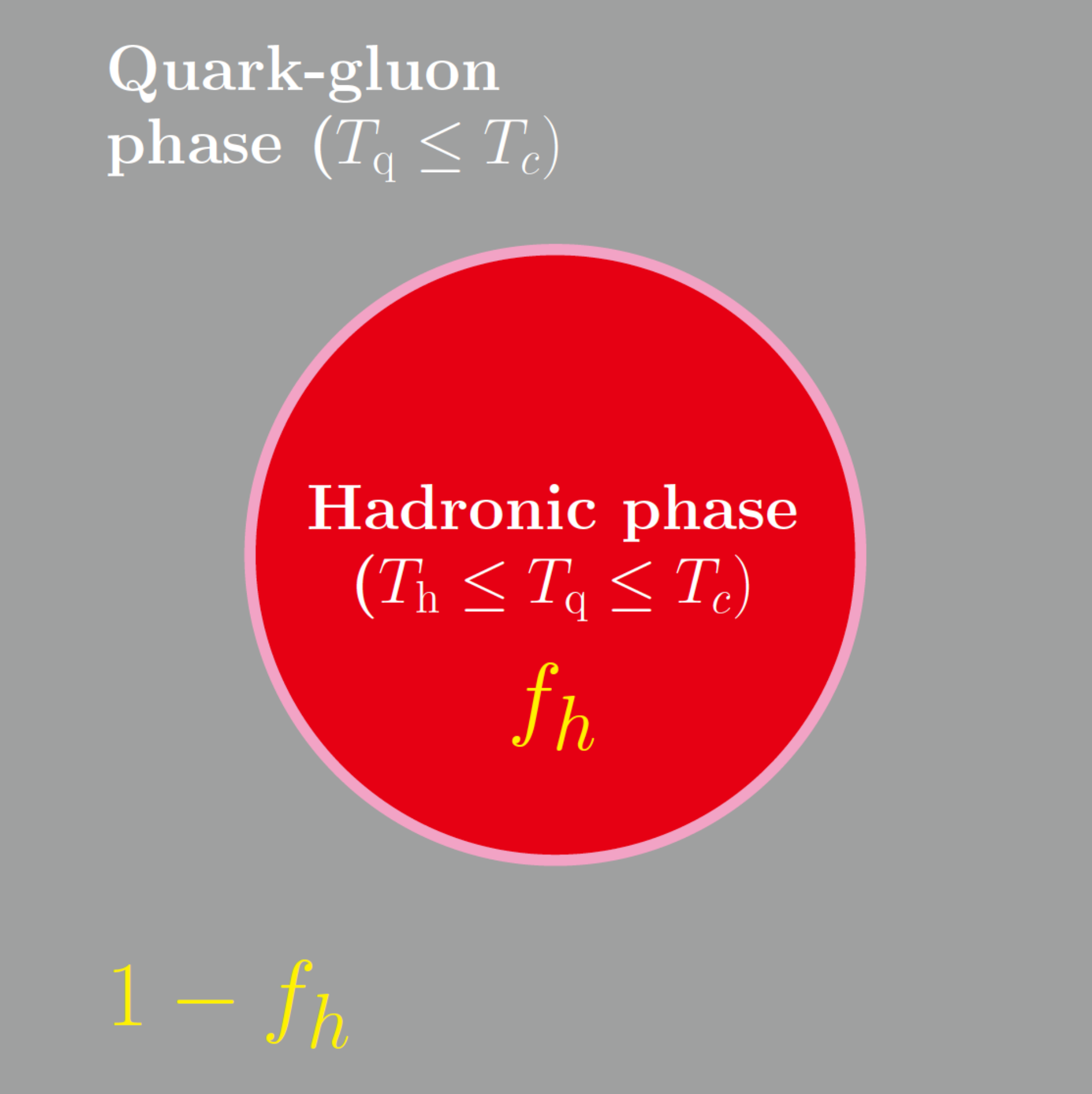}
  \caption{Formation of a h-phase supercooled bubble.} \label{fig:hBubble} 
\end{figure}
 
In this paper, we do not use the result of Ref.   \cite{DeGrand84}, but adopt a cosmological model, taking into a field theoretic account of `supercooling' of the light degrees by forming bubbles  in the RD universe as shown in Fig. \ref{fig:hBubble}, presented recently \cite{KimSKim18}.
We adopt two basic principles:  (i) during the first order phase transition the Gibbs free energy is conserved   and (ii) the phase transition is completed into the homogenious h-phase by the time $t_f$ in the evolving Universe.  
Contributions of light degrees to energy density and entropy, in the initial and final states, are 
\dis{
\textrm{Before}&\left\{\begin{array}{l}\rho  =\frac{\pi^2}{30}\,g_*^i T^4\\[0.5em]
s  =\frac{2\pi^2}{45}\,g_*^i T^3 \end{array}\right.
\\[0.5em]
\textrm{After}&\left\{\begin{array}{l}\rho  =\frac{\pi^2}{30}\,g_*^f T^4\\[0.5em]
s  =\frac{2\pi^2}{45}\,g_*^f T^3 \end{array}\right.\label{eq:BeAf}
}
where we will use
\dis{
g_*^i=51.25,~g_*^f=17.25.\label{eq:gif}
}
It is discussed in detail in Ref. \cite{KimSKim18}. The numerical calculation  of $\chi$ during the QCD phase transition \cite{KimSKim18} along this line (by fitting the tunneling rate with two parameters) is presented 
 as the blue curve in Fig.  \ref{fig:rhoa}. 
 The tunneling rate from the q-phase to the h-phase undergoes at the same temperature and pressure,  \ie with the conserved Gibbs free energy. In the Universe evolution, the bubble formation fraction $f_h$ illustrated in Fig. \ref{fig:hBubble} is calculated numerically, satisfying the above two conditions.

The current energy density of axions is determined by the axion evolution equation with the Hubble expansion included,
\dis{
 \ddot\theta+3H\dot\theta+\frac{m_1^2}{2}\sin\theta=0.\label{eq:diffeq}
}   
where the angle $\theta=A/f_a $ is the axion field and $m_1$ is the axion mass at zero temperature, $T=T_1$. At a cosmic time scale  $3H\sim m_1$, $\ddot\theta$ is negligible and Eq. (\ref{eq:diffeq}) determines an angle $\theta_1$ which is called the initial misalinement angle. The cosmic temperature determining $\theta_1$ is $T_1$ which is known to be $T_1\approx 1\,\gev$. This will be presented in Subsect. \ref{subsec:rhoa} in the cosmology section.

\section{Cosmology with an ``invisible'' axion}\label{sec:CosAxion}

\subsection{The domain wall problem in ``invisible'' axion models}\label{subsec:DWs}

It is well-known that if a discrete symmetry is spontaneously broken then there results   domain walls in the course of the Universe evolution \cite{Okun74}. For the ``invisible'' axion models, it was pointed out that the domain wall number $\ndw$ different from 1 must have led to  serious cosmological problems in the standard Big Bang cosmology \cite{Sikivie82,Vilenkin82}. Therefore, the standard DFSZ models with $\ndw=6$ has not worked successfully in our Universe.  For the ``invisible'' axions, we consider only $\ndw=1$ models.  
\begin{figure}[!t]
\centerline{\includegraphics[width=8cm]{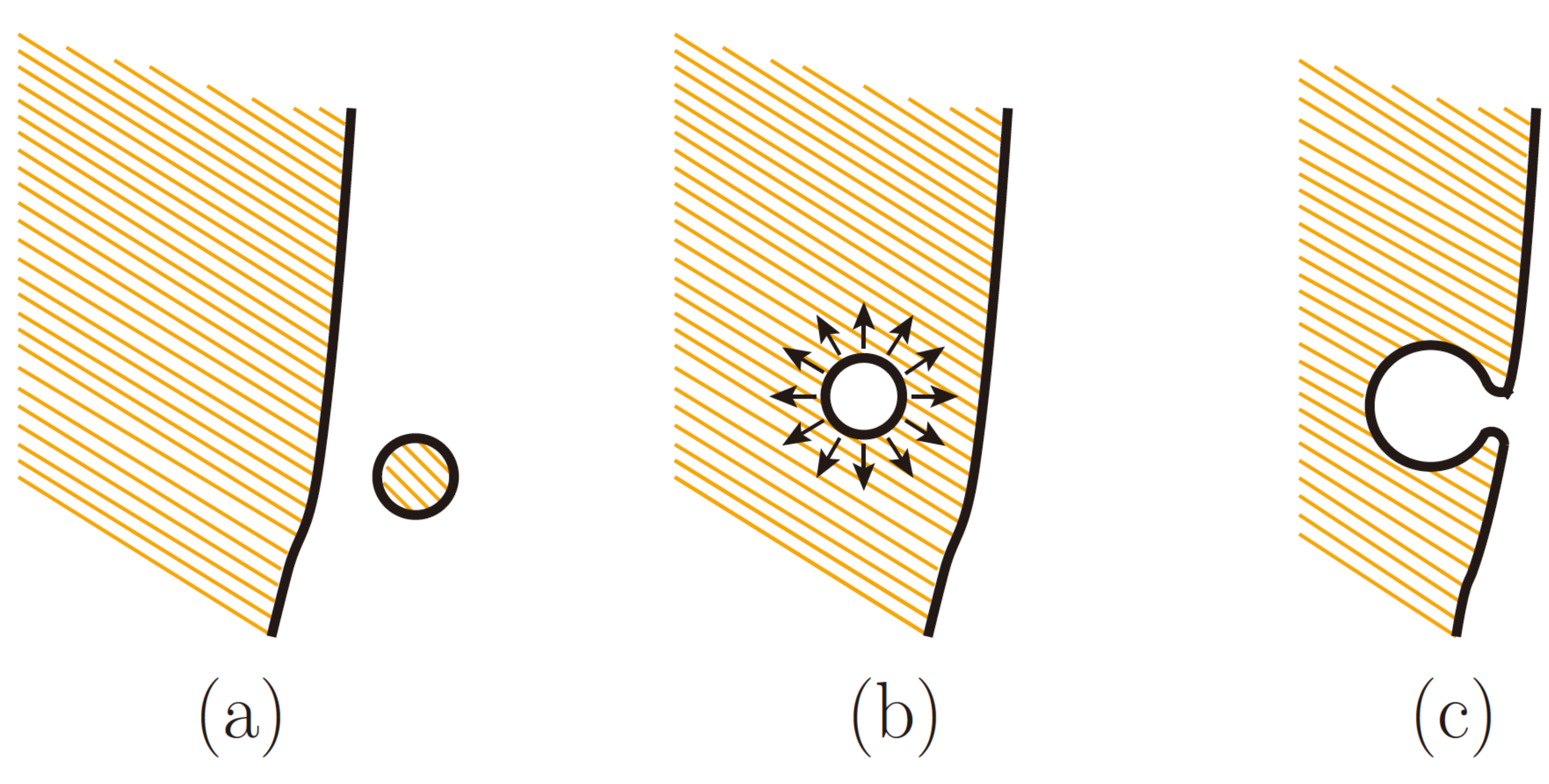} }
\caption{A horizon scale string-wall  for  $\ndw=1$ with a small membrane bounded by string.}\label{Fig:DWon}
\end{figure}
\begin{figure}[!t]
\centerline{\includegraphics[width=5cm]{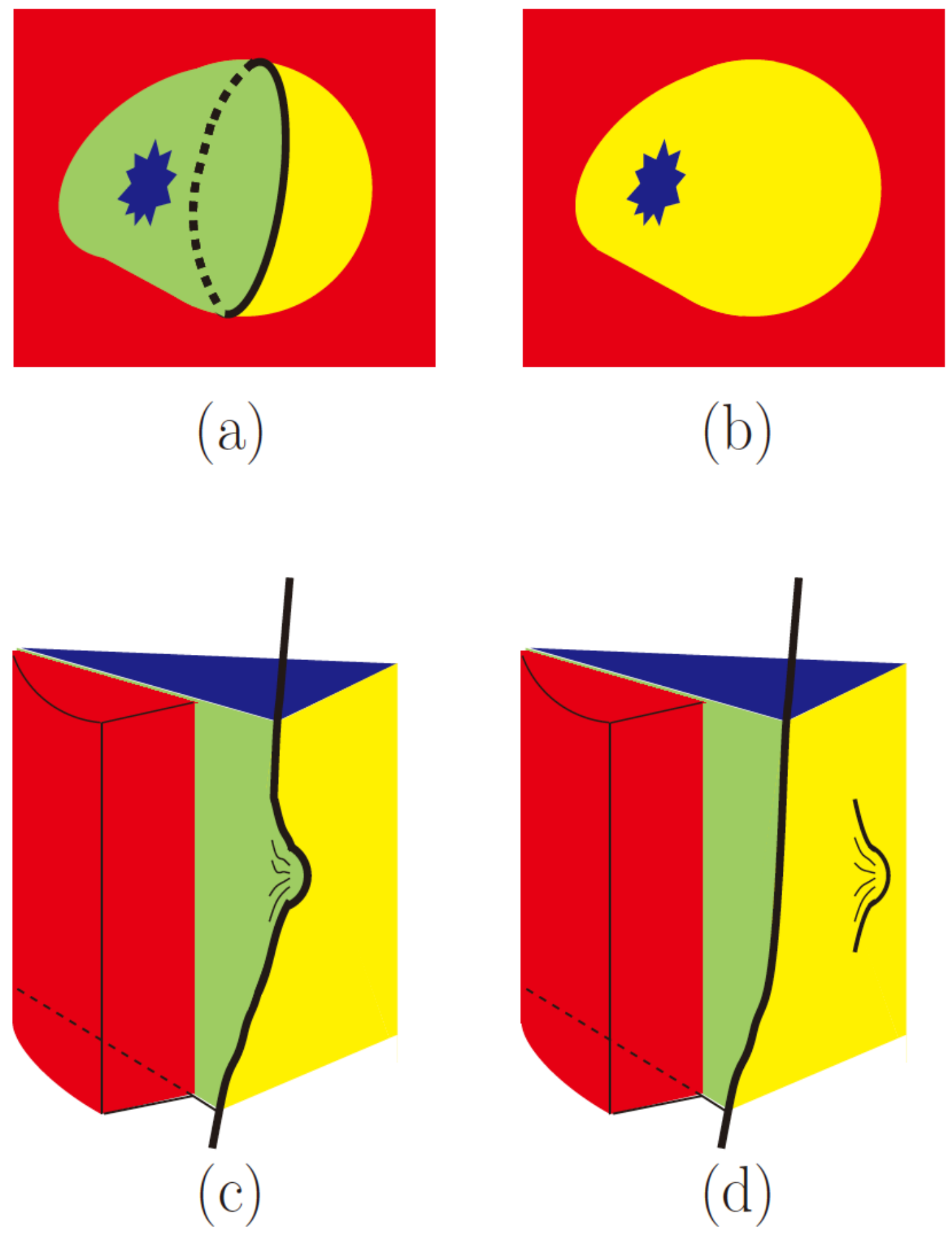} }
\caption{Small DW balls ((a) and (b), with punches showing the inside blue-vacuum) and the horizon scale string-wall system ((c) and (d))  for $\ndw=2$. Yellow walls are $\theta=0$ walls, and  yellow-green walls are $\theta=\pi$ walls.  Yellow-green walls of type (b) are also present.}\label{Fig:DWtw}
\end{figure}
The argument goes like this. In the evolving Universe, there always exists a (or a few) horizon scale string(s) and a giant wall attached  to it \cite{Vilenkin82} is present as shown in
Fig. \ref{Fig:DWon}\,(a). There are a huge number of small walls bounded by an axionic string which punch holes in the giant walls as shown in Fig. \ref{Fig:DWon}\,(b). The punched holes expand with light velocity and eat up the  giant string-walls as shown in
Fig. \ref{Fig:DWon}\,(c). This is the scenario that ``invisible'' axion models with $\ndw=1$ are harmless in cosmolgy  \cite{BarrChoi86}.
However, ``invisible'' axion models with $\ndw\ge 2$ have cosmological problems. 
For example for $\ndw=2$, a horizon scale string and wall system has the configurations shown in Figs. \ref{Fig:DWtw}\,(a), (b), (c), and (d). Figures \ref{Fig:DWtw}\,(a) and (b) show small balls, and Figs. \ref{Fig:DWtw}\,(c) and (d) show a horizon scale string-wall system after absorbing these small balls. Certainly, the horizon scale string-wall system is not erased, which is a cosmological disaster in the standard Big Bang cosmology.

\begin{figure}[!h]
\centerline{\includegraphics[width=5cm]{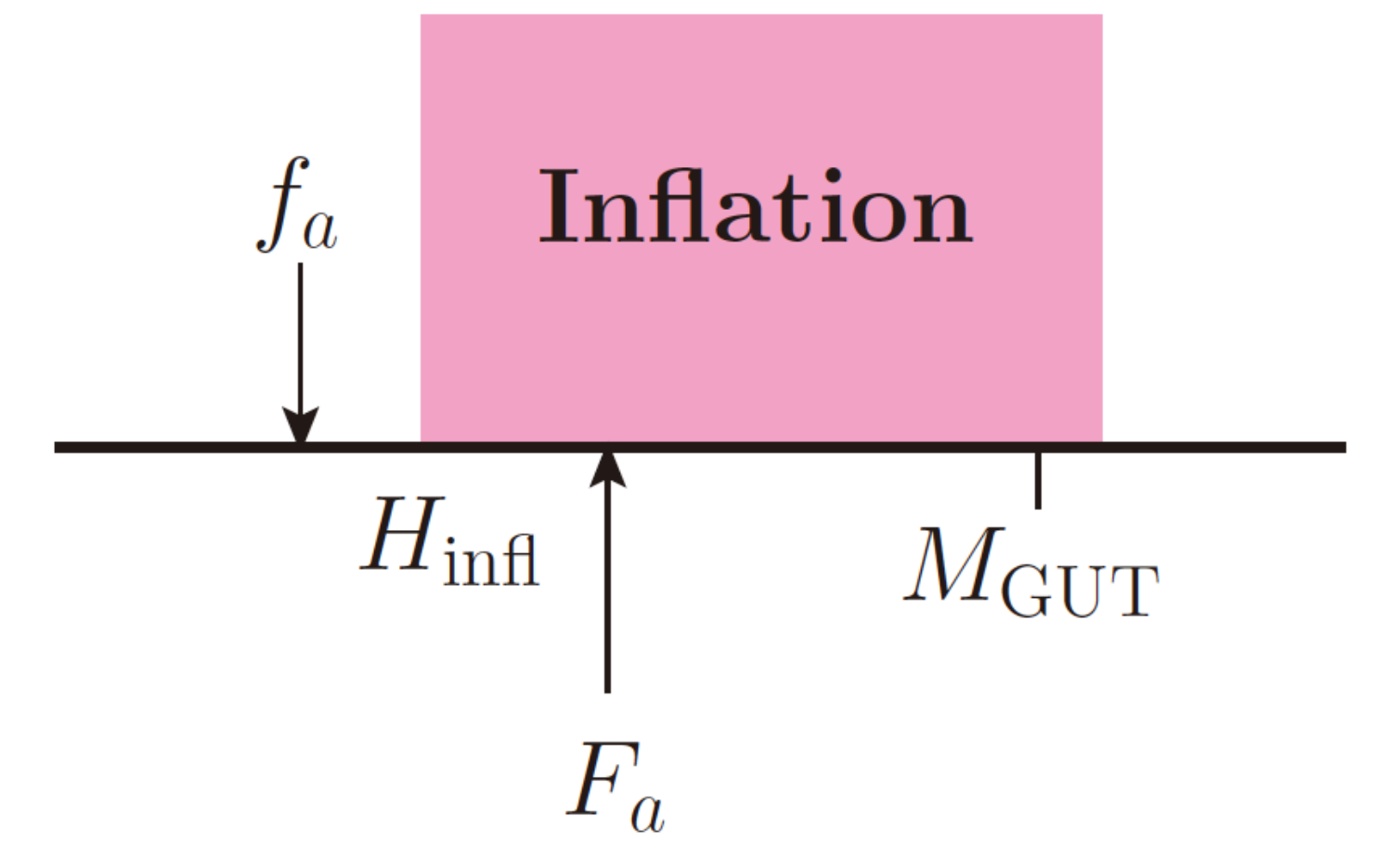} }
\caption{ $f_a$ in the axion window and $F_a$ in the inflationary region.}\label{Fig:faInfl}
\end{figure}

 With inflation, the domain wall problem has to be reconsidered. During inflation the Hubble parameter $H_{\rm infl}$ remains constant.  If the axion decay constant is during inflation as shown with $F_a$ in Fig. \ref{Fig:faInfl}, then a choice of $\langle a\rangle$ is the same in the whole inflating bubble. 
 So there will be no domain wall problem for the axion fields inside the bubble.  At this temperature far above the electroweak scale, quarks are massless and the ``invisible'' axion is also massless. Massless fields influence the evolution over the whole bubble and there can be an influence on the {\bf E} and  {\bf B} mode fluctuations in the inflating bubble. From the early BICEP2 report \cite{BICEP14}, this effect was considered and there resulted a constraint on the axion decay constant.\footnote{However, the data is not strong enough to give a large $r=O(10^{-1})$ with dust contamination carefully included \cite{BICEP16}.} The allowed range is marked yellow in the lower box of  Fig. \ref{Fig:expbound} \cite{Gondolo14,Marsh14}.  However, if the axion decay constant is smaller than  $H_{\rm infl}$ as shown with $f_a$ in Fig. \ref{Fig:faInfl}, then any choice of the VEV $\langle a\rangle$ is possible after the end of inflation and domain walls become a serious cosmological problem \cite{Sikivie82}.  
In this case, the axionic domain problem is better to be resolved with models of $\NDW=1$. One obvious model is the KSVZ axion with one heavy quark.
  
Another, a more sophiscated, solution is the Lazarides--Shafi mechanism in which the seemingly different vacua are identified by gauge transformation \cite{LS82}. The original suggestion was to  use the centers of extended-GUT groups for this purpose \cite{LS82}, which has been used in extended GUT models \cite{ZeeDW82,Frampton82}. 

\begin{figure}[!t]
\begin{center}
\includegraphics[width=0.6\textwidth]{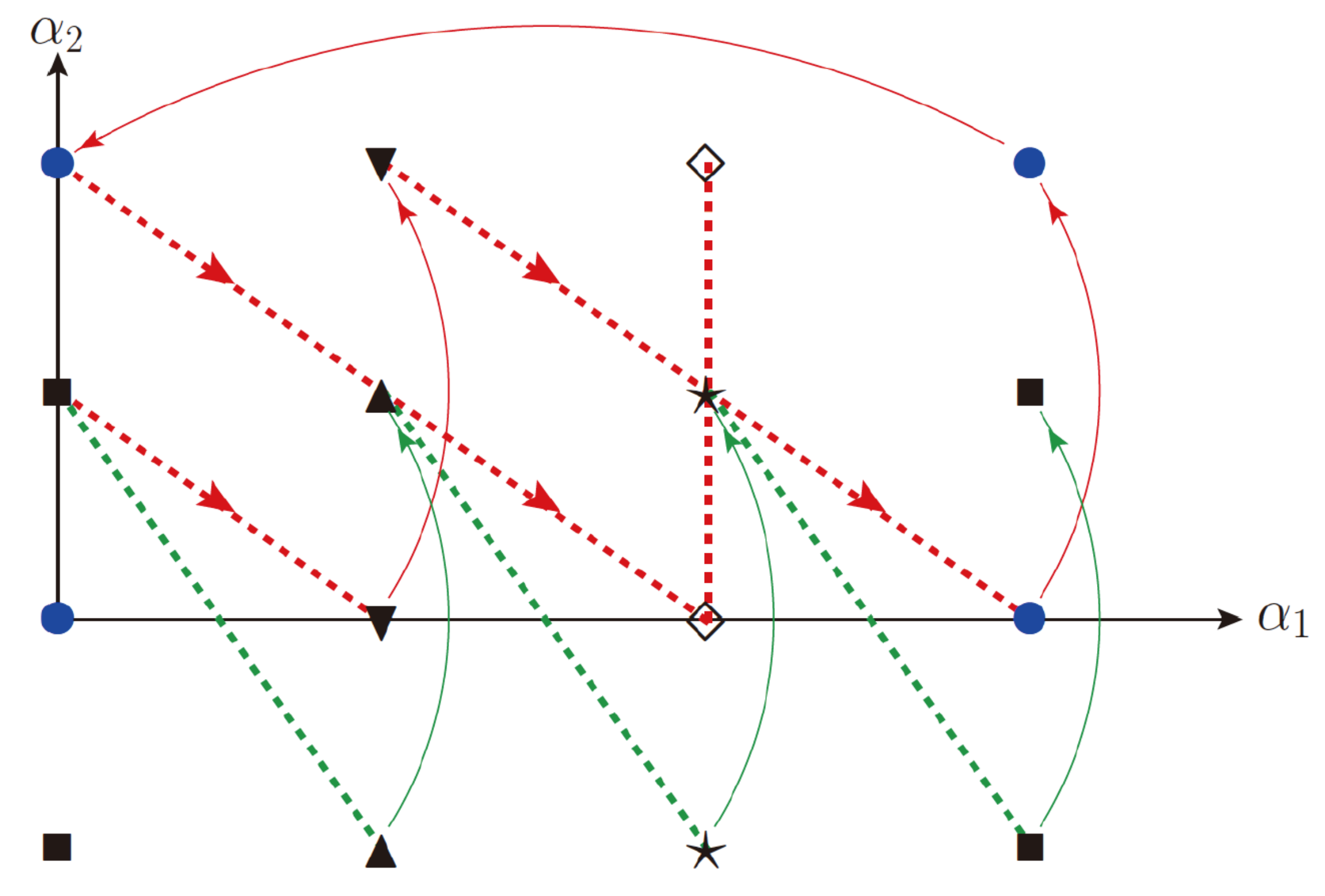} 
\end{center}
\caption{The Goldstone boson direction out of two pseudoscalars for $N_1=3$ and $N_2=2$ \cite{ChoiKimDW85}.} \label{fig:GoldDirection}
\end{figure}

But, more practical solutions come from using two discrete symmetry groups. This method can be extended to the Goldstone boson directions of spontaneously broken global symmetries \cite{ChoiKimDW85,KimPLB16}.  In Fig. \ref{fig:GoldDirection}, we consider two discrete symmetries with $\Z_3$ and $\Z_2$. $\alpha_1$ and $\alpha_2$ are the identifying directions. In case of Goldstone directions, two pseudoscalars are $a_1=f_1\alpha_1$ and $a_2=f_2\alpha_2$. If the potential is nonvanishing only in one direction,  then there is another orthogonal direction in which direction the potential is flat. This flat direction is the Goldstone boson direction. For the discrete symmetry $\Z_3\times\Z_2$ of Fig. \ref{fig:GoldDirection}, there are six inequivalent vacua, marked as  blue bullet ($\color{blue}\bullet$), black down triangle ({\small$\blacktriangledown$}), diamond ($\diamond$), black square ({\tiny$\blacksquare$}), black triangle ({\small$\blacktriangle$}), and star ($\star$). Identifying along the red dash-arrow directions encompass all six vacua, where discrete group identifications along $\alpha_1$ and $\alpha_2$ are shown as horizontal and vertical arcs, respectively. Notice, however, if we identify along the green dash-arrow directions, then only {\tiny$\blacksquare$}, {\small$\blacktriangle$}, and $\star$ are identified. Parallel to the  green dash-arrow, there is another identification of $\color{blue}\bullet$,  {\small$\blacktriangle$}, and  $\diamond$. Thus, there remains $\Z_2$.
This example shows that all Goldstone boson directions are not necessarily identifying all vacua, even if $N_1$ and $N_2$ are relatively prime. The difference of red and green directions is a possibility of allowing a discrete group or not. In the red dash arrow case, when one unit of $\alpha_1$ is increased, one unit of $\alpha_2$ is increased. In this case, if  $N_1$ and $N_2$ are relatively prime, then all vacua are identified as the same vacuum, forbidding any remaining discrete group.  In the green dash arrow case, two units of $\alpha_2$ is increased while one unit of $\alpha_1$ is increased, and $\Z_2$ remains unbroken. This fact was not noted in the earlier papers  \cite{ChoiKimDW85,KimPLB16}. Since the Goldstone boson is derived from VEVs of two Higgs fields in the above example, the possibility of identifying all vacua depends on the ratio of two VEVs. 

Global symmetries are anyway broken at some level \cite{Barr92,Kam92,holman92}. Thus, using the Goldstone boson direction has to check whether the global symmetry breaking is safe from its DW solution. There also exists a cosmological solution, using a new confining gauge group, where the evolution history chooses $\NDW=1$ at high temperature  and then is made safe from the later cosmic evolution \cite{BarrKim14}.

But, the most appealing solution is to use an exact global symmetry, which is spontaneously broken. Indeed, there exists an appealing example. It is physics related to the model-independent axion (MI-axion)  arising from string compactification, which is known to have $\NDW=1$ \cite{Witten85,KimPLB16}. There is a problem on the scale of $f_a$ with the MI-axion. Because of the QCD axion contribution to the energy density of the Universe as discussed in  Subsec. \ref{subsec:rhoa}, $f_a$ is allowed only in the window if the initial misalignment angle is O(1),  $10^{10\,}\gev\le f_a\le 10^{12\,}\gev$. Anyway, this window is the region for the popular axion detection experiments are going on now \cite{cappsite}. [A previous study of ``invisible'' axion energy density before Ref. \cite{KimSKim18} was given in Ref. \cite{Bae08}.]

The origin of the MI-axion degree is the antisymmetric tensor gauge field $B_{MN}\,(M,N=1,2,\cdots,10)$ where both indices $M$ and $N$ take the tangential Minkowski space direction $\mu,\nu=1,\cdots,4$ \cite{Witten84}.
For this tangential direction of $D=4$, the number of degrees is 1 because $ _{D-2} C_{2}=1$. Thus, it is a spin-0 field and pseudoscalar, and hence is called the MI-axion. However, the corresponding decay constant is in the region $10^{15-16\,}\gev$ \cite{ChoiKimfa85}. So, it cannot be the ``invisible'' axion we hope for BCM dark matter with $f_a$ in the axion window.

At the GUT scale, we need a global symmetry rather than the above pseudoscalar degree \cite{ChoiKimfa85}, MI-axion, such that the global symmetry is broken at the intermediate scale.  Understanding this exact global symmetry without the cosmological disaster comes from the elementary mechanism called 't Hooft mechanism \cite{Hooft71} discussed in Subsec. \ref{subsec:Hooft} and briefly commented in Subsec. \ref{subsec:GlFromSt}.
  
\begin{figure}[!t]
\begin{center}
\includegraphics[width=0.6\textwidth]{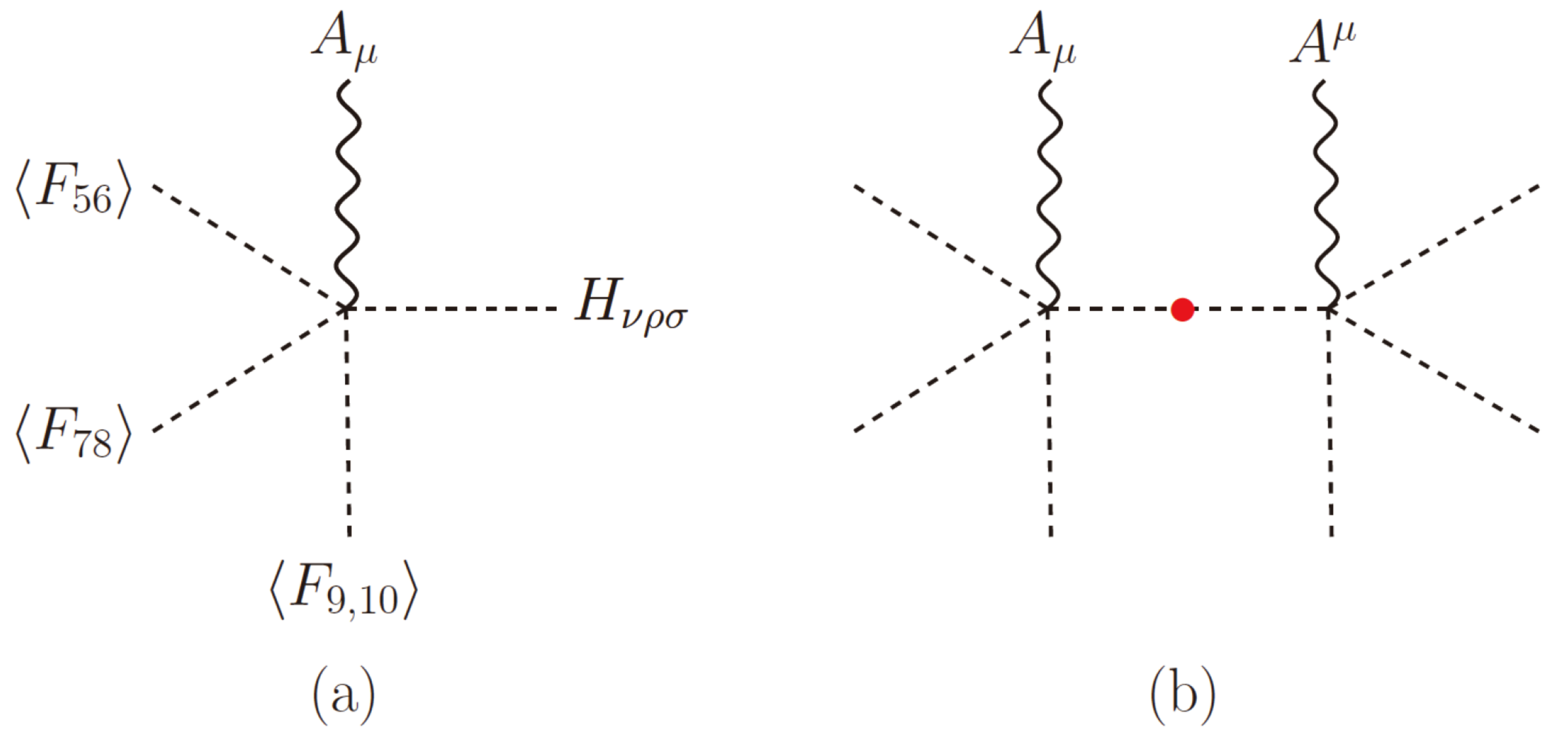} 
\end{center}
\caption{The GS terms.} \label{fig:GS}
\end{figure}

The standard way to represent the pseudscalar direction is by the dual of its field strength \cite{Witten84}
\dis{
 H_{\mu\nu\rho}=M_{\rm MI}
 \epsilon_{\mu\nu\rho\sigma}\,\partial^\sigma a_{\rm MI}.  \label{eq:afromH}
}
$H_{\mu\nu\rho}$ couples to the non-Abelian gauge fields by the Green--Schwarz (GS) term \cite{GS84}. The GS term is composed of a product with one $B_{MN}$ and four non-Abelian gauge fields, $F_{PQ}, $ etc., contracted with $\epsilon^{MNPQ\cdots}$.
 Figure \ref{fig:GS}\,(a) shows the coupling of $H_{\mu\nu\rho}$  with $A_\sigma$ by transferring one derivative of $F_{\mu\nu}$ to $B_{\rho\sigma}$.  Figure \ref{fig:GS}\,(b) shows the gauge boson mass from this term.
Thus, we obtain  
\dis{
\frac12 \partial^\mu  a_{\rm MI}\partial_\mu a_{\rm MI}+    M_{\rm MI}A_\mu\partial^\mu a_{\rm MI}+\frac12 M_{\rm MI}^2A^2_\mu.  \label{eq:HooftMI}
}
The question is when the GS term is generating the coupling $A_\mu\partial^\mu a_{MI}$. It is when there appears an anomalous U(1) gauge symmetry from compactification of 10D $\EE8$ down to a 4D gauge group \cite{Anom87,ichinose87}. The  anomalous gauge symmetry \Uanom~is a subgroup of $\EE8$ with gauge field $A_\mu$ and $a_{\rm MI}$ is from $B_{\mu\nu}$, both of which are local symmetries in 10D. In the heterotic string compactification, there is only one \Uanom. One phase, \ie $\alpha=a_{\rm MI}/M_{\rm MI}$, is working for the 't Hooft mechanism, \ie the coupling $M_{MI}A_\mu\partial^\mu a_{MI}$ in Eq. (\ref{eq:HooftMI}), and hence one global symmetry survives below the compactification scale $M_{\rm MI}$.  The compactification scale $M_{\rm MI}$ is expected to be much larger than the decay constant $f_{\rm MI}$ of the MI-axion estimated \cite{ChoiKimfa85}.
In the orbifold compactification, there appear many gauge U(1)'s which are anomaly free except the \Uanom. After the global \Uanom~is surviving below the scale $M_{\rm MI}$,   the 't Hooft mechanism can be applied repeatedly until all anomaly free gauge U(1)'s are removed around the GUT scale. One anomaly free gauge U(1) which is required down to the electroweak scale is U(1)$_Y$ of the SM. Then, at the intermediate scale $10^{9\sim 11\,}\gev$, one VEV $f_a$ of a SM singlet scalar $\phi$ breaks the global symmetry \Uanom~spontaneously and there results the needed ``invisible'' axion at the intermediate scale. There is another bonus of $\NDW=1$ in this  ``invisible'' axion from \Uanom~global symmetry.
 
Let us first discuss in a hierarchical scheme and then present a general case based on the generalized 't Hooft mechanism.  In the literature, the Fayet--Iliopoulos terms (FI-term) for \Uanom~has been discussed extensively.  In the hierarchical scheme,  the VEVs of scalars is assumed to be much smaller than the string scale. Then, one can consider   the global symmetry \Uanom, surviving down from string compactification. Even if one adds the FI-term  for \Uanom, $|\phi^* Q^a\phi-\xi|^2$ with $\xi\ll M_{string}^2$, it is not much different from considering the global symmetry with the usual D-term,  $|\phi^* \Qanom \phi|^2$ (as if there is a gauge symmetry) since $\xi\ll M_{string}^2$. In fact, there exists a string loop calculation obtaining $\xi$ at string two-loop \cite{AtickTwoLoop88}. If the string two-loop calculation turns out to be hierarchically smaller (because of two-loop) than the string scale, then the above  hierarchical explanation works. Even if the FI parameter $\xi$ is large, still there survives a global symmetry.   It is based on just counting the number of continuous degrees of freedom. Let us consider two phase fields, the MI-axion and some phase of a complex scalar carrying the U(1)$_{\rm anom}$ charge. Since we consider two phases and two terms, one may guess that the gauge boson ($A^\mu_{\rm anom}$) obtains mass and the remaining phase field also obtains mass by the FI D-term. But, it does not work that way, because there is no potential term for the phase field, rendering such mass to the remaining phase field, because the charges of the gauge U(1) from $\EE8$ and the charge operator $Q_a$ in the FI D-term are identical. It is equivalent to that there is no mass term generated because the exact Goldstone boson direction (the longitudinal mode of $A^\mu_{\rm anom}$) coincides with the phase of $\phi$ in the FI D-term.

We illustrate the case with one anomaly free U(1) gauge boson $A_\mu$ and the FI D-term for $\phi$ with generator $\Qanom$. Since $\phi$ carries the gauge charge $\Qanom$, we obtain its coupling to $A_\mu$ from the covariant derivative, by writing  $\phi=(\frac{v+\rho}{\sqrt2})e^{ia_{\phi}/v}$,
\dis{
 |D_\mu \phi|^2 &=|(\partial_\mu -ig
Q_{a}A_\mu)\phi|^2_{\rho=0}= \frac12(\partial_\mu a_{\phi})^2-gQ_a A_\mu \partial^\mu a_{\phi}+\frac{g^2}{2}Q_a^2v^2 A_\mu^2 .\label{eq:AmassPhi}
}
In addition, the gauge boson $A_\mu$ has the coupling to $a_{\rm MI}$ by the GS term, the sum of two terms is
\dis{
 \frac12\left( g^2Q_a^2v^2 \right)(A_\mu)^2+A_\mu( M_{\rm MI}\partial^\mu a_{\rm MI} -g Q_a v \partial^\mu  a_{\phi} )+\frac12\left[ (\partial_\mu a_{\rm MI})^2+(\partial^\mu  a_{\phi})^2 \right].
\label{eq:scaleGl}   }
Thus, we note that $\cos\theta\,a_{MI} -\sin\theta\,a_{\phi} $ becomes the longitudinal degree of $A_\mu$ where
\dis{
 \sin\theta=\frac{g Q_a v}{\sqrt{M_{\rm MI}^2+g^2Q_a^2v^2}},
}
and a new global degree direction is
\dis{
\theta_{\rm QCD}\propto \cos\theta\,a_{\phi} +\sin\theta\,a_{\rm MI}.
 }
 $\theta_{\rm QCD}$ is the QCD vacuum angle direction and breaking \Uanom~at the intermediate scale produces the ``invisible'' axion. We obtained this important result from that only one combination of the phase fields is removed since the longitudinal degree of $A_\mu$ chooses the same generator for the shifts of $a_{\rm MI}$ and $a_\phi$.  Note that choosing $\Qanom$ is not unique because one can add any combination of anomaly free gauge charges to $\Qanom$, without changing physics of \Uanom~\cite{KimKyaeNam17}.

 \begin{figure}[!t]
\begin{center}
\includegraphics[width=0.7\textwidth]{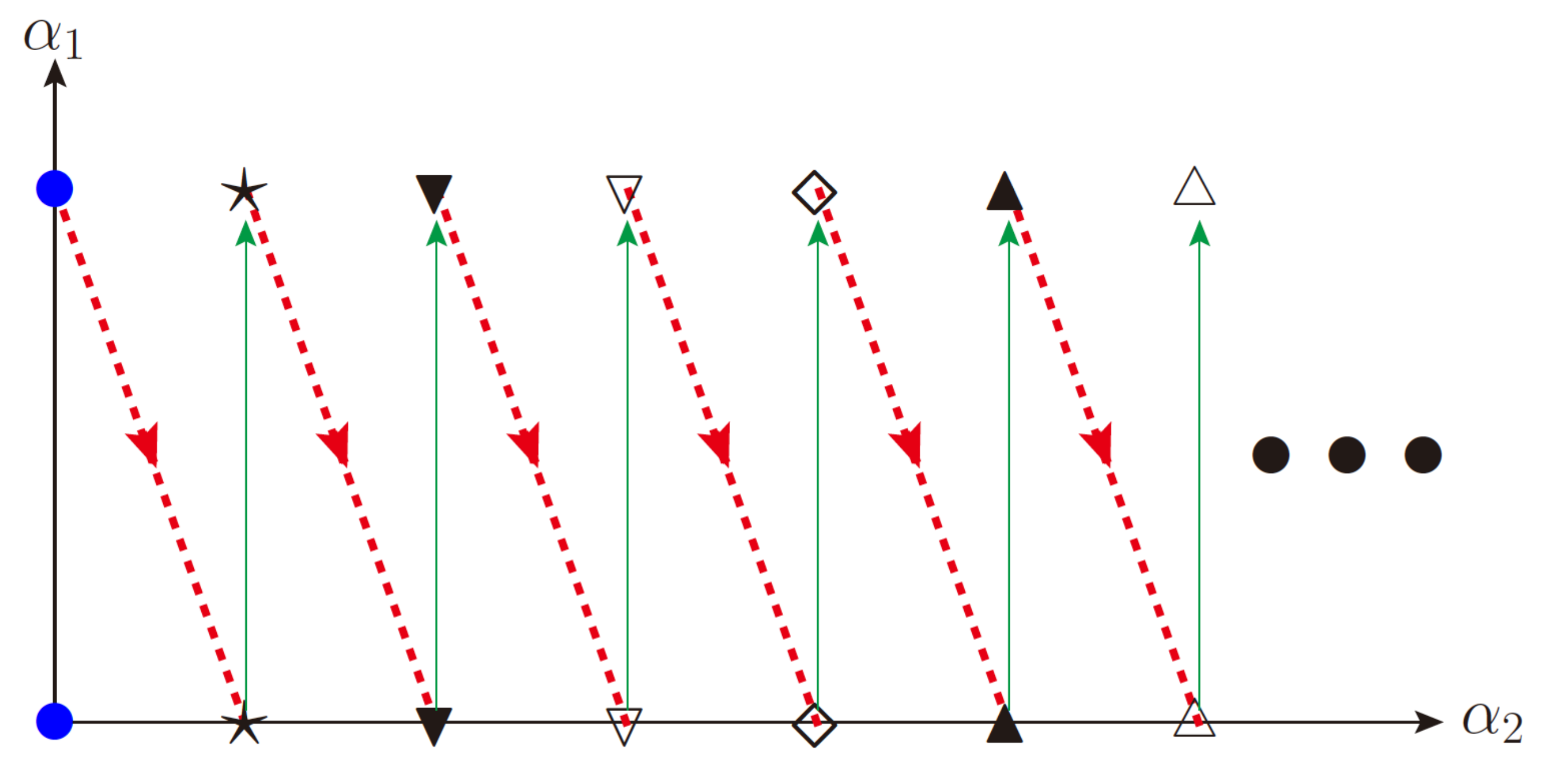} \\ (a)\vskip 1cm
\includegraphics[width=0.7\textwidth]{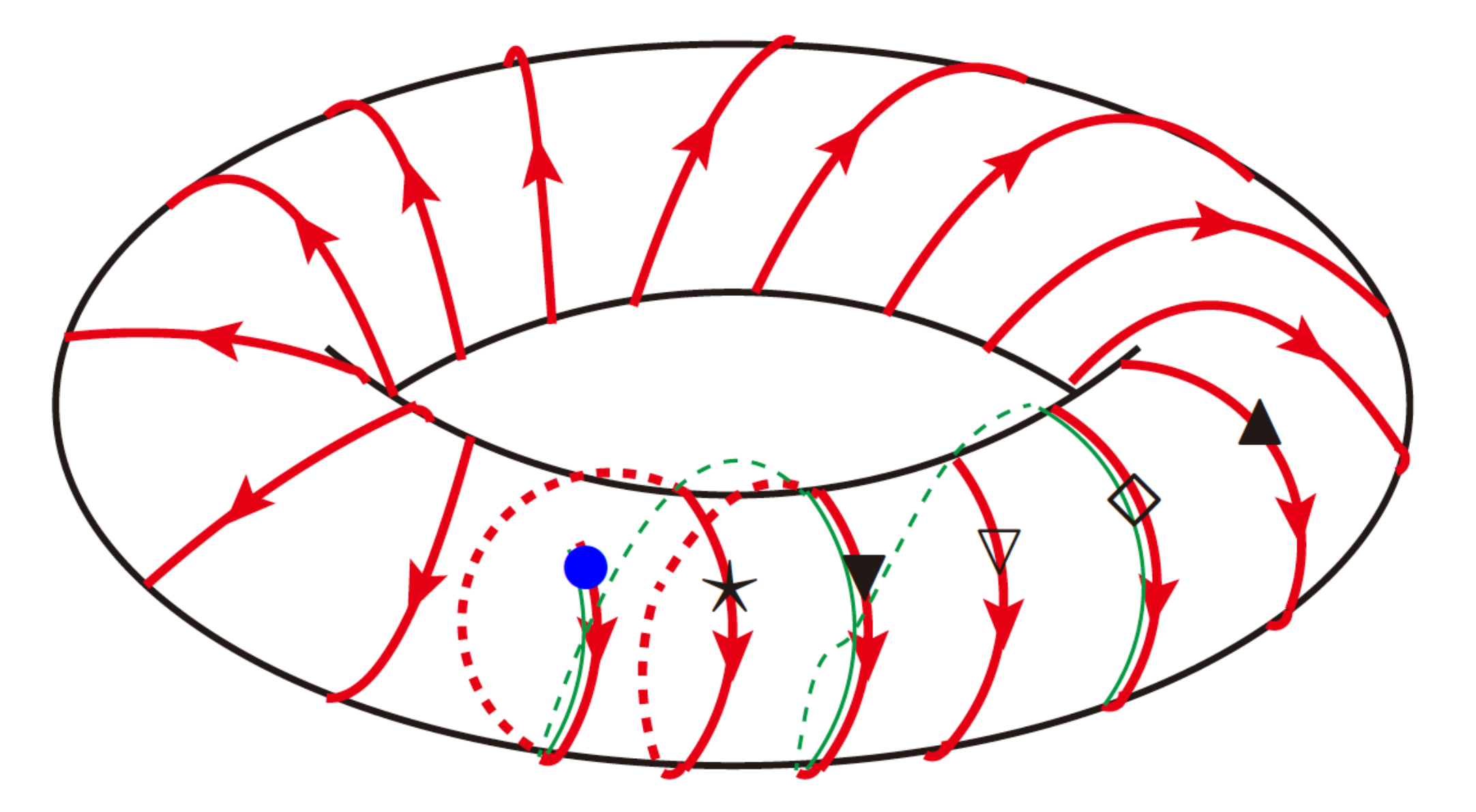} \\(b)
\end{center}
\caption{The MI-axion example of Fig. \ref{fig:GoldDirection}: (a) the standard torus identification, and (b) identification by the winding direction in the torus.} \label{fig:DWMIax}
\end{figure}

Even if the intermediate scale is obtained by the 't Hooft mechanism, if $f_a$ is smaller than the Hubble parameter $H_{\rm infl}$ at the end of inflation, then the DW problem should be avoided. The strategy with the Goldstone boson direction \cite{ChoiKimDW85,KimPLB16} was discussed in Fig. \ref{fig:GoldDirection}. For the Goldstone boson from the global \Uanom, it is repeated in
Fig. \ref{fig:DWMIax}\,(a). Since $\NDW=1$ in the MI-axion direction ($\alpha_1$  in Fig. \ref{fig:GoldDirection}), the red dash arrow direction identifies all vacua. In Fig.  \ref{fig:DWMIax}\,(b), it is re-drawn on the familiar torus. The red arrows show that $\alpha_2$ shifts  by one unit for one unit shift of $\alpha_1$. In this case, all the vacua are identified and we obtain $\NDW=1$. The green lines show that $\alpha_2$ shifts  by two units for one unit shift of $\alpha_1$. If $N_2$ is even, then we obtain $\NDW=2$ since only halves of $N_2$ are identified by green lines. To find out $\NDW$, it is useful to factorize $N_2$ in terms of prime numbers.  Even though $N_2$ is very large \cite{KimKyaeNam17,KimPRD17}, of order $10^3$, there are plenty of relatively prime numbers from those factors in $N_2$.  Figure  \ref{fig:DWMIax}\,(b) is drawn with $N_2=17$ and the green lines also identify all vacua since 1 and 17 are relatively prime. In string compactification, it is easy to find many 
relatively prime numbers  to all prime numbers appearing in the factors  of Tr\,$\Qanom$. For example, in Eq. (\ref{eq:TrQanom}),   Tr\,$\Qanom$ was cited as 3492 which is expressed in terms of prime numbers as $2^2\times 3^2\times 97$. Not to introduce a fine-tuning on the ratio of VEVs, if we consider two VEVs are comparable, let us look for prime numbers relative to 2, 3, and 97. Near 3492, there are 3491, 3493, 3497, 3499, etc., relatively prime to 2, 3, and 97. So, a VEV of $\phi$ near the string scale and the gauge \Uanom~charge render a global symmetry below the compactification scale such that the global current satisfies 
\dis{
\partial_\mu J^\mu_{\rm PQ}=\frac{1}{32\pi^2}
G^a_{\mu\nu} \tilde{G}^{a\,\mu\nu}.
}
If we neglect anomaly free gauge U(1)'s, a scalar field $\sigma$ carrying the PQ charge houses the ``invisible axion'' with decay constant $f_a$ if $\langle\sigma\rangle=f_a/\sqrt2$. The intermediate scale $f_a$ is not considered to be a fine tuning.  $f_a$ can come from another mechanism such as the supergravity scale \cite{KimScale84} or by some solution of the gauge hierarchy problem.
      Thus, forbidding the DW problem for the ``invisible'' axion from the global \Uanom~is not considered as a fine-tuning  on the ratio of VEVs of Higgs fields as explained above. Due to the 't Hooft mechanism we discussed before, the ``invisible'' axion scale can be lowered from the string scale down to an intermediate scale  \cite{KimKyaeNam17}.

This ``invisible'' axion from the global \Uanom~can be proved or disproved to exist in Nature by the cavity experiments \cite{cappsite}.

\subsection{Cosmic energy density}\label{subsec:rhoa}
  
The evolution equation of axion field includes the Hubble expansion,
\dis{
 \ddot\theta+3H\dot\theta+\frac{m^2}{2}\sin\theta=0,\label{eq:diffeq}
}   
where the angle $\theta=A/f_a $ is the axion field and $m$ is the axion mass.
At a cosmic time scale  $3H\sim m$, $\ddot\theta$ is negligible and Eq. (\ref{eq:diffeq}) determines an angle $\theta_1$ which is called the initial misalinement angle. The cosmic temperature determining $\theta_1$ is $T_1$ which is known to be $T_1\approx 1\,\gev$ \cite{Preskill83}.  In addition, the adiabatic condition that the number of axions is conserved was used, which is valid if the potential contains only the axion mass term. However, the anharmonic term $\lambda a^4$ or the higher order terms of the cosine function can be important, in which case the axion number is not conserved. This effect was considered before \cite{Bae08}. Recently, it has been studied carefully with the tunneling idea implemented by formation of true vacuum bubbles \cite{KimSKim18} as described in Sec. \ref{subsec:QCDphase}.

   Naively, one would expect that the lattice calculation might give a very different ``invisible'' axion energy density than the earlier estimate,  but  if we compare the result with the naive high temperature value,  both estimates give almost the same value for the axion energy density as shown in the last two equations of Eq. (\ref{eq:AxPhmasses}), and the  estimated axion window is   $10^{9\,}\gev<f_a<10^{11\,}\gev$.     The QCD phase transition discussed in Subsect. \ref{subsec:QCDphase} must have played a role here. Firstly,  $\Lambda_{\rm QCD}$ used in the high energy scattering is describing a single particle effect and  a convenient expansion parameter is   $\Lambda_{\rm QCD}/|Q|$. On the other hand, the susceptibility $\chi$ obtained from the lattice calculation is a quantity in many body phenomena, which takes into account the strong interactions in terms of the coupling constant $\alpha_3$. So,  $\Lambda_{\rm QCD}$ is not directly comparable to $\chi$. Second,   the quark-hadron phase transition  discussed in Subsec. \ref{subsec:QCDphase}, and the bottle neck period \cite{Bae08,KimSKim18} must be taken into account. So, it is important to calculate the axion energy density in the  present Universe with the following issues treated properly:

\begin{figure}[!t]
 \includegraphics[width=0.75\linewidth]{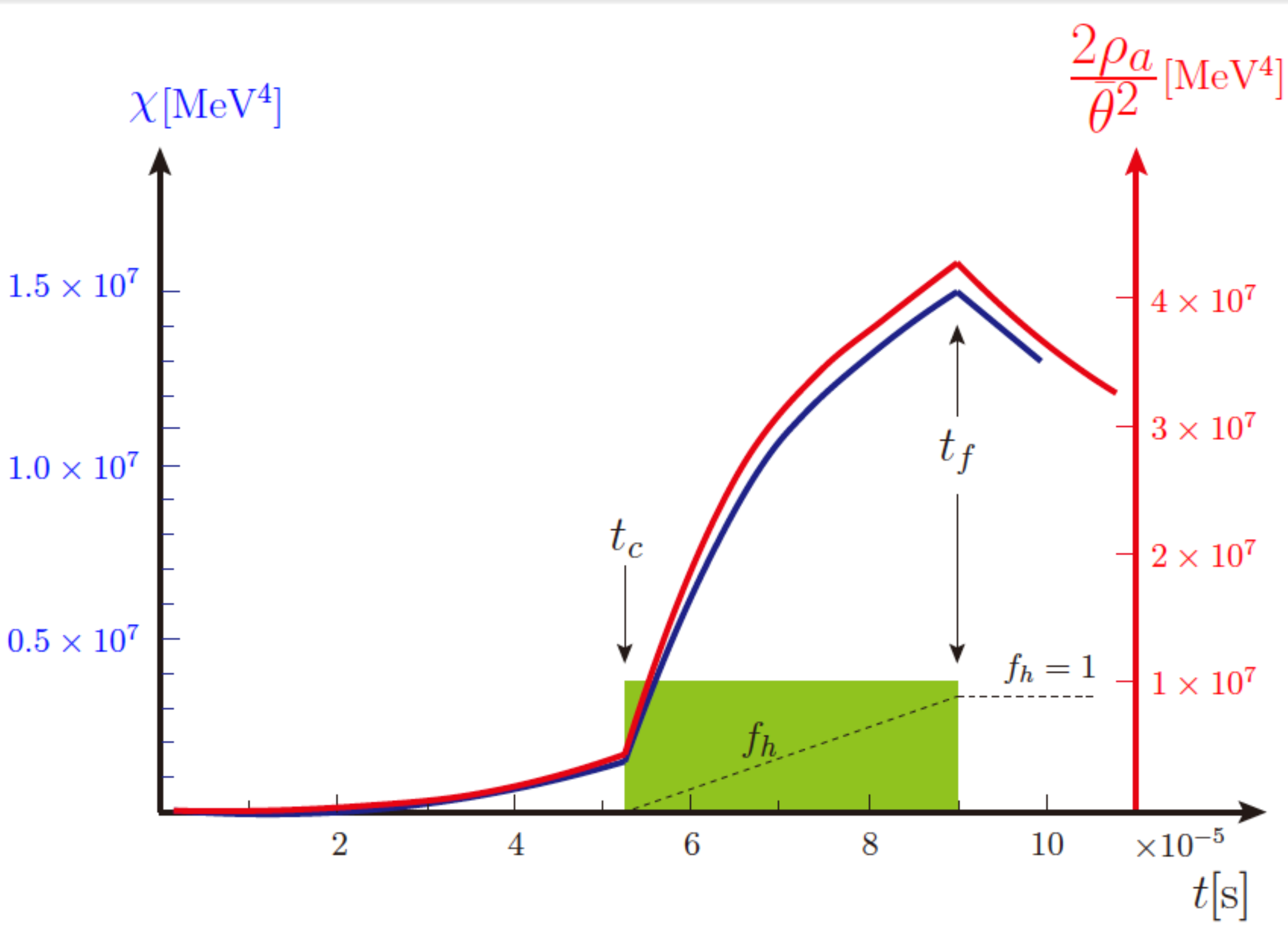}
\caption{A view of susceptibility $\chi$  (blue curve)  during the quark-hadron phase transition, and a typical axion energy density (red curve) for $m_a=100\,\mu\eV$ \cite{KimSKim18}.  }\label{fig:rhoa}
\end{figure}
      \begin{itemize}
\item Include the anharmonic effect carefully. The standard axion window is obtained in the region of the misalignment angle where the anharmonic effect is large.   
\item Use the standard Big Bang cosmology from the cosmic time after which the axion number is (almost) conserved.
\item Include the first order QCD phase transition calculated in field theory as discussed in Sec. \ref{subsec:QCDphase}.
\item Connect smoothly the axion energy density above and below the critical temperature of the QCD phase transition. 
   \end{itemize}
   The axion energy density profile as the function of cosmic time $t$, taking into account the above issues except the first one \cite{KimSKim18}, is summarized in Fig. \ref{fig:rhoa}. Right after the bottle-neck period, the initial mis-alignment angle is denoted as $\thb_f$ \cite{Bae08,KimSKim18}.
After $t_f$ in Fig. \ref{fig:rhoa}, $\thb_f$ decreases as shown in Fig. \ref{Fig:PotAxion}.
 
 \subsection{Searches of ``invisible'' axions}
 
\begin{figure}[!b]
\centerline{\includegraphics[width=8cm]{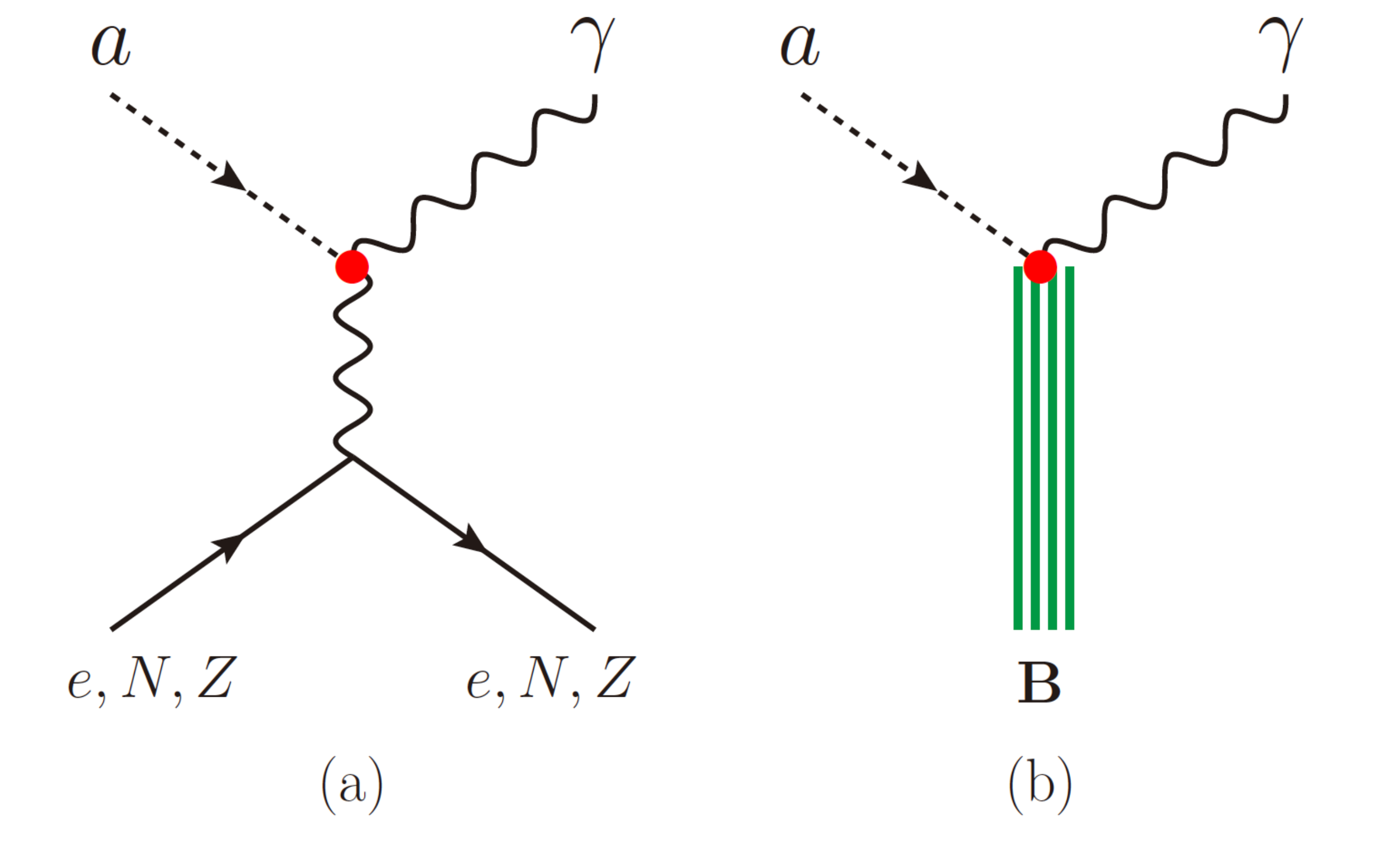}
 }
\caption{The Feynman  diagrams relevant for Solar (a) and cosmic (b) axion detections.}\label{Fig:Primakoff}
\end{figure}

The ``invisible'' axion was pointed out visible, even though the detection rate is very tiny  \cite{Sikivie83}. There are two types of detectors, solar axion detectors (SADs) and cosmic axion detectors (CADs). Both detectors use the Primakoff conversion \cite{Primakoff51} of axion to photon,  $a\,F^{\rm em}_{\mu\nu}
 \tilde{F}^{\rm em\,\mu\nu}$ of Eq. (\ref{eq:EMax}). We show the Feynman diagrams used for SADs and  CADs in  Fig. \ref{Fig:Primakoff}\,(a) and  (b), respectively.
  
In the search of the QCD ``invisible'' axion, the axion--photon--photon coupling  $c_{a\gamma\gamma}$ of Eq. (\ref{eq:EMax}) is the key parameter. The customary numbers presented in most talks are just illustrative examples.    Because the axion decay constant $f_a$ can be in the intermediate scale, the ``invisible'' axions can live up to now ($m_a <24\, \eV$) and constitute DM of the Universe.
 
\begin{figure}[!t]
\centerline{\includegraphics[width=8cm]{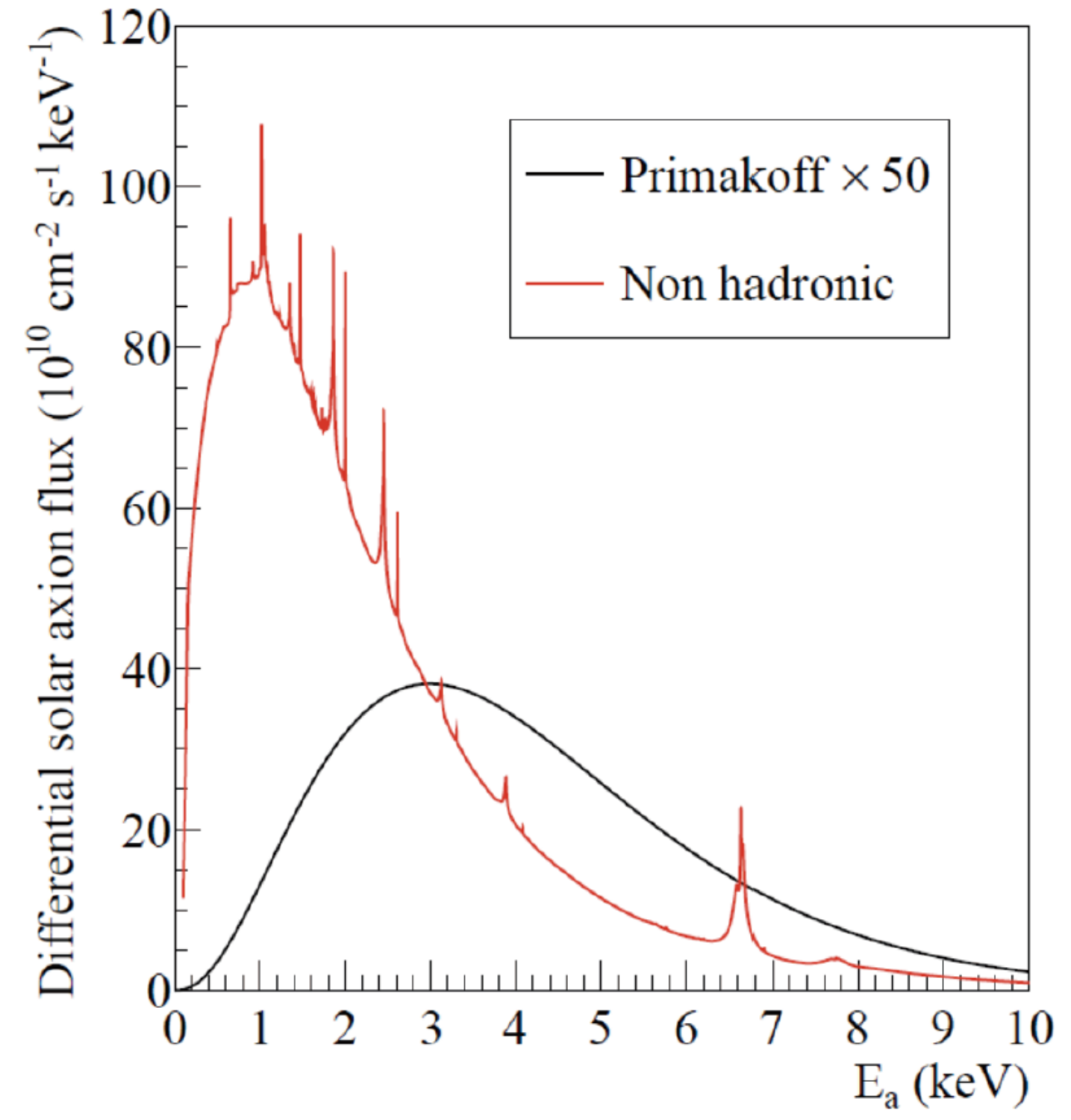}
 }
\caption{The rate of Solar axions given in \cite{IrastorzaVenice17}. The black and red curves are estimated in \cite{vanBibber89} and   \cite{Redondo13}, respectively.}\label{Fig:RateSun}
\end{figure}

Solar axions are hot (x-ray energy) axions produced in the core of Sun at temperature O($10^7$K). So, the energy loss of Sun by solar axions should not exceed that by neutrino loss \cite{Wong13}.  The expected Solar axion spectrum is shown in Fig. 
\ref{Fig:RateSun}, presented in  \cite{IrastorzaVenice17,vanBibber89,
CAST07}. The red curve of  Fig.  \ref{Fig:RateSun} applies to the DFSZ type axions.
The first SAD was  built at Brookhaven using a 2.2 ton iron core dipole magnet oriented to
Sun with a proportional chamber for x-ray detection \cite{Lazarus92}. It was followed by
  the Tokyo group using a 4 tesla superconducting SAD \cite{Minowa98}. The next improvements was by the CERN Axion Solar Telescope (CAST) group \cite{CAST05,CAST11}. These bounds are shown in Fig. \ref{Fig:expbound}. The current effort toward the ultimate SAD is the IAXO which is at the concept design now \cite{IrastorzaVenice17}.

\begin{figure}[!t]
\centerline{\includegraphics[width=6cm]{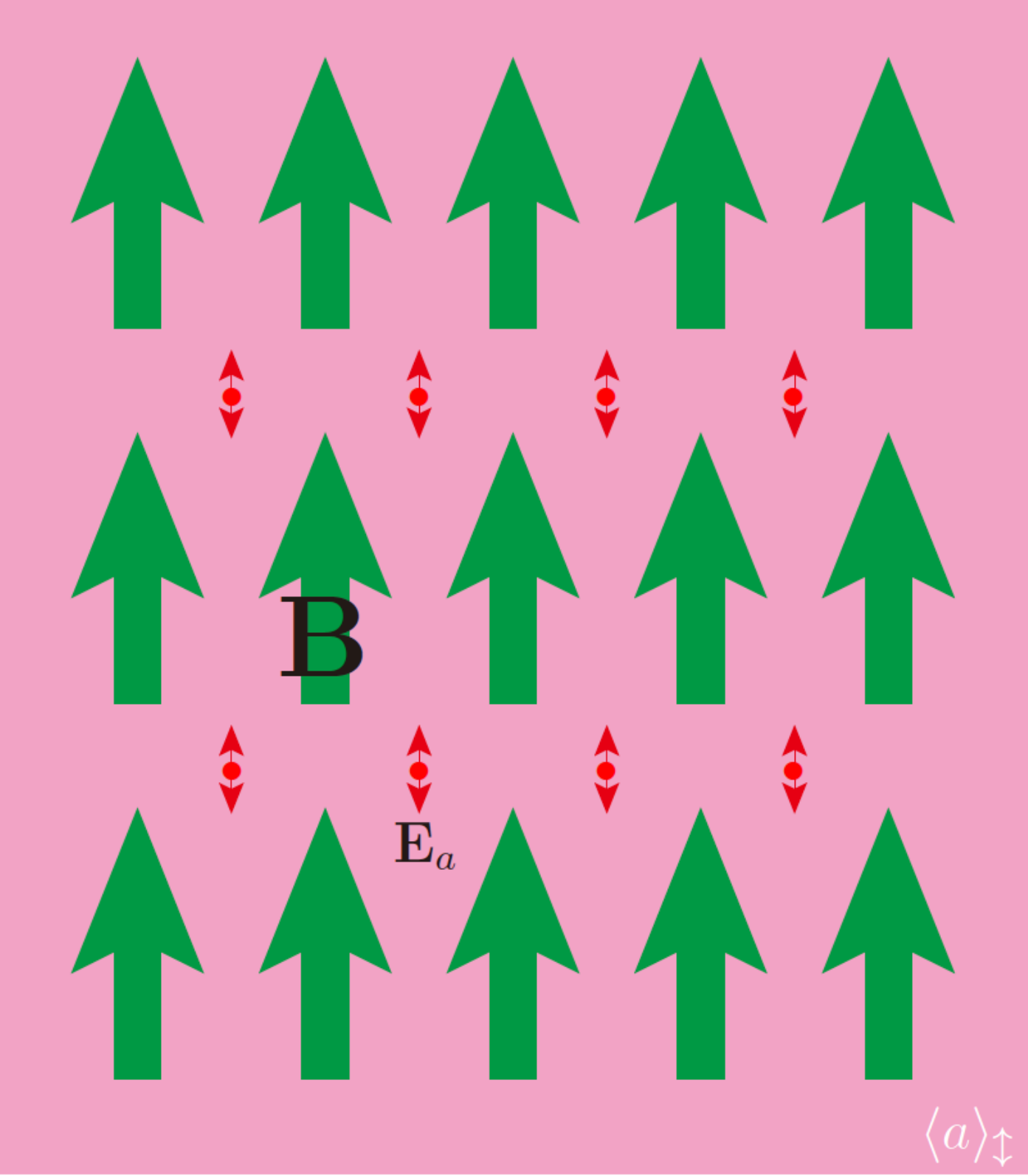}
 }
\caption{The resonant detection idea of the cosmic QCD axion. The {\bf E}-field follows the axion vacuum oscillation.}\label{Fig:AxDetView}
\end{figure}

Cosmic axions oscillate and  the BCM nature in the Universe \cite{KimYannShinji} is the underlying principle of CAD designs.  Equation (\ref{eq:EMax}), \ie  Fig. \ref{Fig:Primakoff}\,(b), gives the form ${\bf E}\cdot {\bf B}$ such that ${\bf E}$ parallel to ${\bf B}$ contributes to the coupling. The usual design \cite{Sikivie83} is a cavity detector  immersed in a {\it strong constant} magnetic field. Then, $ {\bf E}$ follows the cosmic oscillation of the classical axion field $\langle a \rangle$,   generating the axions in the cavity, which is given completely in the axio-electrodynamics in \cite{Hong14}. [Note that similar results with the resonance condition were given earlier \cite{SikKrauss85}.] This idea of using the oscillating {\bf E} is pictorially shown in Fig.  \ref{Fig:AxDetView}.

After the discovery of the Higgs boson which seems to be a  fundamental elementary particle, the possibility of the QCD axion being fundamental gained some weight. The future axion search experiment can detect such a fundamental axion, even its contribution to CDM is as low as 10\% \cite{cappsite}.   Cosmology of the ``invisible'' axions has  started in 1982--1983 \cite{Preskill83,AbbottSik83,DineFish83} with the micro-eV axions \cite{KSVZ1,KSVZ2,DFSZ,DFSZ2}. 
 
For the detection probabilty of cosmic axions, the axion energy density $\rho_a$ as CDM ($\sim 0.3\,\gev\,{\rm cm}^{-3}$) is used.   
Relating $\rho_a$ with $f_a$ depends on the history of the evolving Universe due to the contribution to $\rho_a$ from the annihilating string-wall system of axions.   The recent numerical calculation for an $\ndw=1$ model shows that more than 90\,\% of $\rho_a$ is contributed from the annihilating string-wall system \cite{Kawasaki12}, which however did not include the effects shown in Fig. \ref{Fig:DWon}.
 
The window of $f_a$ near $10^{10}\gev$ is the one obtained from the currently oscillating axion field. It depends on the initial misalignment angle $\theta_1$. Most probably, $\theta_1$ is order 1, but if it is much smaller, the currently oscillating amplitude of axion field would be correspondingly smaller. Then, the $f_a$ window can be wider. If $\theta_1$ is O($10^{-3}$), the window can be open to the GUT scale. 
This region of the GUT scale $f_a$ is discussed as `anthropic window'.
There have been CAD designs for the anthropic window axions \cite{CADanthropic}.
The 2014 BICEP2 report of ``high scale inflation at the GUT scale'' around $\gtrsim(10^{16\,}\gev)^4$ implied the reheating temperature after inflation $\gtrsim 10^{12\,}\gev$ \cite{BICEP14}. Then, studies on the isocurvature constraint with that BICEP2 data pinned down the axion mass in the upper allowed region \cite{Marsh14}.  However, more data, with dust contamination taken into account, has not lived up with this earlier report \cite{BICEP16}. But, it is likely that the energy density during inflation might be near the GUT scale, predicting a somewhat smaller tensor to scalar ratio $r$ at O(0.01) than the initial report of O(0.1) \cite{BICEP14}. Even for this one order smaller value of $r$,  requiring $\NDW=1$ is a necessity. So, it is very important in axion search experiments to know the tensor to scalar ratio $r$.
 
The natural choice of $\theta_1$ prefers the usual axion window for the axion mass, in particular in the $10^{-4}$ eV region. To find axions in this region, one has to design cavities capable of detecting microwaves of wave length O(cm). Making this size of cavities is the most difficult part. There are several ideas in this direction, (i) placing the detector vertically at the geometric center of the toroidal magnet \cite{CADanthropic}, (ii) designing with multiple sheets of dielectric material \cite{Dielectricplates}, (iii) placing multiple cavities inside the same magnetic field \cite{Multiple,Jeong17}, and (iv) designing pizza-cylinder cavities \cite{Pizza}. 

 \begin{figure}[!t]
\begin{center}
\includegraphics[width=0.65\linewidth]{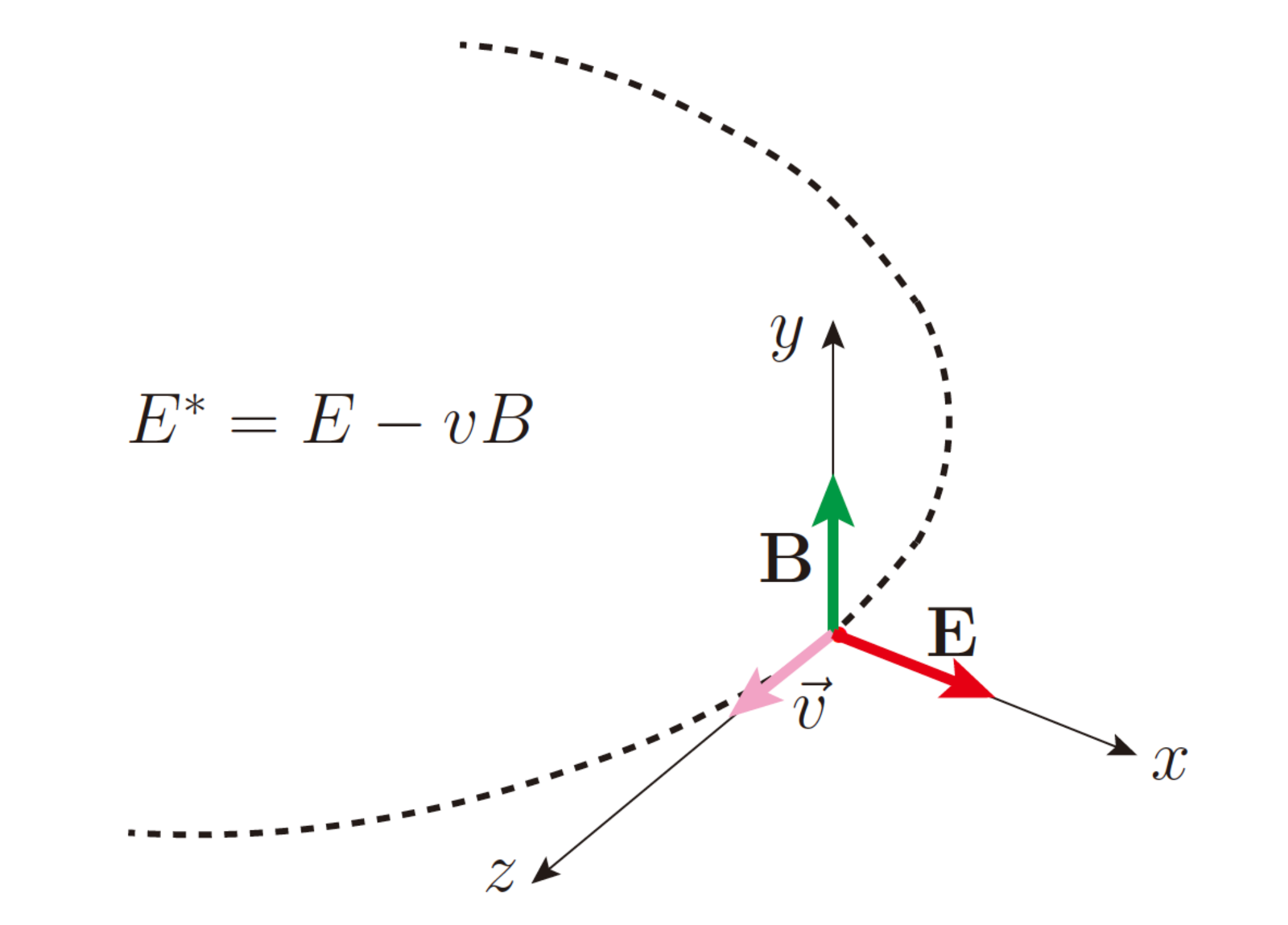} 
\end{center}
\caption{ The coordinate system used for particle storage and {\bf E} and {\bf B} calculation. The EDM precession takes place by the effective electric field {\bf E}$^*$. } \label{fig:pEDM}
\end{figure}

There is an ingenious design to detect $f_a$, far beyond the axion window at $10^{13}\gev\lesssim f_a\lesssim 10^{19}\gev$ in the anthropic window, which uses the  EDM   search method of a particle in a  storage ring \cite{pEDMexp}. The directions of {\bf E},  {\bf B}, and   circulating beam are indicated in Fig.  \ref{fig:pEDM}. With a nonzero $g-2$ frequency, the average EDM precession angle becomes zero for the static  EDM case because the relative {\bf E} field direction to the spin vector {\bf s}  changes within every cycle of $ g-2 $ precession. The spin {\bf s} of proton satisfies
\dis{
\frac{d\bf s}{dt}= {\bf d} \times  {\bf E}^*
}
where ${\bf d}$ is the EDM, and 
\dis{
{\bf E}^*={\bf E}+ {\bf v}\times {\bf B}.
}
  The method discussed in detail in Ref.   \cite{pEDMexp} uses the resonance between the axion frequency and $g-2$ frequency. The EDM accumulates only when the two frequencies match. 
 The tunable frequency range for the example calculated with storage ring bending radius of 10 m was   $10^2-10^7$ Hz for Table I of Ref. \cite{pEDMexp}, which corresponds to $m_a\sim 0.7\times(10^{-13}-10^{-8})$~eV, or $f_a\sim (10^{20\to 19}-10^{15\to 13})$ GeV. [Note added in proof: The recent report given in Ref.  \cite{NMRtransPl}, excluding $f_a=[0.6\times 10^{24}\gev, 0.6\times 10^{31}\gev]$], is far off to the trans-Planckian region.] 
The EDM follows the  oscillation of the classical axion field, $\langle a(t)\rangle$. Thus, if {\bf s} changes the direction every half cycle of the $g-2$ precession, then EDM oscillation could add up with appropriate condition. This storage ring  
method uses a combination of {\bf B} and
{\bf E} fields to produce a resonance between the $g-2$ precession frequency and the background axion field oscillation to greatly enhance the sensitivity to it \cite{pEDMexp}. The field directions shown here are applicable to the deuteron storage ring case. The merit of this storage ring method is the type used in standard particle physics experiments and proves the frequency of the EDM  oscillation, implying the existence of ``invisible'' axion with mass corresponding to that precession frequency.
   
\subsection{Bose--Einstein condensation}\label{subsec:BEcond}
 
\begin{figure}[!b]
\begin{center}
\includegraphics[width=0.85\linewidth]{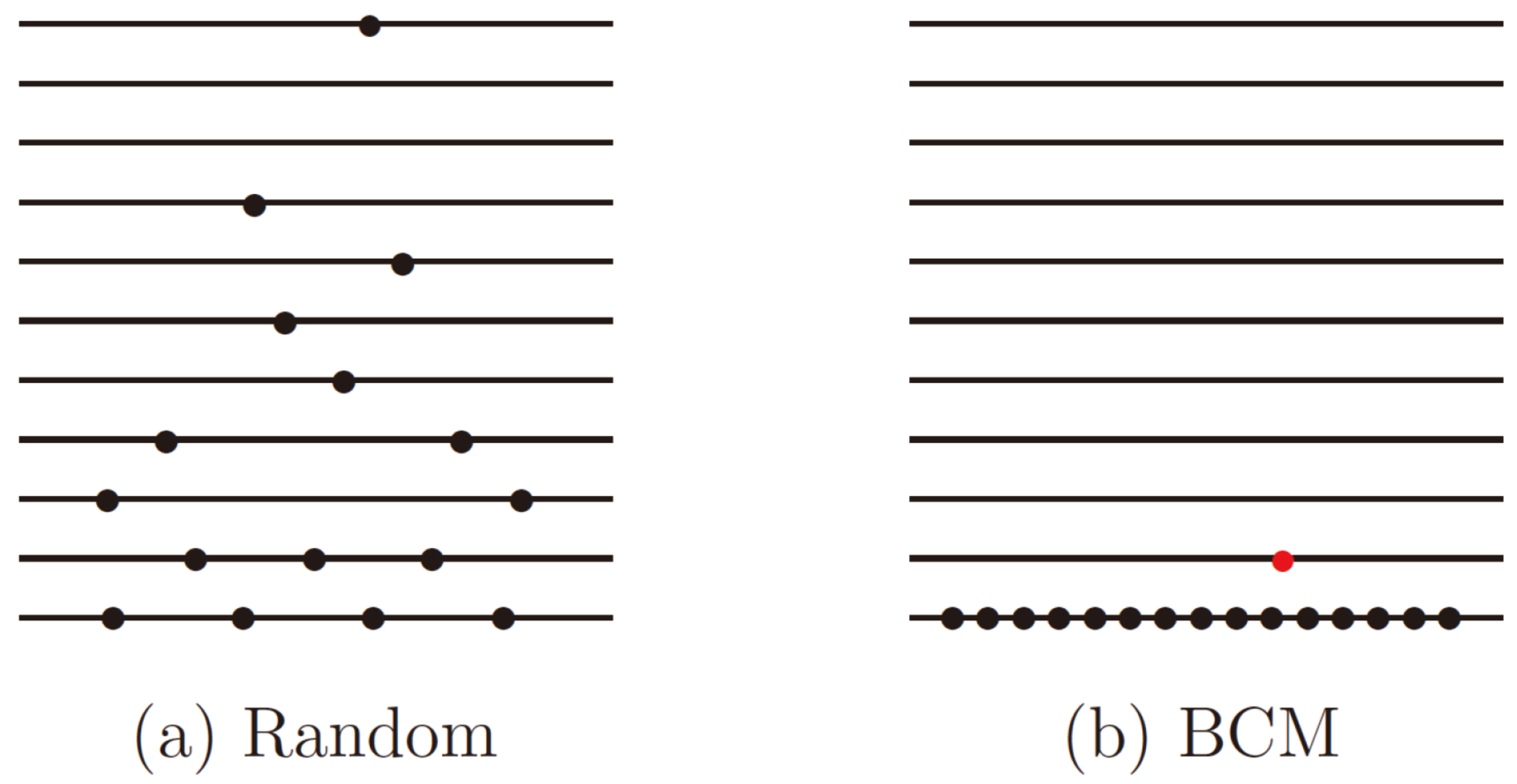}
\end{center}
\caption{Energy levels of harmonic oscillator. (a) Levels are filled randomly. (b) At very low temperature, most particles fill the ground states.
} \label{fig:BCMenergy}
\end{figure}

In quantum mechanics with many particles, the ground state level with $E=\frac12 \hbar\omega$ is filled by particles. In our case, the ground state is filled by bosonic particles. Being boson, the ground state of $\phi$ invites more bosons to be filled with the same quantum numbers, e.g. with the same momentum.  This is the reason we call their motion {\it collective}.\index{collective motion}\index{BCM!collective motion}   Even though the energy is split into potential and kinetic energies, their sum remains the same in the harmonic motion. Asuming that thermalization of the particles is achieved, this situation is shown in Fig. \ref{fig:BCMenergy}. In Fig. \ref{fig:BCMenergy}\,(a), higher levels are  filled due to the temperature effect. At a sufficiently low temperature, most particles fill the ground state as shown in Fig. \ref{fig:BCMenergy}\,(b), which is the realization of the Bose--Einstein condensate.\index{Bose--Einstein condensate} In this case, the particles in the ground state have the same momentum.

For the case of Fig. \ref{Fig:PotAxion}, the sum is the value of $V$ at the bullet. This sum is the energy within the horizon. The horizon scale $R(t)$ expands, the energy inside the horizon remains the same (within the volume $\propto R(t)^3$), and hence our BCM behaves like CDM. In this case, BCM is the collective motion of Goldstone bosons with the same magnitude of momentum. So, the whole motion may be called coherent motion.\index{BCM!collective motion}  
The BCM has the oscillation frequency $m_0$, \ie the vacuum value $\langle\phi\rangle$ oscillates with  frequency $m_0$. Bosons in the horizon have the same momentum. Because the BCM behaves like CDM, the $\phi$ particle momenta of BCM are negligible compared to  mass $m_0$ and we can set it zero in the homogenious cosmic background. But our Galaxy and Sun moves with zero velocity in the homogenious background, and hence the $\phi$ particle velocity can be taken as $-(\vec{v}_{\rm\, Galaxy}+\vec{v}_{\rm\, Sun})$, whose magnitude is usually taken as $\approx 220\,$km/s.\index{BCM!current velocity} If temperature is low enough, the dispersion due to the Boltzmann distribution is smaller than the dispersion due to the velocity of the measuring apparatus, which is practically of order $220\,$km/s.  
 
\subsubsection{Classical limits:}
The  classical limits are obtained by taking $\hbar\to 0$.\index{classical limit!WIMP}\index{classical limit!BCM}
But, there is a difference between  the WIMP and BCM dark matters, in taking another limit:  
\begin{eqnarray}
 {\rm WIMP:}&&E=\hbar\omega={\rm fixed},~{\bf p}=\hbar\bf{k} ={\rm fixed}\label{eq:WIMPLim}\\
 {\rm BCM:}&&E={\cal N} \hbar\omega={\rm fixed},~{\bf p}={\cal N} \hbar\bf{k}={\rm fixed},\label{eq:BCMLim}
\end{eqnarray}
where ${\cal N} $ is the number of bosons which is related to the phase space number density by ${\cal N} =\hbar n_a$,
\begin{eqnarray}
&&{\cal N} = \frac{2\pi^3 n(t)}{\frac{4\pi}{3}(m_a\delta v)^3}\simeq 2\cdot 10^{58} \left(\frac{f_a}{10^{11\,}\rm GeV} \right)^{8/3},\\
&&~~{\rm with}\quad \delta v(t)\simeq \frac{1}{m_a t_2}\,\frac{R(t_2)}{R(t)}.\nonumber
\end{eqnarray}
For the BCM case, viz. Eq. (\ref{eq:BCMLim}),  $\hbar\omega$ is extremely small for a very large ${\cal N} $. 
In this case, we can proceed to discuss the Bose--Einstein condensation if another condition (thermalization condition) is satisfied.
This difference of classical limits is stressed in  \cite{Sikivie17}, in relation to the caustic rings and the  angular momenta of spiral galaxies.

\subsubsection{Collective motion:}
The ground state  corresponds to the bottom level of \ref{fig:BCMenergy}\,(b). But Goldstone bosons created by the vacuum motion are not necessarily at the ground state. For the Bose-Einstein condensation to be realized, ``thermalization'' \index{thermalization} has to be effective.  Since the creation of   QCD axions, thermalization has not occurred  because\index{axion!thermalization}
thermalization requires statistical equilibrium. Creation of QCD axions by the $\langle a\rangle$ oscillation, by the condition $H=m_a/3$, during the QCD phase transition means that axions are not in the thermal equilibrium. At the creation time of axion quanta $a_{\rm quantum}$, the axion vacuum $\langle a\rangle$  is at the red bullet of Fig.  \ref{Fig:PotAxion}, with   momentum of $a_{\rm quantum}$ vanishing. The number density $n_a$ of $a_{\rm quantum}$ at the creation time is the required CDM energy density $\rho_a$ which is the value $V$ at the red bullet point, \ie  $\rho_a=n_a m_a$ with momentum ${\bf p}_a=0$. When the axion vacuum rolls down the hill, the axion quanta obtain momenta which are not necessaily correlated. In this situation, axions are called in the collective motion. This phenomenon on axion vacuum applies to any BCM.

\section{Models with discrete symmetries, global symmetries,  and non-Abelian gauge groups}\label{sec:GlAndNonA}

Discrete gauge symmetries are defined to be subgroups of gauge symmetries \cite{KraussWil88} in the top-down approach such that spontaneous breaking of the gauge symmetries to those discrete groups do not lead to any
unsatisfactory gravitational effects \cite{Barr92,Kam92,holman92}. Here, we argue that in the bottom-up approach also discrete symmetries can allow hairs and unsatisfactory gravitational effects do not arise   \cite{KimKyaeNam17}. The original worry,  originating from the danger of introducing global symmetries together with gravitational interactions, led to consider that some discrete symmetries might  be required not to be subgroups of global symmetries. String theory also is not favorable to global symmetries \cite{ChoiKimBk06}.

The observed fermions in the low energy effective theory are better to arise from chiral models. The observed standard model(SM) is a chiral model. There is another hidden-sector chiral model discovered recently \cite{KimPRD17}. These defined `chirality' only from gauge symmetries, without resorting to any global or discrete symmetries. Except these gauge models with the chiral spectra,
some global and discrete symmetries are still used in the low energy effective theories for the rationale of cold dark matter (CDM).  For WIMPs, the exact or almost exact discrete symmetries are needed. For collective motion of light bosons \cite{KimYannShinji}, the QCD axion is most popular toward a CDM candidate.

 \begin{figure}[!b]
\begin{center}
\includegraphics[width=0.55\linewidth]{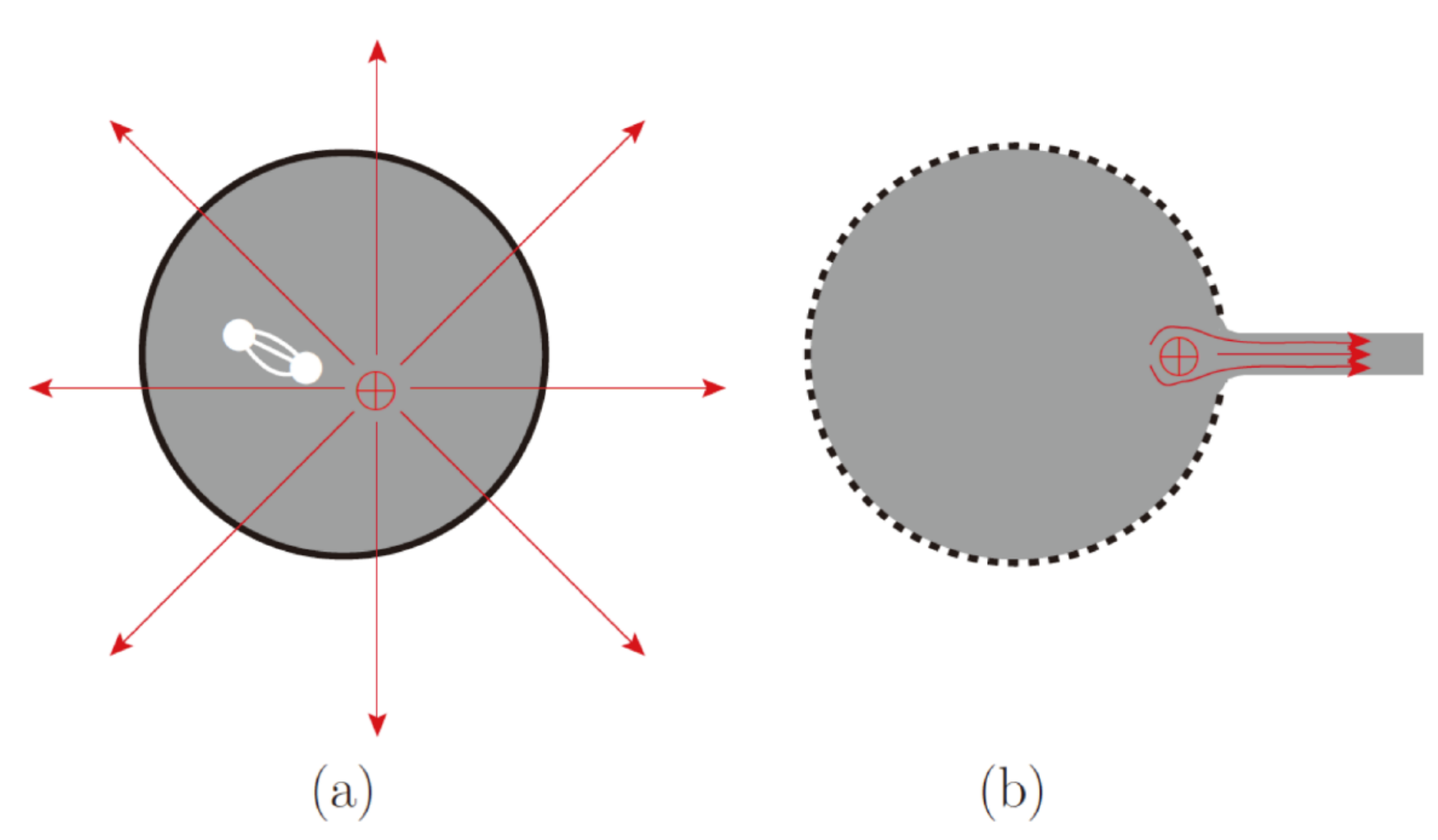} 
\end{center}
\caption{ A charged black hole. } \label{fig:ChBH}
\end{figure}

Black holes are known to have hairs of gauge symmetries. These are field lines ({\bf E} field) around the black holes. For example, consider a  charged black hole with + charge inside as shown in Fig. \ref{fig:ChBH}\,(a). The larger event horizon (out of two solutions) of the Reissner-Nordstr$\ddot{\rm o}$m black hole  occurs at
$
r_+=\frac12\left(r_S+ \sqrt{r_S^2-4r_Q^2} \right)\label{eq:RNhor}
$
where $r_S$ is the Schwarzschild radius without the charge. The event horizon takes into account the energy inside it. It is the black hole of graviton fields, for the graviton fields to be impossible to go out of the horizon. It is because graviton couples only to positive energy, and hence graviton fields leave and end at positive energy sources as shown in Fig. \ref{fig:ChBH}(a) with white lines. On the other hand,  U(1) gauge field lines leave + charges and end at -- charges.  Within the closed boundary, enclosing a net + charge shown in Fig. \ref{fig:ChBH}\,(a), the electromagnetic field cannot end inside the blackhole horizon as shown with red arrows.  The {\bf E} field lines go out of the horizon of the gravity field, which looks like hairs from the black hole.  Namely, when one considers the mass of a  charged particle inside a black hole, it includes the outside field energy also. So, to consider the charged particle energy just inside the horizon, one should subtract the electromagnetic field energy permeating from the horizon to infinity.  In Fig. \ref{fig:ChBH}\,(b), one can move the {\bf E} field lines to a small pinched hole such that at most of the surface of the sphere the horizon of graviton fields is also the horizon of the gauge fields except at the pinched hole. Through this hole,  the {\bf E} field lines can be connected to the outside from the black hole.
 
 \begin{figure}[!b]
\begin{center}
\includegraphics[width=0.8\linewidth]{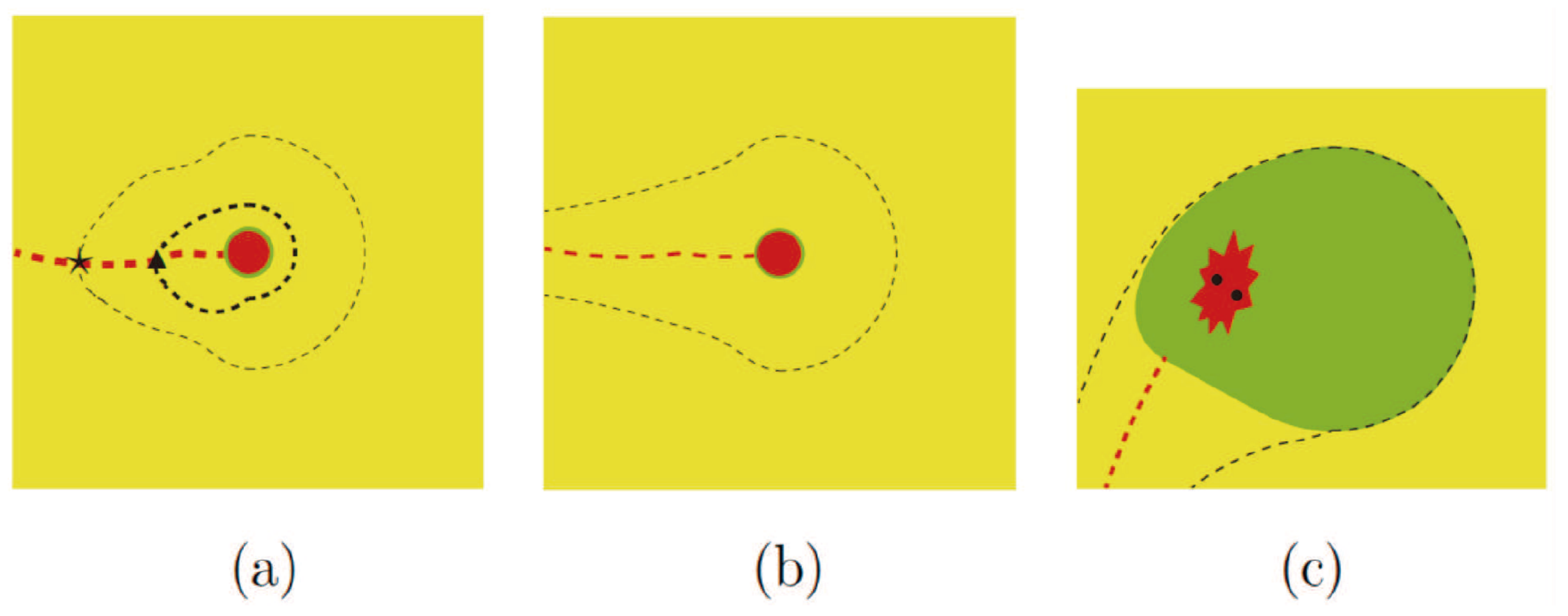} 
\end{center}
\caption{Closed surfaces around a tadpole, (a) two different surfaces, (b) hairs packed to a bundle, and (c) closing in the boundary near the tadpole. } \label{fig:TadBound}
\end{figure}
 
Discrete symmetries can allow non-equivalent vacua if they are distinguished by VEVs of scalar fields. In the evolving Universe, the discrete symmetries spontaneously broken by the VEVs of spin-0 fields create domain walls \cite{Okun74}. Consider $\Z_2$ and its two distinct vacua, which will be colored  yellow $(q=0)$ and red $(q=1)$ in Fig. \ref{fig:TadBound}. Between them, there is a domain wall as shown in Fig.  \ref{fig:TadBound}\,(a), (b), and (c). The domain wall is colored limegreen. One can consider discrete charges: sum of the charges of particles and  vacuum. Suppose, the yellow vacuum carries 0 discrete charge. If we enclose the particles in the red vacuum by a closed surface, there can be a definite discrete quantum number. Let us consider  the total discrete charge within the closed surface passing through the star in Fig. \ref{fig:TadBound}\,(a).
Moving the boundary of this closed surface to the closed surface passing through the triangle, the discrete charge inside the surface remains the same. Therefore, one can define a `tadpole' of discrete charges, through the tail of which a discrete quantum number goes out, as shown in Figs.  \ref{fig:TadBound}\,(b). Even this closed surface can be moved to touch the domain wall as shown in Figs.  \ref{fig:TadBound}\,(c).  The dashed lines end at the tadpole or at an infinity point. This tadpole works like flux lines discussed in Fig. \ref{fig:ChBH}. Thus, it was argued that the discrete symmetry is not broken by wormholes in the bottom-up approach \cite{KimKyaeNam17}.
  
We have discussed continuous parameters in spontaneously broken global symmetries and their manifestations as pseudo-Goldstone bosons. The pseudo-Goldstone boson mass $m_{\rm pseudo}$ corresonding to a global U(1) depends on the explicit breaking scale $\delta^4$, the red part in Fig. \ref{Fig:discrete}, and the decay constant  $f$: $\delta^2/f$. The explicit symmetry breakings are broadly distinguished to two classes: (i) by small terms in the potential $V$ and (ii) by non-Abelian gauge interaction. The case (i) can be studied for the Yukawa couplings ${\cal L}_Y$ and terms in $V$. If all terms in  ${\cal L}_Y$  and $V$  respect  the global symmetry, then loop contributions cannot break the global symmetry. So, the breaking effect must be considered at the tree level.
The case (ii) is known as the instanton effect  and the $\thb$ direction discussed in Subsect. \ref{subsec:Inv} is the pseudo-Goldstone boson direction. This breaking arises at the one loop level as the (global symmetry)--(non-Abelian group)$^2$ anomaly.

Here, a few BCMs are commented. For the N-flation and \Ude, the breaking terms correspond to the case (i),  obtaining contribution only from $V$. For the ``invisible'' QCD axion, it belongs to the case (ii) and there should be no contribution from $V$. We discussed the top-down scenario for obtaining \UPQ~from \Uanom~global symmetry from string compactification. For this to remain as a solution of the strong CP problem, however, QCD should be the  only unbroken non-Abelian gauge group.  The reason is the following.

We will present in (\ref{eq:twoNonA}) an effective potential of two axions $a_1$ and $a_2$ with two non-Abelian groups below their confining scales. Note that $p,q,h,$ and $k$ are parameters given in the model and $f_1$ and $f_2$ are two decay constants (or VEVs of scalar fields). Diagonalization of the mass matrix of $a_1$ and $a_2$ gives the heavy and light pseudoscalar masses presented in Eqs. (\ref{eq:HLamasses})  and (\ref{eq:AaBin}) \cite{KimFRP15}. Here, we take different limits of parameters from the KNP inflation.
Suppose that there are hierarchies $f_2\gg f_1$ and $\frac{\Lambda_2^4}{\Lambda_1^4}\gg \frac{f_2^2}{f_1^2}$. Then, the heavy and light masses are
\dis{
 m_H^2\simeq \Lambda_2^4\left(\frac{h^2}{f_1^2} +\frac{k^2}{f_2^2} \right),~~m_L^2\simeq \Lambda_1^4\left(\frac{(pk-qh)^2}{k^2f_2^2+h^2f_1^2} \right).\label{eq:Mtwoa}
}
From Eq. (\ref{eq:Mtwoa}), we note that the decay constant of the heavy pseudoscalar is the smaller decay constant $f_1$ and  the decay constant of the light pseudoscalar is the larger decay constant $f_2$. So, if we try to have $\Lambda_1\to\Lambda_{\rm QCD}$, then we must choose the larger decay cosntant which can be a GUT or string scale. So, we fail in obtaining an intermediate scale $f_a$ from the \Uanom~global symmetry for the ``invisible'' axion. For the \Uanom~global symmetry to lead to the ``invisible'' axion, we should not have any other non-Abelian gauge group above the TeV scale  from string compactification. If another confining gauge group above the TeV scale  is introduced, then a scenario must be introduced such that it is broken after achieving the objective \cite{BarrKim14}.

\section{A BCM as dark energy: Quintessential axion }
\label{sec:axionde}

As for the QCD axion case, the DE scale can arise via a BCM of Fig. \ref{Fig:discrete} \cite{KimNilles14}.   The global symmetry violating terms belong to the red part in Fig. \ref{Fig:discrete}.   In the Higgs portal scenario, the BCM pseudoscalar for DE must couple to the color anomaly since it couples to the Higgs doublets and the Higgs doublets couple to the SM quarks. On the other hand,  mass of the BCM pseudoscalar for DE is in the range $10^{-33}\sim 10^{-32}\,\eV$ \cite{Carroll98}. Therefore, the QCD anomaly term of the BCM must be forbidden to account for the DE scale of $10^{-46\,}\gev^4$. There must be a global symmetry free of the QCD-anomaly. It is shown that such two global U(1) symmetries are achieved in general \cite{KimDE14}. 
Out of the two U(1)'s, let us pick up the global symmetry  \Ude~for the  DE BCM. The other U(1), carrying the color anomaly, is for the ``invisible'' axion. Now, we can take the following window for the axion decay constant \cite{Bae08,Kawasaki12},
\begin{equation} 
10^{9\,}\gev<f_a<10^{11\,}\gev.\label{eq:fawindow}
\end{equation}
On the other hand, the \Ude~breaking scale $\fde$ is of order of the Planck scale \cite{Carroll98},
\begin{equation} 
 \fde\approx \Mp.\label{eq:fde}
\end{equation}

From string compactification, we argued that the \Uanom~global symmetry via the 't Hooft mechanism is for the ``invisible'' QCD axion \cite{Kim88,KimKyaeNam17} whose decay constant $f_a$ corresponds to   (\ref{eq:fawindow}). The global symmetry \Ude~cannot have a QCD anomaly \cite{KimDE14} and in the compactification of the heterotic string it must be anomaly free since there is only one \Uanom~from the heterotic string. Thus, the VEVs $\fde$ and  $f_a$ breaking \Ude~and \Uanom~global symmetries, respectively, are clearly distinguished in the compactification of the heterotic string. But, their hierarchical values given in Eqs. (\ref{eq:fde}) and (\ref{eq:fawindow}) are not considered to be a fine-tuning problem as pointed out in Fig. \ref{fig:DWMIax}.

\begin{figure}[!t]
\centerline{\includegraphics[width=12cm]{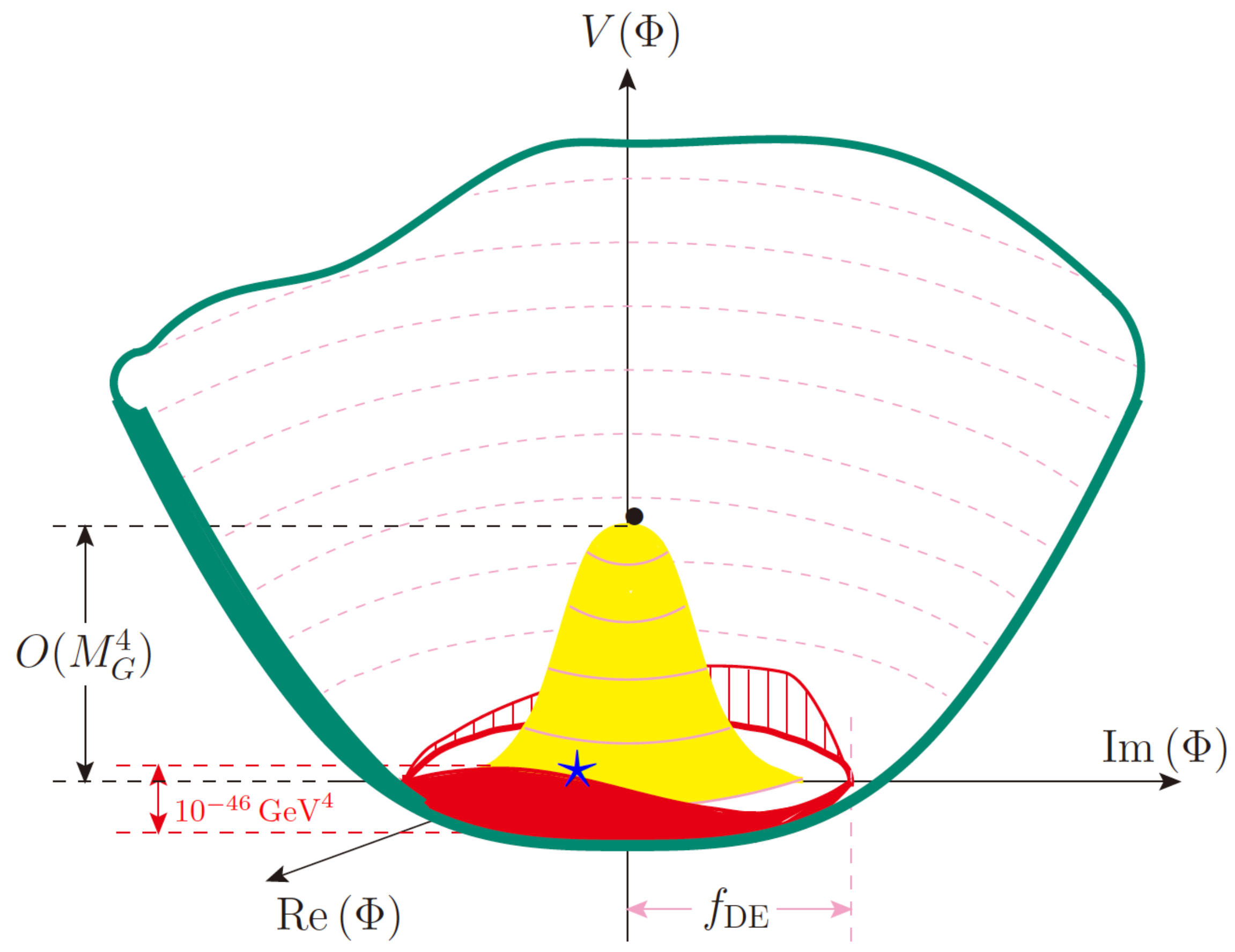}
 }
\caption{The DE potential in the red 
angle-direction in the valley of radial field of height  $\approx \Mg^4$.}\label{Fig:DEV}
\end{figure}
     
 It is known that the anti-symmetric tensor field $B_{MN}$ in string theory leads to  string axions  having GUT scale decay constants \cite{ChoiKimfa85,Svrcek06}. So, it was argued in the SUSY framework that the ``invisible'' QCD axion from string theory is better to arise from matter fields  \cite{KimPRL13}. 
 
By introducing  two global symmetries, we can remove the \Ude--SU(3)$_c$--SU(3)$_c$  where SU(3)$_c$ is QCD and the \Ude~charge is a linear combination of two global symmetry charges. The decay constant corresponding to \Ude~is $\fde$.  Introduction of two global symmetries is inevitable to interpret the DE scale and hence in this scenario the appearance of \UPQ~is a natural consequence. The height of DE potential  is so small, $10^{-46\,}\gev^4$, that the needed discrete symmetry breaking term of Fig. \ref{Fig:discrete} must be small, implying the discrete symmetry is of high order.  
 For the QCD axion, the height of the potential for the ``invisible'' axion is $\approx\Lambda_{\rm QCD}^4$. For the DE pseudo-Goldstone boson, the height of the potential of the radial field is $\approx\Mg^4$, according to the Higgs portal idea,  shown at the central top in  Fig. \ref{Fig:DEV}. With \UPQ\,and \Ude, one can construct a DE model with the potential in the valley in  Fig. \ref{Fig:DEV} from string compactification \cite{KimDE14}. Using the SUSY language, the discrete and global symmetries below $\Mp$ are the consequence of the full superpotential $W$. So, the exact discrete symmetries related to string compactification are respected by the full $W$, \ie the vertical column of Fig. \ref{Fig:discrete}. For \Ude, the global symmetry is not exact and   the red part at the tree level contributes.\footnote{For the \UPQ, the symmetry is better to be exact or almost exact toward the solution of the strong CP problem with $|\thb|<10^{-10}$ as discussed in Subsect. \ref{subsec:Inv}.} Definition of the symmetry \Ude~in the lavender part is at a high dimensional level and the breaking term in the red part is even more high dimensional \cite{KimDE14}. A typical  discrete symmetry ${\bf Z}_{10\,R}$  is considered in \cite{KimDE14}, whose level is high enough such that the defining and breaking terms appear at the high dimensional levels. The ${\bf Z}_{10\,R}$  charges descend from a gauge U(1) charges of the string compactification \cite{Kim13worm}.   In this scheme with the Higgs portal, we introduced three scales for the VEVs, TeV scale for $H_uH_d$, the GUT scale \Mgt~for singlet VEVs, and the intermediate scale for the ``invisible'' QCD axion. The other fundamental scale is $\Mp$. The trans-Planckian decay constant $\fde$ of (\ref{eq:fde}) can be a derived scale, which can be applied also to the inflation \cite{KNP05}.
 
\section{ULA}
\label{sec:ULA}

\begin{figure}[!t]
\includegraphics[width=1\textwidth]{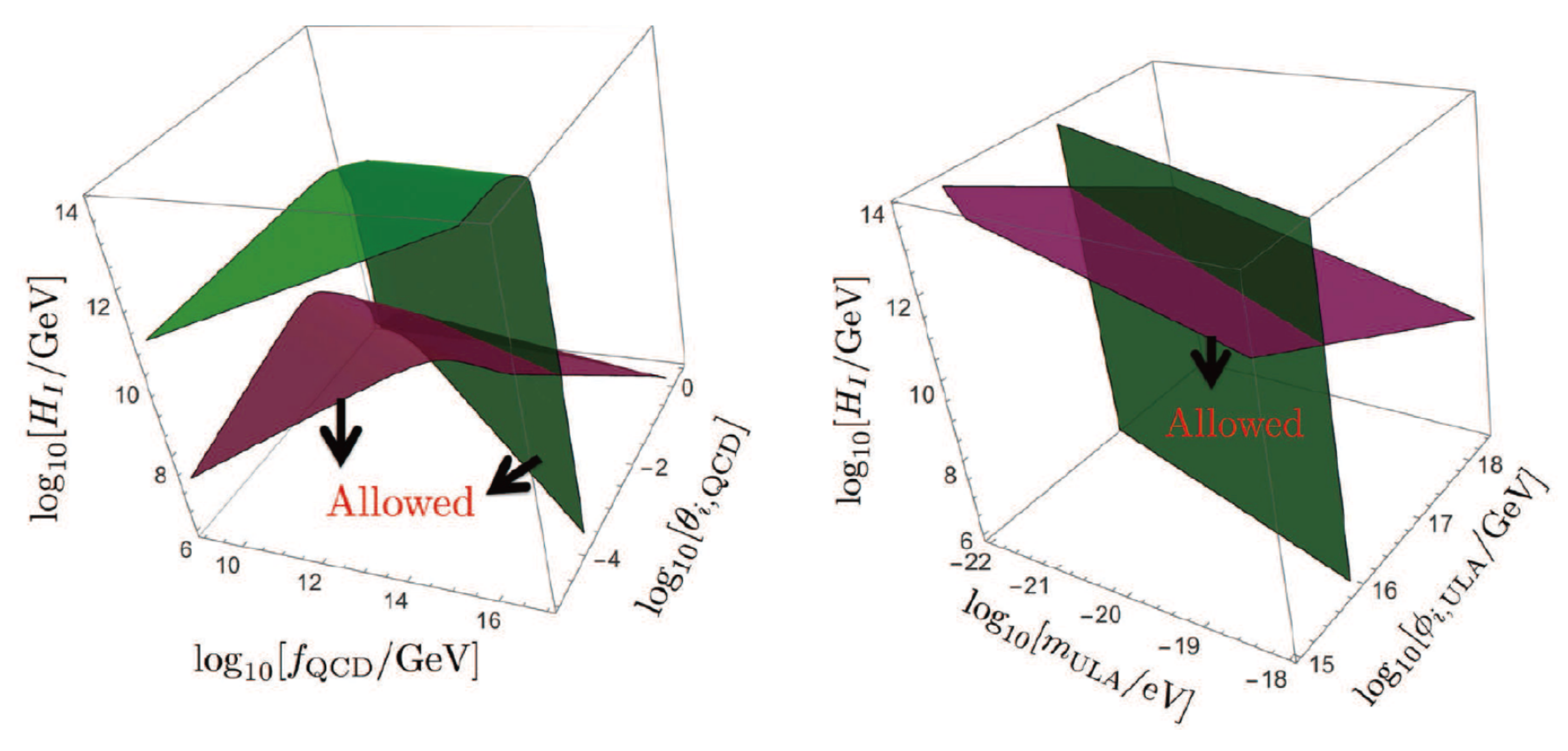}
  \caption{Parameters allowed for 90\,\% and  10\,\% of CDM by ULA and  ``invisible'' axion, respectively. The green bound is from the energy density and the purple bound is from the isocurvature. Left panel for QCD axion: All $f_a\le 10^{17\,}\gev$ can be accomodated.   Right panel for ULA: Isocurvature perturbations constrain $H_{\rm infl}\lesssim 10^{12\,}\gev$ can be accomodated.   Figure taken from Ref.  \cite{MarshKim}. } \label{fig:ULAax} 
\end{figure}

Cold dark matter suffers from the ``small-scale crises'' \cite{SmallCrises85}  which stem from the scale invariance of CDM structure formation.  There are several ideas for alleviating the problem on   the ``small-scale crises'' such as  by the effects of star formation \cite{StarFormSmallCr} and by warm dark matter scenario \cite{WDMsmallCr,Turok01}.  Ultra light axion (ULA), a kind of pseudo-Goldstone boson, also offers an elegant solution \cite{Hu00,Marsh14}. ULAs differ
from CDM essentially because of the large de Broglie
wavelength, which imprints a scale on structure formation, suppressing linear density perturbations \cite{Marsh10,Noh11}. An example with two pseudo-Goldstone bosons, an ULA with mass $10^{-22}$\,eV constituting 90\,\% of CDM mass and QCD axion constituting 10\,\% of CDM mass, has been studied without   the ``small-scale crises'' \cite{MarshKim}.
We required the following cosmological scenario:
\begin{itemize}
\item[(i)] \UPQ~and \ULA~are broken during inflation.
\item[(ii)] $m_{\rm ULA}\approx 10^{-22\,}$eV.
\item[(iii)] $f_{\rm ULA}\approx 10^{17\,}\gev$.
\item[(iv)] $10^{9\,}\gev<f_{\rm QCD}< 10^{14\,}\gev$. This allows the ``invisible'' QCD axion to compose about 10\,\% of the DM with fine-tuning no worse than $|\thb|\ge 10^{-2}$.
\item[(v)] $H_{\rm infl}< 10^{9\,}\gev$ which is the maximum Hubble scale allowed by isocurvature if the ``invisible'' QCD axion is not fine-tuned.
\end{itemize}     
In Fig. \ref{fig:ULAax}, we present the allowed ``invisible'' axion and ULA parameter regions.
 
Now, there is a question on possible scales for the ULA parameters. The basic philosophy is  again	 Fig. \ref{Fig:discrete} where the underlying symmetry is a discrete symmetry allowed in some ultra-violet completed theory.
Withiout requiring another fine-tuning,\footnote{For the decay constant of ``invisible'' QCD axion, we accept a fine-tuning as in the SM. Maybe, supersymmetry breaking can accompany this ``invisible'' axion scale \cite{KimScale84} or one may use some kind of intermediate scale string theory \cite{Quevedo99}.} the ULA decay constant $f_{\rm ULA}$ is assumed at the Planck or GUT scale. 
Then, the question is what is the ULA breaking scale $\Delta V $. The first guess is a non-Abelian gauge group SU(2)$_W$ in the SM. In this case, the magnitude of the height of $\Delta V $ is  of order
\dis{
\Delta V_{\rm W}\approx e^{- 2\pi /\alpha_2(\Mg)} \Mg^4\simeq  (0.592\cdot 10^{-68} \,\Mp^4)\simeq  (0.22\,\gev)^4,
}
where we used $\Mp=2.5\times 10^{16\,}\gev$ and $\alpha_2(\Mg)\simeq 1/25$. For $f_{\rm ULA}\simeq \Mp$, the ULA mass would be $4\times 10^{-12\,}$eV. It is ten orders of magnitude off from  the needed mass of $10^{-22\,}$eV. 

This leads us to consider another possibility for generating a shallow $\Delta V$. Starting from a discrete symmetry $\Z_4\times\Z_3$, a simple model was constructed \cite{MarshKim}. The first consideration is to consider a defining interaction term which is the lowest order term in $V$.  Since the ``invisible'' axion is in the SM singlet field, SUSY is a prototype example to couple $\sigma^2$ to $H_uH_d$ by the so-called the $\mu$-term \cite{KimNillesPLB84}.[Without SUSY, a fine tuning of order $10^{-18}$ is needed between the mass parameters in $V$.] The superpotential $W$ gives the information on discrete and global symmetries. The lavender part of Fig. \ref{Fig:discrete} respects the  discrete symmetry $\Z_4\times\Z_3$. The charges in the example are  presented in Table \ref{tab:PQULA}. Both global symmetries \UPQ~and \ULA~are broken by the red parts, by the QCD anomaly for \UPQ~and not by  the QCD anomaly for \ULA.  Out of two directions, only one combination can be broken by the QCD anomaly. So, to introduce \ULA~in general, \UPQ~must be introduced. Toward a solution of the strong CP problem, we require that \UPQ~is not broken (or very feebly if broken) by terms in $W$ but is required to be broken by the nonvanishing color anomaly. Quantum numbers of  Table  \ref{tab:PQULA} fulfil these requirements. We take the VEV of $X_1$ to fix the decay constant of the
QCD axion, $f_a/\sqrt2=\langle X_1\rangle\approx O(10^{11}\gev)$.  

\begin{table}[!t]
\caption{A set of global charges for vanishing ULA-color anomaly with  $\Z_4\times\Z_3$. $\Z_{12}$ charges unifying $\Z_4\times\Z_3$ are also shown.}
{\begin{tabular}{@{}lccccccc@{}} \toprule
   & $q_L$  & $u_L^c$& $d_L^c$& $H_u$& $H_d $ & $X_1$ & $X_2 $  \\ \colrule
Q$_{\rm PQ}$ &$1$&$1$&$1$&$-2$&$-2$& $2$ &0 \\[0.5em]
Q$_{\rm ULA}$ &$1$&$-3$&$1$&$2$&$-3$& $0$ &1 \\[0.1em] \colrule
${\bf Z}_{4}$ &$1$&$-3$&$1$&$2$&$-3$& $0$ &$-3$ \\[0.5em]
${\bf Z}_{3}$ &$2$&$0$&$0$&$1$&$1$& $2$ &$0$ \\[0.5em]  
${\bf Z}_{12}$ &$5$&$9$&$9$&$10$&$1$& $8$ &$9$ \\[0.1em] \botrule
\end{tabular} \label{tab:PQULA}
}
\end{table}
  
The approximate global symmetry  U(1)$_{\rm ULA}$ is broken by some $\Delta V$ if we consider all the terms in $W$ allowed by the discrete symmetries.
 U(1)$_{\rm ULA}$ is broken by
 \dis{
 \frac{H_uH_d X_1^2X_2^5}{M_{\rm UV}^6}\label{eq:breakULA}
 } 
which respects the discrete symmetry of Table \ref{tab:PQULA} and the \UPQ~symmetry, but breaks the  U(1)$_{\rm ULA}$ symmetry
of Table  \ref{tab:PQULA}. Because of high powers of fields in Eq. (\ref{eq:breakULA}) due to a large $\Z_N$ ($N=12$), it is possible to make $\Delta V$ sufficiently small to make the ULA mass in the region of $10^{-22\,}$eV \cite{MarshKim}.
  
\section{Gravity waves from inflation}
\label{sec:Inflation}

\begin{figure}[!t]
\centerline{\includegraphics[width=11cm]{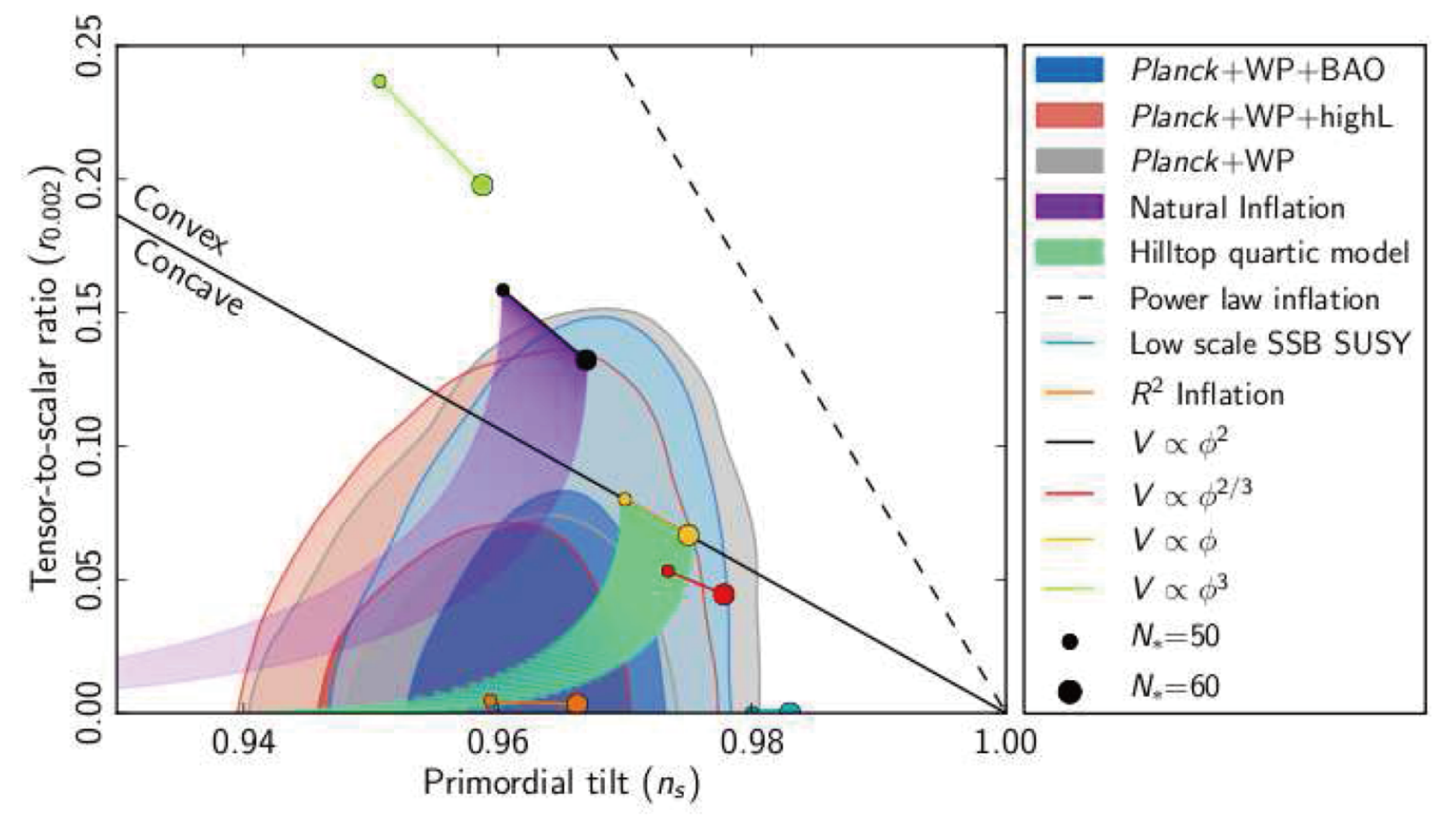}
 }
\caption{Tensor to scalar ratios $r$ for several models, presented in \cite{NillesCorfu17}.}\label{Fig:ratior}
\end{figure}

The inflationary idea \cite{LeMaitre31} was so attractive in understanding the horizon, flatness, homegeneity and isotropy problems \cite{Guth81}, numerous inflationary models have been considered since the early 1980's \cite{Linde82,Albrecht82,LindeChaotic}. Even there exists a calculable framework \cite{Hawking82,GuthPi82}  in inflationary models for the ``quantum'' density perturbation \cite{Mukhanov81}.  For some time, the chaotic inflation \cite{LindeChaotic} attracted a great deal of attention because it can lead to a large tensor/scalar ratio, $r$, in particular with the earlier report of the BICEP2 collaboration \cite{BICEP14}. Figure \ref{Fig:ratior} shows predictions of $r$ for several potentials and the limits given by the Planck data.
  
The inflationary models need almost flat potentials \cite{Lyth93}. The logarithmic form near the origin is very flat.  The $\mu^2\phi_I^2$ potential with very small $\mu^2/\Mp^2$ is very flat even at some trans-Planckian values of $\phi$. Sinusoidal forms are almost flat near the top of the potential. The magnitude of $\Delta T/T$ measured by COBE excluded the logarithmic form among the new inflationary scenarios \cite{COBE91}.  It seems that the $\phi_I^2$ inflation, where $\phi_I$ is the inflaton, is  ruled out by more than $1\,\sigma$ by the refined $r$ measurements \cite{BICEP16}, as shown in Fig. \ref{Fig:ratior}. 

Thus, the sinusoidal forms are the remaining attractive possibility which belongs to the class of ``natural'' inflation \cite{Freese90}. These sinusoidal forms of  $a_I$ appear in the potential as\footnote{If the symmetry breaking is only by the non-Abelian anomaly, then we obtain $\delta=0$.}   
\begin{eqnarray}
V\propto 1-\frac12(e^{i\delta}e^{ia_I/f_I} +{\rm H.c.})=1-\cos\left(\frac{a_I}{f_I}+\delta\right).
 \end{eqnarray} 
So, near $\langle a_I\rangle=f_I(\pi+\delta)$, the potential is very flat and the natural inflationary potential works for inflation. 

This is one of ``hill-top'' inflationary models. Lyth noted that, for a large $r$, the field value $\langle\phi\rangle$ must be larger than $15\,\Mp$, which is known as the Lyth bound \cite{Lyth97}.  So, the natural inflation needs the decay constant at a trans-Planckian scale \cite{KNP05}. Also, a trans-Planckian scale is introduced for a quintessential DE \cite{Carroll98}. Introducing a natural inflation with a  trans-Planckian decay constant is possible if one considers two spontaneously broken global symmetries \cite{KNP05}, which is called the Kim--Nilles--Peloso (KNP) models. With two confining non-Abelian gauge groups, the global charges can be assigned such that an enough trans-Planckian decay constant results. 

This is illustrated with an effective potential of two axions.  Let us consider two  non-Abelian gauge groups with scales $\Lambda_1$ and $\Lambda_2$ and two pseudoscalars $a_1$ and $a_2$ resulting from breaking two global symmetries. The effective potential of $a_1$ and $a_2$ below the confining scales is \cite{KNP05}
\begin{eqnarray}
V\propto \Lambda_1^4\left(1-\cos\left[p\frac{a_1}{f_1}+q\frac{a_2}{f_2} \right]\right) +\Lambda_2^4\left(1-\cos\left[h\frac{a_1}{f_1}+k\frac{a_2}{f_2} \right]\right)\label{eq:twoNonA}
\end{eqnarray}
where $p,q,h,$ and $k$ are parameters given by the model and $f_1$ and $f_2$ are two decay constants (or VEVs of scalar fields). Diagonalization of the mass matrix of $a_1$ and $a_2$ gives the heavy and light pseudoscalar masses as \cite{KimFRP15},
\begin{eqnarray}
m_H^2=\frac12(A+B),~m_L^2=\frac12(A-B),\label{eq:HLamasses}
\end{eqnarray}
where
\begin{eqnarray}
A=\frac{p^2\Lambda_1^4+h^2\Lambda_2^4}{f_1^2}+\frac{q^2\Lambda_1^4
+k^2\Lambda_2^4}{f_2^2},~~B=\sqrt{A^2-4(pk-qh)^2
\frac{\Lambda_1^4\Lambda_2^4}{f_1^2f_2^2}}.\label{eq:AaBin}
\end{eqnarray}
If $pk=qh$, there results a massless mode with $m_L=0$. Then, for $pk=qh+\Delta$, there can result an effectively large decay costant $f_L$ as shown in Ref. \cite{KNP05}. For simplicity, we can glimpse this phenomenon for the parameters $\Lambda_1=\Lambda_2=\Lambda, f_1=f_2=f$ and $p=q=h=k$ \cite{KimFRP15},
\begin{eqnarray}
 f_{L}\simeq \frac{2|p|}{|\Delta|}f.\label{eq:falight}
\end{eqnarray}
In this two axion case, the height of the pontential is increased roughly to $2\Lambda^4$ and the decay constant $f_L$ is increased by the level of  the quantum number discrepancy, $|p/\Delta|$.
 
However, for inflation, not needing the vacuum angle to be zero, the explicit breaking terms of two global symmetries need not arise from gauge anomalies but can arise from $\Delta V$ in the red part of Fig. \ref{Fig:discrete}.
This possibility generalized the KNP mechanism to N spontaneously broken global symmetries under the name ``N-flation'' \cite{Nflation06} and has been studied extensively in string theory \cite{Silverstein14,Mirbabayi15}. The $N$-flation attempted to soften the problem  $|p/\Delta|\gg 1$ of the KNP model but the height of the potential increases by a factor $N$.

Nevertheless, this idea of increasing $N$ has been further pursued in obtaining a kind of cosmological generation of the weak scale from axion-type potentials, the so-called ``relaxion'' idea \cite{Graham15}, which however seems not completely satisfactory \cite{ChoiK16,Kaplan16,Im16,Flacke17}. 
  
The hilltop potential of Fig. \ref{Fig:DEV} is a Mexican hat potential of \Ude, \ie obtained from some discrete symmetry, allowed in string compactification \cite{KimNilles14}. The discrete symmetry may provide a small DE scale. The trans-Planckian decay constant, satisfying the Lyth bound, is obtained by a small quartic coupling $\lambda$ in the hilltop potential $V$. The requirement for the vacuum energy being much smaller than $\Mp^4$ is achieved by restricting the inflaton path in the radial direction in the hilltop region, $\langle\phi \rangle \lesssim \fde$, to converge to an appropriate phase direction of Fig. \ref{Fig:DEV}.   

\begin{figure}[!t]
\centerline{\includegraphics[width=9cm]{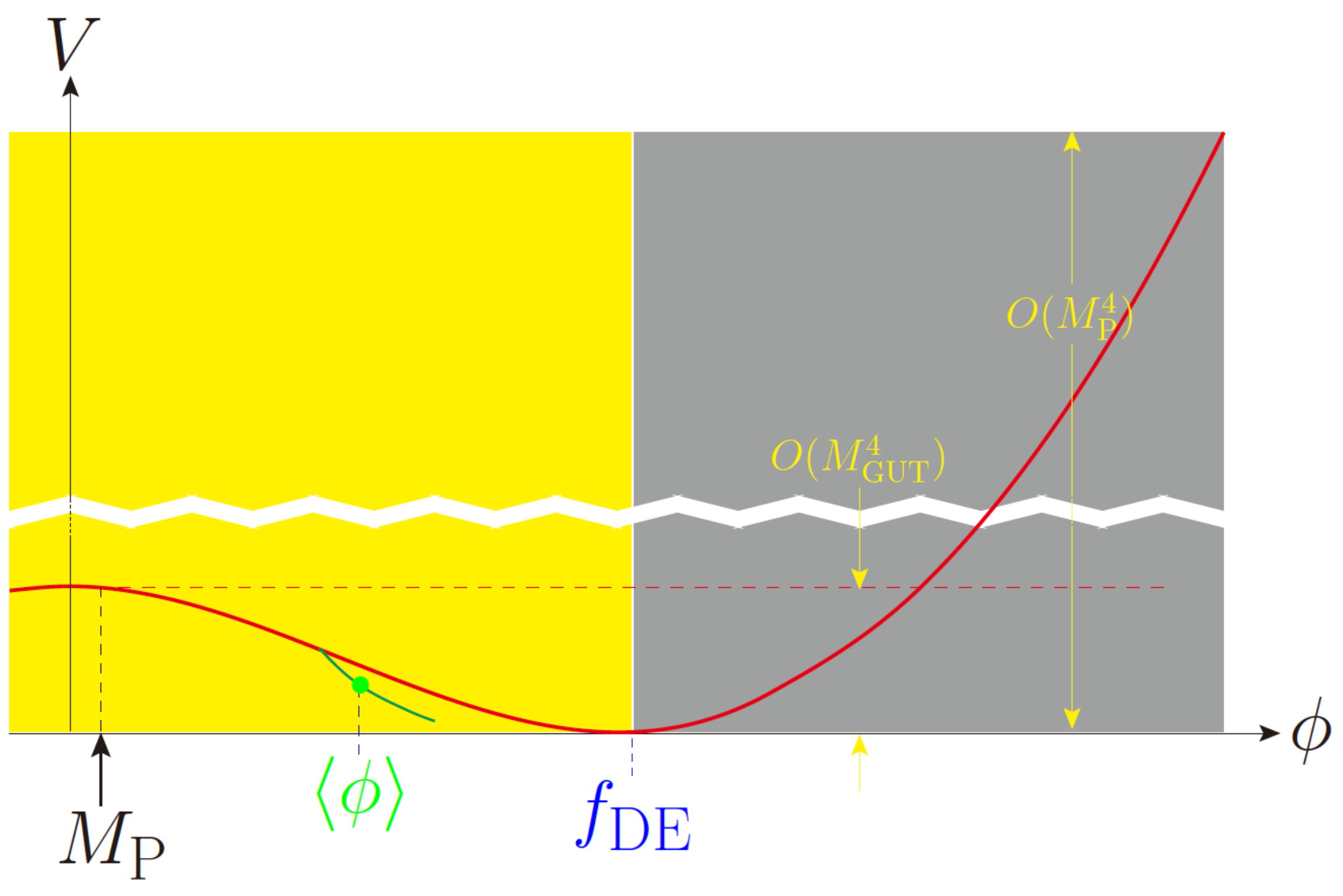}
 }
\caption{The trans-Planckian decay costant in the hilltop inflation.}\label{Fig:Trans}
\end{figure}

We can compare this hilltop inflation with the $\mu^2\phi^2$ chaotic inflation. The hilltop inflation is basically a consequence of discrete symmetries \cite{Kim13worm,KimNilles14,KimDE14}, allowed in string compactification. If some conditions are satisfied between the discrete quantum numbers of the GUT scale fields and trans-Planckian scale fields, the hilltop potential of Fig.  \ref{Fig:Trans} can result. On the other hand, the $\mu^2\phi^2$ chaotic inflation does not have such symmetry argument, and lacks a rationale, forbidding higher order $\phi^n$ terms.  

In contrast to the gravity waves generated during inflation, the gravity waves from  binary black hole coalescence observed recently \cite{GW1509,GW216} occured late in the cosmic time scale. In a recent observation of a coalescence, graviton mass is bounded as  $<0.77\times 10^{-22\,}\eV$ \cite{GW1701}. Related to our MI ``invisible'' axion from superstring, for the MI axion creation to be of any observable effect along the  coalescence, the black hole mass is required to be small, $\sim 2\times 10^{14\,}$kg \cite{KimJKPSgw}.

\section{Discussion and conclusion}
\label{sec:DisConl}

 We discussed the pseudoscalar particles used in cosmology systematically from  symmetry principles. These include QCD axions, quintessential axions, ULAs, and ALPs. Light pseudoscalars can be generically considered to be pseudo-Goldstone bosons.      Goldstone bosons $a_i$ are particle realization of continuous parameters $\alpha_i$ of spontaneously broken  global symmetries.   The spontaneous symmetry breaking scales are parametrized by the decay constants $f_i$. Any global symmetry is known to be broken at least by quantum gravitational effects, and hence all Goldstone bosons are called `pseudo-Goldstone' bosons. Starting from exact discrete symmetries, we may study a systematic classification of these approximate global symmetries, because compactification of an ultra-violet completed theory can lead to exact discrete symmetries.  This strategy is depicted in Fig. \ref{Fig:discrete}. The discrete symmetry (together with gauge symmetries in consideration) allows all possible terms 
 in the potential $V$ (or superpotential $W$ in case of SUSY extension). These terms are symbolized by the far left column in Fig. \ref{Fig:discrete}. If we consider only a few terms in $V$, as depicted as the lavender square, the discrete symmetry is enhanced to a global symmetry. All possible terms allowed by the global symmetry   is depicted as the bottom row in Fig. \ref{Fig:discrete}. The red colored terms in the far left column of  Fig. \ref{Fig:discrete} violates this global symmetry. Specifically, let us call this global symmetry U(1)$_{\rm global}$. The well-known example is the PQ symmetry. In Eq. (\ref{eq:WeinV}), let us keep all terms except the $c_{IJ}$ term; then the potential $V$ has a global symmetry which is the PQ symmetry. Considering fermions in the theory, the global symmetry may  have an anomaly through the fermion loops with some non-Abelian gauge group, say  $G$: U(1)$_{\rm global}$-$G$-$\tilde{G}$. If the global symmetry has this kind of gauge anomaly, it is depicted as the red columns not touching the bottom row of Fig. \ref{Fig:discrete}. The breaking strength by this anomaly term is represented by $\Delta\Lambda^4_G$.  Peccei and Quinn used QCD as $G$ and suggested the global symmetry for a solution of the strong CP problem. For the PQ mechanism  for a solution of the strong CP problem to work, there should be no red column terms in $V$, \ie the global symmetry must be exact or almost exact such that $\Delta V$ is sufficiently small compared to  $\Delta\Lambda^4_G$.
 
The mass scale of a pseudo-Goldstone boson $a_i$ is determined by $f_i$ and the explicit breaking terms $\Delta V\oplus \Delta\Lambda^4_G$. But, the zeroth rule is finding out the defining terms of the global symmetry in question, \ie the lavender part of Fig. \ref{Fig:discrete}. For example, SUSY extension of the SM with an ``invisible'' axion can define the PQ symmetry only at dimension 5 level by the $\mu$-term \cite{KimNillesPLB84}. If the global symmetry breaking term occurs at a sufficiently high order, then $\Delta V$ can be extremely small but $\Delta\Lambda^4_G$ is given just by the scale of $G$. This is the reason that, without allowing any non-Abelian anomaly, one can obtain a quintessential axion with mass of order $10^{-33}$eV and decay constant around the Planck scale, by defining a global symmetry by a sufficiently high order term in $V$ \cite{KimJKPS14}. Along this line  another example, obtaining a ULA of mass $10^{-22}$eV,  has been presented also \cite{MarshKim}. Note, however, that one must remove the color anomaly in these DE and ULA models, implying that \UPQ~must accompany these models.

Much of this review has been devoted to one example among these BCMs: the ``invisible'' QCD axion. This is because it acts importantly in particle physics (as a solution of the strong CP problem) and in cosmology (providing a CDM candidate in the Universe). Also, in experimental and theoretical physics,  the ``invisible'' QCD axion is important. It can be detected if it provides some portion of the CDM density in the Universe. Furthermore, an exact global symmetry can survive down to the intermediate scale through the anomalous U(1) symmetry from compactification of string theory. All these are discussed at some length together with the solution of the domain wall problem in axion models.
 
\section*{Acknowledgments}

J. E. K. and Y. K. S. are supported in part by Institute of Basic Science of Korea (IBS-R017-D1-2017-a00). J.E.K. is also supported in part by the National Research Foundation (NRF) grant funded by the Korean Government (MEST) (NRF-2015R1D1A1A01058449).


\end{document}